\documentclass[11pt]{article}
\usepackage{amsmath}
\usepackage{amssymb}
\usepackage{times}
\usepackage{graphicx}

\begin{document}
\newtheorem{theorem}{Theorem}
\newtheorem{lemma}{Lemma}
\newtheorem{corollary}{Corollary}
\newtheorem{definition}{Definition}
\newtheorem{conjecture}{Conjecture}
\newtheorem{axiom}{Axiom}
\newtheorem{problem}{Problem}
\newcommand{\Real}{{\bf R}}
\newcommand{\tr}{{\mbox{tr}}}
\newcommand{\real}{{\bf R}}

\title{On Dark Matter, Spiral Galaxies, \\ and the Axioms of General
Relativity}
\author{Hubert L. Bray
\thanks{Mathematics Department, Duke University, Box 90320, Durham, NC  27708, USA, bray@math.duke.edu}}
\date{April 22, 2010}
\maketitle

\begin{abstract}

We define geometric axioms for the metric and the connection of a
spacetime where the gravitational influence of the connection may be
interpreted as dark matter.  We show how these axioms lead to the
Einstein-Klein-Gordon equations with a cosmological constant, where
the scalar field of the Klein-Gordon equation represents the
deviation of the connection from the standard Levi-Civita connection
on the tangent bundle and is interpreted as dark matter.

This form of dark matter, while not quantum mechanical, gives
virtually identical predictions to some other scalar field dark
matter models, including boson stars, which others have shown to be
compatible with the $\Lambda CDM$ model on the cosmological scale.
In addition, we quantify the already known fact that this scalar
field dark matter, unlike the WIMP model, is automatically cold in a
homogeneous, isotropic universe, sufficiently long after the Big
Bang.

With these motivations in mind, we show how this scalar field dark
matter, which naturally forms dark matter density waves due to its
wave nature, may cause the observed barred spiral pattern density
waves in many disk galaxies and triaxial shapes with plausible
brightness profiles in many elliptical galaxies.  If correct, this
would provide a unified explanation for spirals and bars in spiral
galaxies and for the brightness profiles of elliptical galaxies. We
compare the results of preliminary computer simulations with photos
of actual galaxies.

\end{abstract}

\section{Overview}

There are three ideas, each of independent interest, which, if
correct, would create a new connection between differential geometry
and astronomy, very much in the tradition of general relativity's
previous successes at describing the large-scale structure of the
universe.  We begin by discussing these ideas in general terms.

\underline{Idea 1}:  Natural geometric axioms motivate studying the
Einstein-Klein-Gordon equations with a cosmological constant.  The
Klein-Gordon equation is a wave type of equation for a scalar field
which we propose as a model for dark matter.  While this geometric
motivation is new, modeling dark matter with a scalar field
satisfying the Klein-Gordon equation is not (see \cite{JWLee} for a
survey of scalar field dark matter and boson stars).  Hence, ideas 2
and 3 apply to these other works as well.

\underline{Idea 2}:  Wave types of equations for matter fields, such
as the Klein-Gordon equation, naturally form density waves in their
matter densities because of constructive and destructive
interference, like waves on a pond, or the Maxwell equations for
electromagnetic radiation. Unlike waves on a pond or the Maxwell
equations, however, the group velocities of wave solutions to the
Klein-Gordon equation (with positive ``mass'' term) can be anything
less than the speed of light, and can be arbitrarily slow for
wavelengths which are long enough. This allows for the possibility
of gravitationally bound ``blobs'' of dark matter to form.  In this
paper we make some conjectures about these scalar field dark matter
density waves.

 \underline{Idea 3}:  Density waves in
dark matter, through gravity, naturally form density waves in the
regular baryonic matter.  In the case of disk galaxies, where
friction in the interstellar medium of gas and dust is important
\cite{BT}, we exhibit examples where barred spiral density wave
patterns form in the regular matter, as seen in figures 1, 2, 3, and
4. In the case of elliptical galaxies, where the interstellar medium
is believed to be mostly irrelevant \cite{BT}, we show how these
dark matter density waves tend to produce triaxial ellipsoidal
shapes for the regular matter with plausible brightness profiles, as
seen in figures 5, 6, and 7.

\begin{figure}\label{NGC1300}
   \begin{center}
   \includegraphics[height=75mm]{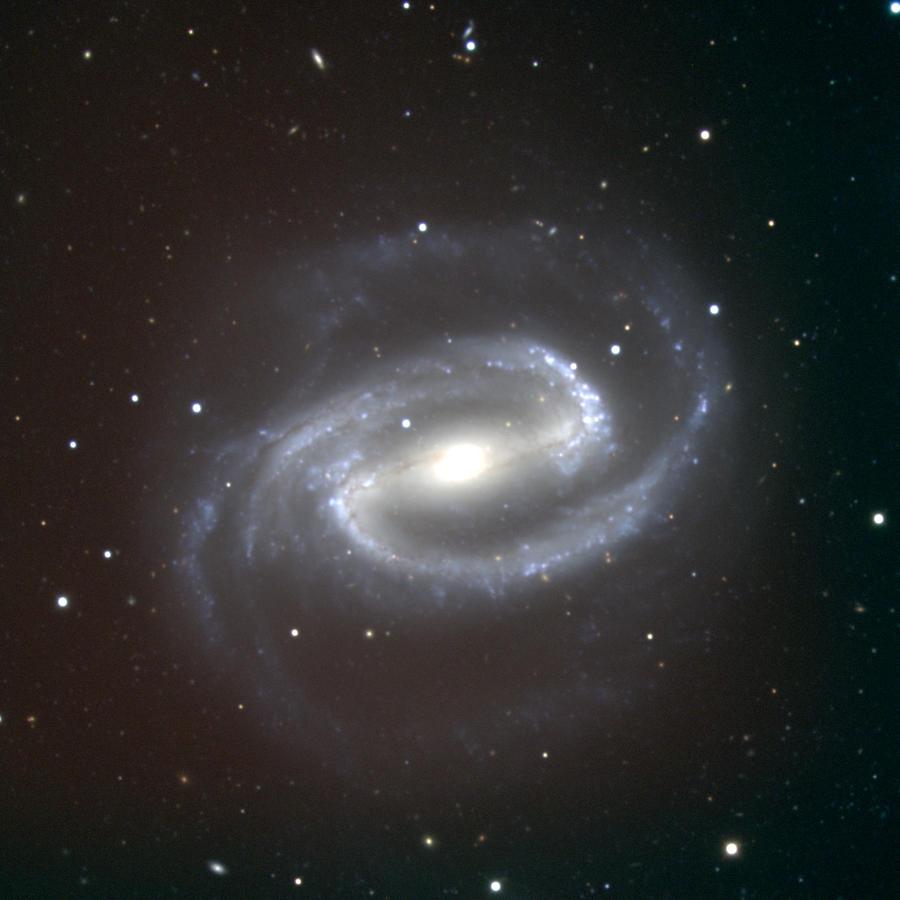}
   \includegraphics[height=75mm]{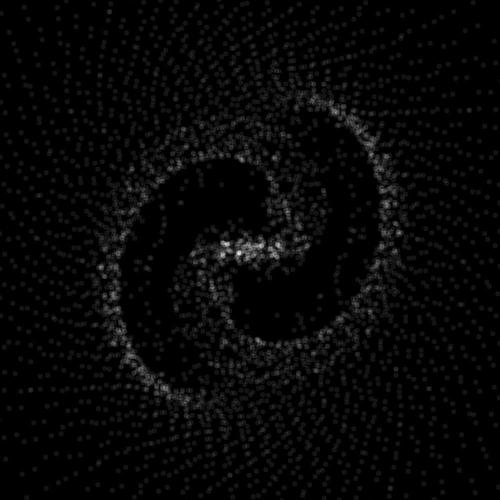}
   \end{center}
   \caption{NGC1300 on the left, simulation on the right.  The simulated image on the right results from running the Matlab function
   spiralgalaxy(1, 75000, 1, -1, 2000, 1990, 25000000, 8.7e-13, 7500, 5000, 25000000, 10000) described in section \ref{SGS1}.
   Left photo credit: Hillary Mathis/NOAO/AURA/NSF.  Date:  December 24, 2000. Telescope:  Kitt Peak National Observatory's 2.1-meter telescope.
   Image created from fifteen images taken in the BVR pass-bands. }
\end{figure}

This paper is an attempt to put the above three ideas together in as
precise a way as possible.  The rules of the game we choose to play
here are strict:  define a concise set of geometric axioms, and then
try to understand the implications of those axioms.  The axioms we
choose, stated in the next section, are more fundamental than
defining an action. Instead, we declare the spacetime metric and
connection on the tangent bundle as the fundamental objects of our
universe, and then define the properties that the action for these
two objects must have.  In this paper we make the case that the
simulated images in figures 1-6 as well as the computed brightness
profile described by figure 7 are all consequences of these axioms.

The resulting theory is a generalization of the vacuum Einstein
equations with a cosmological constant, which already famously
explains gravity, 73\% of the mass of the universe as dark energy
\cite{WMAPobservations}, the accelerating expansion of the universe,
black holes, and all other vacuum general relativity effects. If
successful, this generalized theory would then, by describing dark
matter, account for 95\% of the mass of the universe
\cite{WMAPobservations} and explain some portion of the structures
of galaxies.

\begin{figure}\label{NGC4314}
   \begin{center}
   \includegraphics[height=75mm]{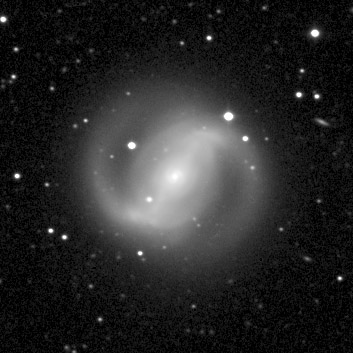}
   \includegraphics[height=75mm]{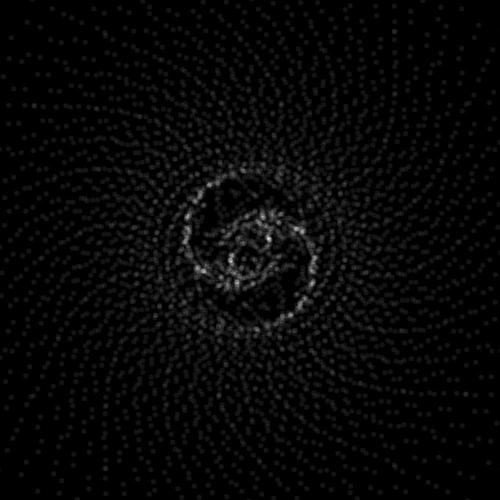}
   \end{center}
   \caption{NGC4314 on the left, simulation on the right.  The simulated image on the right results from running the Matlab function
   spiralgalaxy(1, 75000, 1, -0.15, 2000, 1990, 25000000, 8.7e-13, 7500, 5000, 30000000, 50000) described in section \ref{SGS2}.
   Left photo credit:  G. Fritz Benedict, Andrew Howell, Inger Jorgensen, David Chapell (University of Texas),
   Jeffery Kenney (Yale University), and Beverly J. Smith (CASA, University of Colorado), and NASA.  Date:  February 1996.
   Telescope:  30 inch telescope Prime Focus Camera, McDonald Observatory.}
\end{figure}

However, strictly speaking, quantum mechanics is not part of the
theory we propose here, as should be expected since general
relativity and quantum mechanics have yet to be unambiguously
unified.  Hence, our theory is clearly incomplete.  This is not so
bad considering that every theory known today is, in the strictest
sense, incomplete. However, it is still a reasonable question to
wonder if our theory does a good job of describing dark matter, even
if it does not describe regular particulate matter. Hence, as a way
of testing our dark matter model, we treat the remaining 4\% of the
mass of the universe, composed of particles of various kinds, in the
traditional manner similar to test particles, but with mass, which
arguably makes this theory compatible with $\Lambda CDM$ models, as
will be explained in section \ref{cosmologicalpredictions}.  The
simulated images in figures 1-6 are pictures of the effect of the
dark matter on the regular matter that we have sprinkled into the
theory. Finding a less contrived way of getting regular matter into
the next theory, while still respecting the idea of keeping our
axioms as simple as possible, is an important open problem

So, does the Klein-Gordon equation accurately describe dark matter
and predict some observed properties and structures of galaxies? In
this paper we present evidence of this possibility by trying to
understand the effect that this model of dark matter would have on
the structure of galaxies, which is a reasonable idea since galaxies
have large components of dark matter.

In doing so we have had to make approximations and educated guesses,
so the comparisons in figures 1-6, while encouraging, should be
taken in this context. Also, our ``simulations'' of galaxies only
simulate the effect of the dark matter on the regular matter and
hence are very primitive.  Perhaps a better name would be
``numerical experiments.'' However, one has to start somewhere, and
it is already interesting that compelling patterns very much
resembling actual galaxies have emerged.  We will describe the
models we have used in sections \ref{GalacticScale},
\ref{SpiralGalaxies}, and \ref{EllipticalGalaxies} and the
assumptions we have made. We will do our best to clearly label where
we have had to make approximations and educated guesses, as well as
rigorous arguments, so that readers may make their own judgments
about what is presented here.

\begin{figure}\label{NGC3310}
   \begin{center}
   \includegraphics[height=75mm]{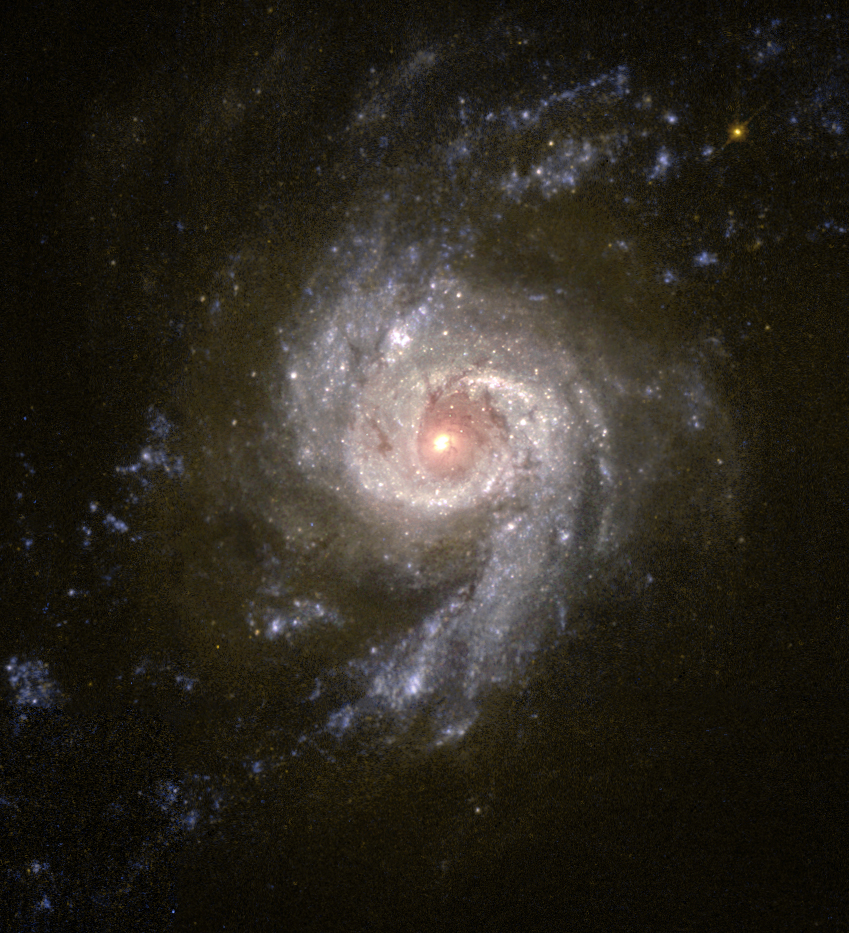}
   \includegraphics[height=75mm]{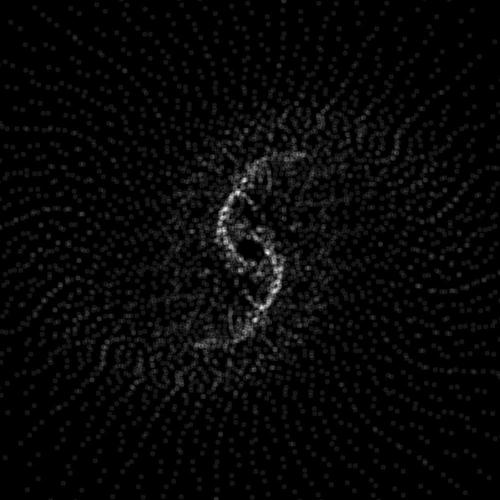}
   \end{center}
   \caption{NGC3310 on the left, simulation on the right.  The simulated image on the right results from running the Matlab function
   spiralgalaxy(1, 75000, 1, -0.15, 2000, 1990, 100000000, 8.7e-13, 7500, 5000, 45000000, 50000) described in section \ref{SGS3}.  Left photo credit:
   NASA and The Hubble Heritage Team (STScI/AURA).  Acknowledgment: G.R. Meurer and T.M. Heckman (JHU), C. Leitherer, J. Harris and
   D. Calzetti (STScI), and M. Sirianni (JHU).  Dates:  March 1997 and September 2000.  Telescope:  Hubble Wide Field Planetary Camera 2.}
\end{figure}

\section{Geometric Motivation}\label{GeometricMotivation}

Einstein's theory of general relativity was made possible by Gauss
and Riemann who, decades before, began the field of mathematics now
called differential geometry. Since then, advances in differential
geometry have played a crucial role in understanding the
implications of Einstein's theory. Einstein used differential
geometry to make the qualitative statement ``matter curves
spacetime'' precise, thereby showing that gravity results as a
consequence of this fundamental idea. In contrast, Newton's inverse
square law for gravity, while a great approximation in the low-field
limit, has been shown to be false by measuring the precession of the
orbit of Mercury, for example, as well as the bending of light
around the Sun, which is twice what is predicted by Newtonian
physics and exactly what is predicted by general relativity. Hence,
understanding gravity would appear to require differential geometry.
In light of this rich history of differential geometry playing a
vital role in understanding gravity and the large scale structure of
the universe, it seems reasonable, among other ideas, to look for
geometric motivations for dark matter.

\begin{figure}\label{NGC488}
   \begin{center}
   \includegraphics[height=75mm]{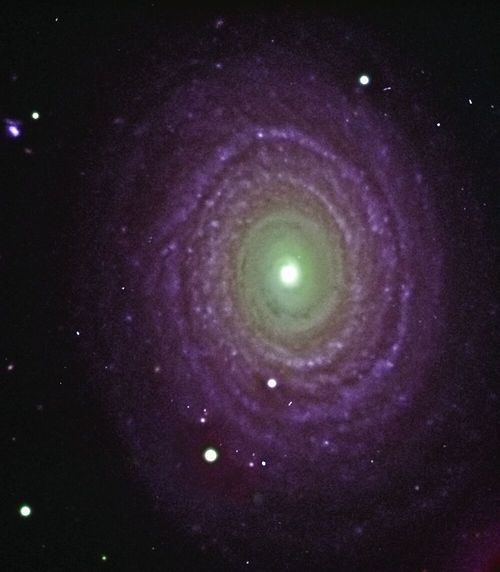}
   \includegraphics[height=75mm]{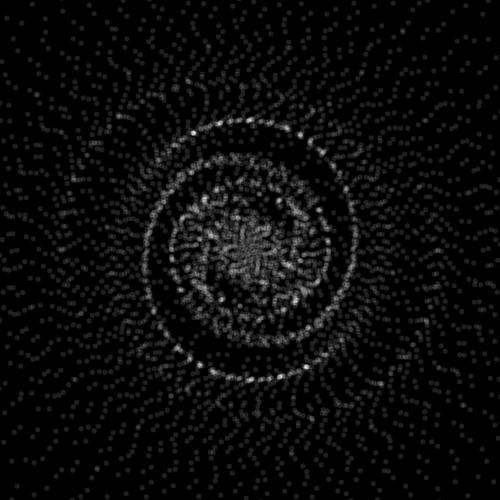}
   \end{center}
   \caption{NGC488 on the left, simulation on the right.  The simulated image on the right results from running the Matlab function
   spiralgalaxy(0.1, 100000, 1, -0.5, 20000, 19900, 50000000, 8.7e-15, 15000, 5000, 82000000, 20000) described in section \ref{SGS4}.  Left photo credit:
   Johan Knapen and Nik Szymanek.  Telescope:  Jacobus Kapteyn Telescope. B, I, and H-alpha bands.}
\end{figure}

The beginning point for our theory is to remove the assumption that
the connection on the tangent bundle of the spacetime, an intrinsic
geometric object second only to the metric in importance, is the
standard Levi-Civita one.  We compare this step to the jump from
special relativity to general relativity, where the assumption that
the metric of the spacetime is the standard flat one is removed. Our
axioms then define the geometric properties that our action, which
is now a function of the metric and the connection, must have.

We note that Einstein and Cartan famously played around with ideas
similar to these by removing the assumption that the connection was
torsion free, while still assuming metric compatibility.  However,
as our beginning point, we make neither assumption. Also, Einstein
and Cartan were not trying to describe dark matter and thus had
different objectives in mind.

Throughout this paper, the fundamental objects of our universe will
be a spacetime manifold $N$ with a metric $g$ of signature $(-+++)$
and a connection $\nabla$.  We will assume that $N$ is a smooth
manifold which is both Hausdorff and second countable, which, while
standard, deserves contemplation, as do all assumptions. We refer
the interested reader to \cite{ONeill} as an excellent reference for
the fundamentals of differential geometry.  We will also assume that
$g$ and $\nabla$ are smooth.  These preliminary assumptions could be
considered our ``Axiom 0.''

A smooth manifold $N$ is a Hausdorff space with a complete atlas of
smoothly overlapping coordinate charts \cite{ONeill}.  Hence, we see
that coordinate charts are more than convenient places to do
calculations, but are in fact a necessary part of the definition of
a smooth manifold. Given a fixed coordinate chart, let
$\{\partial_i\}$, $0 \le i \le 3$, be the tangent vector fields to
$N$ corresponding to the standard basis vector fields of the
coordinate chart. Let $g_{ij} = g(\partial_i, \partial_j)$ and
$\Gamma_{ijk} = g( \nabla_{\partial_i} \partial_j,
\partial_k )$, and let
\[
M = \{g_{ij}\} \;\;\;\mbox{ and }\;\;\; C = \{\Gamma_{ijk}\}
\;\;\;\mbox{ and }\;\;\; M' = \{g_{ij,k}\} \;\;\;\mbox{ and }\;\;\;
C' = \{\Gamma_{ijk,l}\}
\]
be the components of the metric and the connection in the coordinate
chart and all of the first derivatives of these components in the
coordinate chart. We are now ready to state our central geometric
axiom which motivates the remainder of this paper.

\begin{figure}\label{M87}
   \begin{center}
   \includegraphics[height=65mm]{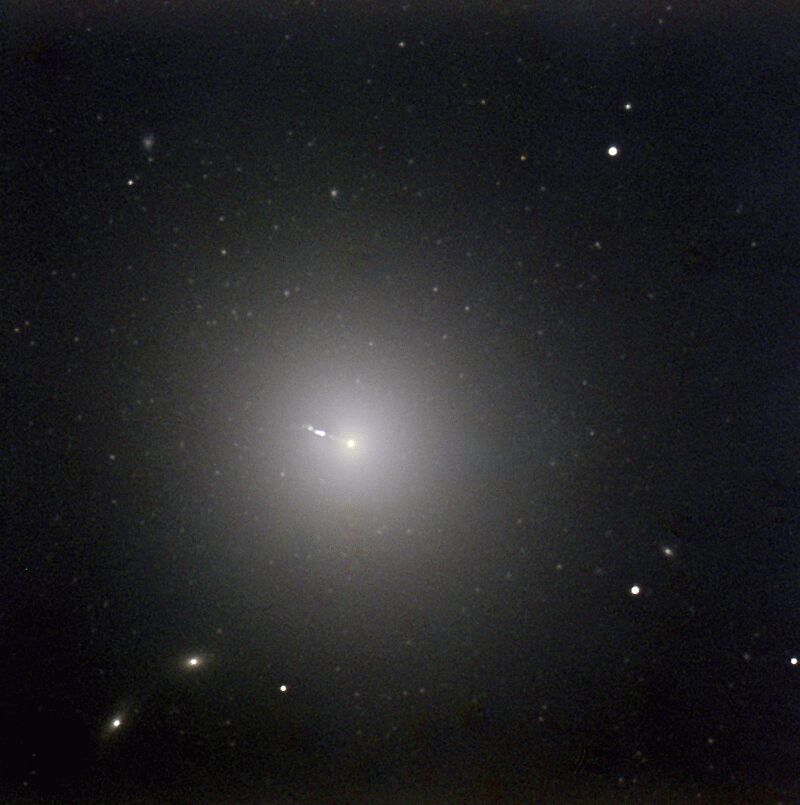}
   \includegraphics[height=70mm]{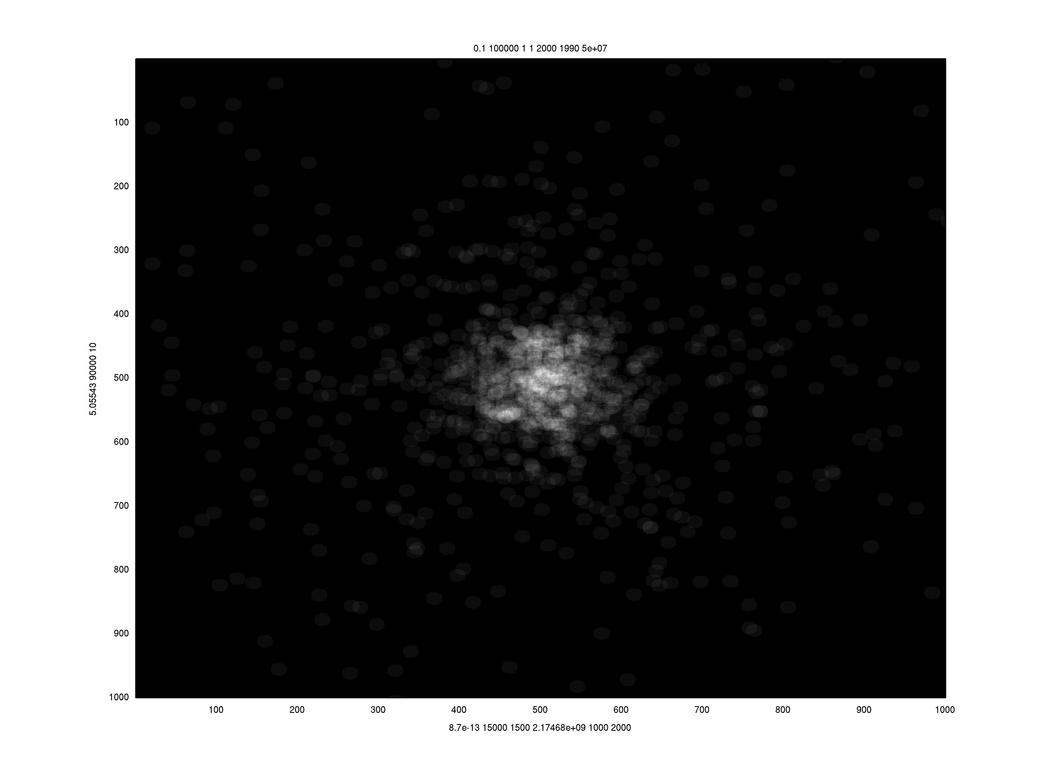}
   \end{center}
   \caption{M87 on the left, simulation on the right.  The simulated image on the right results from running the Matlab function
   ellipticalgalaxy(0.1, 100000, 1, 1, 2000, 1990, 50000000, 8.7e-13, 15000, 1500, 2174670000, 1000, 2000) described in section \ref{EGS1}.
   Note that this simulated image is the top view of the same simulation in the next figure
   which shows the side view.  Left photo credit: ING Archive and Nik Szymanek.  Date: 1995.  Telescope:  Jacobus Kapteyn Telescope.
   Instrument: JAG CCD Camera.  Detector: Tek.  Filters B, V, and R.}
\end{figure}

\begin{axiom}\label{A1}
For all coordinate charts $\Phi : \Omega \subset N \rightarrow R^4$
and open sets $U$ whose closure is compact and in the interior of
$\Omega$, $(g,\nabla)$ is a critical point of the functional
\begin{equation}
   F_{\Phi,U}(g,\nabla) = \int_{\Phi(U)}
   \mbox{Quad}_M(M' \cup M \cup C' \cup C)
   \; dV_{R^4}
\end{equation}
with respect to smooth variations of the metric and connection
compactly supported in $U$, for some fixed quadratic functional
$Quad_M$ with coefficients in $M$.
\end{axiom}
Note that we have not specified the action, only the form of the
action.  As is standard, we define
\begin{equation}
   \mbox{Quad}_{Y}(\{x_\alpha\}) = \sum_{\alpha,\beta}
   F^{\alpha\beta}(Y) x_\alpha x_\beta
\end{equation}
for some functions $\{F^{\alpha\beta}\}$ to be a quadratic
expression of the $\{x_\alpha\}$ with coefficients in $Y$.

The implications of the above axiom when the connection is removed
have been long understood.  When the integrand in Axiom \ref{A1} is
reduced to $Quad_M(M')$, vacuum general relativitiy generically
results. When the integrand is reduced to $Quad_M(M' \cup M)$,
vacuum general relativity with a cosmological constant generically
results. Here ``generically'' means for a generic choice of
quadratic functional, so that the zero quadratic function, for
example, is not included in these claims. These two results were
effectively proved by Cartan \cite{Car22}, Weyl \cite{Wey22}, and
Vermeil \cite{Ver17} and pursued further by Lovelock \cite{Lov72}.
The point to keep in mind is that since $(g,\nabla)$ must be a
critical point of this functional in \emph{all} coordinate charts,
then something geometric, that is, not depending on a particular
choice of coordinate chart, must result. Hence, if we remove the
assumption that the connection is the standard Levi-Civita
connection, the above axiom seems like a reasonable place to start.

\begin{figure}\label{NGC1132}
   \begin{center}
   \includegraphics[height=65mm]{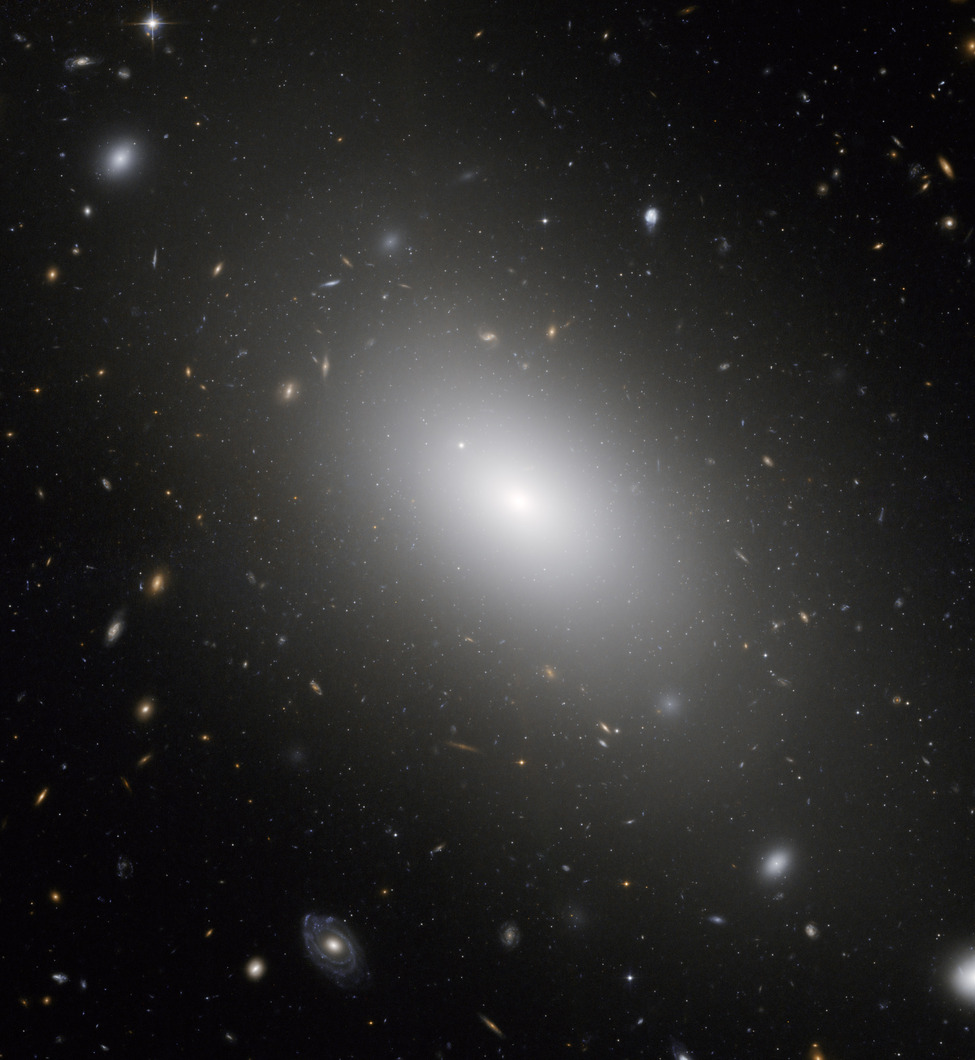}
   \includegraphics[height=70mm]{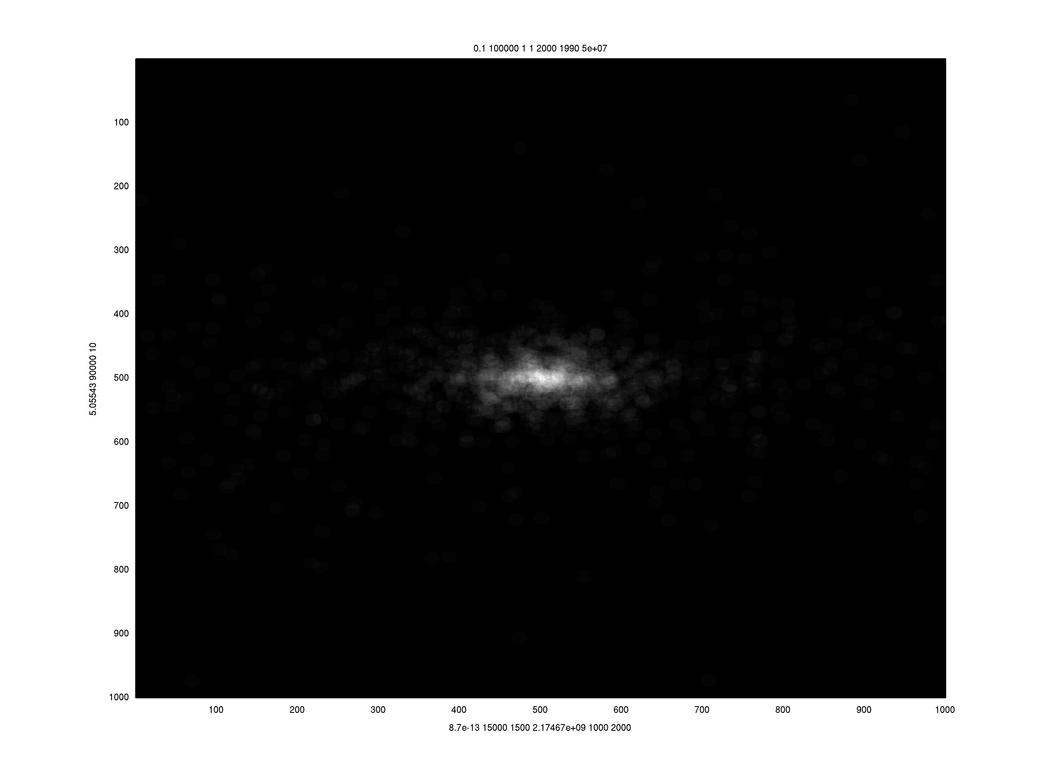}
   \end{center}
   \caption{NGC1132 on the left, simulation on the right.  The simulated image on the right results from running the Matlab function
   ellipticalgalaxy(0.1, 100000, 1, 1, 2000, 1990, 50000000, 8.7e-13, 15000, 1500, 2174670000, 1000, 2000) described in section \ref{EGS1}.
   Note that this simulated image is the side view of the same simulation in the previous figure
   which shows the top view.  Left photo credit:  NASA, ESA, and the Hubble Heritage (STScI/AURA)-ESA/Hubble
   Collaboration.  Acknowledgment: M. West (ESO, Chile).}
\end{figure}

For organizational reasons we have placed the bulk of the geometric
calculations in the appendices where we have provided a detailed
geometric discussion of the implications of Axiom \ref{A1}. In this
section we are content to report that the Einstein-Klein-Gordon
equations with a cosmological constant result from quadratic
functionals compatible with Axiom \ref{A1}.  More generally, we
conjecture this same outcome for a generic choice of quadratic
functional compatible with Axiom \ref{A1}.  Explicitly, the
Einstein-Klein-Gordon equations with a cosmological constant (in
geometrized units with the gravitational constant and the speed of
light set to one) are
\begin{eqnarray}\label{eqn:EE}
   G + \Lambda g &=& 8 \pi \mu_0 \; \left\{ 2 \frac{ df \otimes df}{\Upsilon^2}
   - \left(\frac{|df|^2}{\Upsilon^2} + f^2 \right)g \right\} \\
   \Box f &=& \Upsilon^2 f    \label{eqn:KG}
\end{eqnarray}
where $G$ is the Einstein curvature tensor, $f$ is the scalar field
representing dark matter, $\Lambda$ is the cosmological constant,
and $\Upsilon$ is a new fundamental constant of nature whose value
has yet to be determined. The other constant $\mu_0$ is not a
fundamental constant of nature as it can easily be absorbed into
$f$, but simply is present for convenience and represents the energy
density of an oscillating scalar field of magnitude one which is
solely a function of $t$.  It is perfectly fine to set $\mu_0$ equal
to one, just as we have done to the speed of light and the
gravitational constant.  As a final comment, note that since
\begin{equation}
   \Box f = \nabla \cdot \nabla f = \frac{1}{\sqrt{|g|}} \partial_i
   \left(\sqrt{|g|} \; g^{ij} \partial_j f \right) = g^{ij} \left(
   \partial_i \partial_j f - \bar\Gamma_{ij}^{\hspace{.1in}k} \partial_k f \right)
\end{equation}
equation \ref{eqn:KG} is hyperbolic in $f$ when the metric has
signature $(-+++)$.

\begin{figure}
   \begin{center}
   \includegraphics[height=59mm]{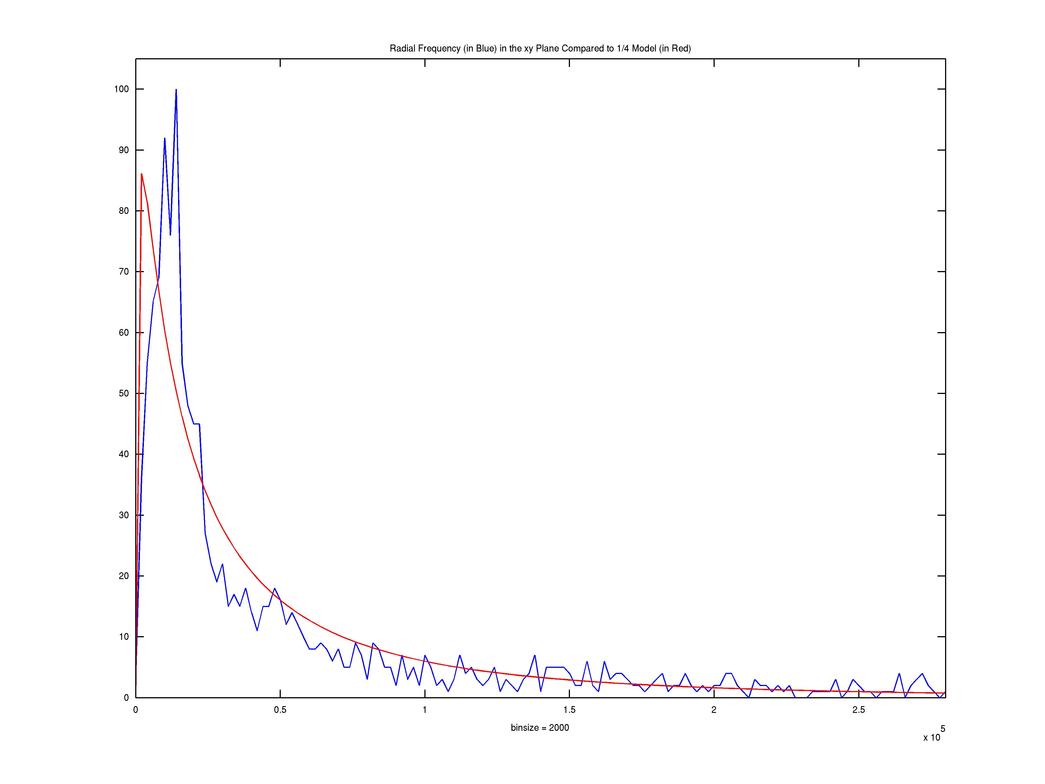}
   \includegraphics[height=59mm]{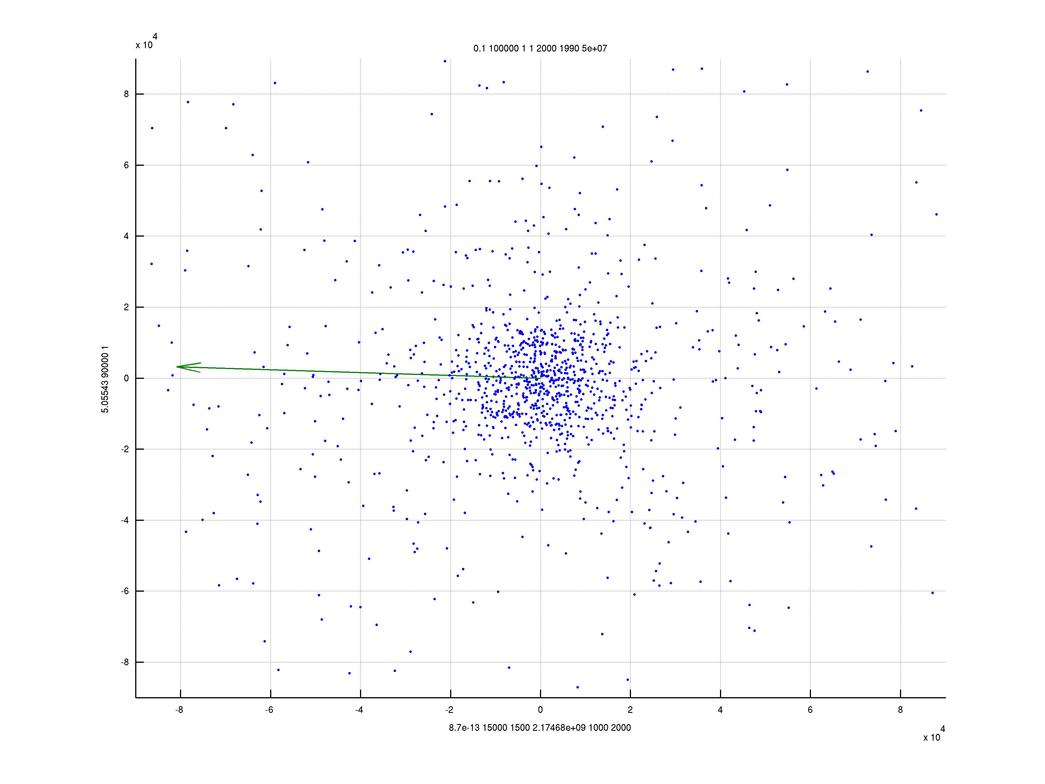}
   \end{center}
   \caption{The right image is the simulated elliptical galaxy image
   from figure 5, where the blue dots represent stars.  If one
   computes the distance of each of the stars from the origin in
   the viewing plane, the histogram of those computed radii is the
   blue curve on the left (and is closely related to the brightness profile
   of the galaxy).  The red curve on the left represents
   what is typically actually observed in elliptical galaxies
   according to the ``$R^{1/4}$ model.''  This is explained in
   more detail in section \ref{EllipticalGalaxies}.}
\end{figure}

As is explained in the appendices, the scalar field represents the
deviation of the connection from the standard Levi-Civita
connection.  Hence, when the scalar field is zero, the connection is
the Levi-Civita connection.  More generally, there is a formula for
the connection on the tangent bundle of the spacetime in terms of
the metric and the scalar field, which is equation
\ref{ConnectionFormula} in the appendices.

We must admit that we do not have a definitive idea of how the
connection manifests itself physically, other than gravitationally,
which is made explicit in the Einstein-Klein-Gordon equations.  For
example, since light rays follow null geodesics, and the geodesic
equation involves the connection, it may be possible that light rays
are affected by the connection in addition to the metric. However,
one can also think of light rays as being along paths which are
critical points of the geodesic energy functional which does not
involve the connection.  This latter view of null geodesics
guarantees that these curves, once null, stay null. Hence, there is
no guarantee that the connection affects light rays, and there is
even a reasonable argument that it does not.  The question of how
the connection might be detected is a very interesting open problem.
In this paper we will simply study the huge gravitational impact
that the connection, through the scalar field $f$, has in our
theory.

At this point an aside is warranted.  We have called equation
\ref{eqn:KG} the Klein-Gordon equation, which it is, except that the
constant in the equation is not called mass as is usual, nor is
Planck's constant present.  Hence, some might want to call the above
equation a modified Klein-Gordon equation to be precise.  The point
we wish to make is that there are no quantum mechanical implications
being made here.  We have ``coincidentally'' derived an equation
which also comes up in quantum mechanics.  Of course this is
actually not much of a coincidence, since the Klein-Gordon equation
is one of the simplest relativistic equations which one could
consider.

On the other hand, this is where our discussion intersects with the
fascinating works of many others who have studied ``scalar field
dark matter'' and ``boson stars.''  In the boson star case, the
motivation is quantum mechanical, so the above scalar field $f$ is
supposed to represent the overall wave function for a very large
number of very tiny bosons with masses on the order of $10^{-23}$ eV
(see \cite{JWLee} for a survey).  We relate our constant $\Upsilon$
to the mass of the Klein-Gordon equation by noting that the Compton
wavelength in both cases is
\begin{equation}
   \lambda = \frac{2\pi}{\Upsilon} = \frac{h}{m} \approx 13 \mbox{ light
   years}
\end{equation}
if we take $m \approx 10^{-{23}} eV$, or equivalently, $\Upsilon
\approx 1/(2 \mbox{ light years})$.  These other motivations are
interesting as well but should be distinguished from the purely
geometric motivations provided in this paper.  For example, in our
context it seems most natural to define $f$ to be real valued, and
we do not mean to suggest an interpretation of $f$ as a probability
density.  At the same time, we refer the reader to the survey
article \cite{JWLee} as well as the many works cited in that survey
for many excellent discussions and ideas, most of which apply to
this work here also, insofar as they apply to the
Einstein-Klein-Gordon equations with a cosmological constant.
Similarly, the results of this paper, from this point on, apply to
boson stars and these other theories as well.

\section{Cosmological Predictions}\label{cosmologicalpredictions}

In this section we discuss the compatibility of this model for dark
matter with the $\Lambda CDM$ (Lambda Cold Dark Matter) model of the
universe \cite{Weinberg}. As our theory only hopes to explain dark
matter and dark energy, we artificially ``add in'' the remaining
quantum mechanical particles and electromagnetic radiation.
Equivalently, in this section we mean to consider the standard
$\Lambda CDM$ model, but with our scalar field model for dark matter
(which satisfies the Klein-Gordon equation) instead of the
traditional WIMP (Weakly Interacting Massive Particles) model.

Fortunately for us, much analysis of the cosmological features of a
scalar field model of dark matter has already been done.  Our exact
case is treated in \cite{MGM2008}, referred to in that paper and
elsewhere as the $V(\phi) = \phi^2$ case (which gives rise to the
Einstein-Klein-Gordon equations).  More general scalar field
potentials are considered in that paper as well as \cite{JWLee},
\cite{FlatCentralDensityProfiles}, \cite{MUL2000}, and
\cite{MUL2001}, but these potentials, when they are even functions
with positive second derivative at zero, are equivalent to the
$\phi^2$ potential in the low field limit.

In \cite{MGM2008} (see also \cite{FlatCentralDensityProfiles},
\cite{MUL2000}, \cite{MUL2001}), the authors explain that scalar
field dark matter has the same cosmology as the the standard CDM
model with WIMP dark matter particles, with the main differences
only becoming apparent on the scale of galaxies. We refer the reader
to \cite{WK1993} for a related discussion where the large scale
behavior of the Einstein-Klein-Gordon equations is compared to the
evolution of collisionless matter.  It would seem reasonable that
there are many more interesting problems to study in these areas,
but the author is not expert enough to comment further.

Furthermore, scalar field dark matter has some advantages over the
WIMP model.   Scalar field dark matter gives a nearly flat density
of dark matter in the centers of galaxies  \cite{MGM2008},
\cite{FlatCentralDensityProfiles}, as opposed to a cusp of density
as predicted by the WIMP model, which has not been observed.  Also,
numerical simulations show that WIMPs have a tendency to clump
forming structures smaller than those observed, the smallest
observations being dwarf galaxies \cite{JWLee}. Scalar field dark
matter, on the other hand, has a natural length scale defined by our
fundamental constant $1/\Upsilon$.  For the group velocities of
scalar field dark matter waves to be much less than the speed of
light in order to be able to form gravitationally bound systems, the
wavelengths of our solutions must be much larger than $1/\Upsilon$.

Finally, scalar field dark matter is automatically cold,
sufficiently far after the big bang, assuming the universe is
homogeneous and isotropic. In general terms, the reason for this is
the wave nature of the scalar field. Said another way, when two
solutions to the Einstein-Klein-Gordon equation are added to one
another, their stress-energy tensors do not add, as can be seen by
the nonlinearity of the stress-energy tensor on the right hand side
of equation \ref{eqn:EE}.  That is, these scalar field solutions,
which can equivalently be thought of as wave solutions, interfere
with one another both constructively and destructively.  The result
is a pressure which oscillates rapidly between being positive and
negative with the average pressure being very close to zero,
sufficiently far after the big bang, in the homogeneous, isotropic
universe case.  These statements are made precise by the following
theorem, proved in Appendix \ref{SFCDM}.

\begin{theorem}\label{cosmothm}
Suppose that the spacetime metric is both homogeneous and isotropic,
and hence is the Friedmann-Lema\^{i}tre-Robertson-Walker metric
$-dt^2 + a(t)^2 ds_\kappa^2$, where $ds_\kappa^2$ is the constant
curvature metric of curvature $\kappa$. If $f(t,\vec{x})$ is a
real-valued solution to the Klein-Gordon equation (equation
\ref{eqn:KG} with mass term $\Upsilon$) with a stress-energy tensor
which is isotropic, then $f$ is solely a function of $t$.
Furthermore, if we let $H(t) = a'(t)/a(t)$ be the Hubble constant
(which of course is actually a function of $t$), and $\rho(t)$ and
$P(t)$ be the energy density and pressure of the scalar field at
each point, then
\begin{equation}
   \frac{\bar{P}}{\bar{\rho}} = \frac{\epsilon}{1 + \epsilon}
\end{equation}
where
\begin{equation}
   \epsilon = - \frac{3\overline{H'}}{4\Upsilon^2}
\end{equation}
and
\begin{equation}
   \bar{\rho} = \frac{1}{b-a} \int_a^b \rho(t) \;dt , \hspace{.2in}
   \bar{P} = \frac{1}{b-a} \int_a^b P(t) \;dt, \hspace{.2in}
   \overline{H'} = \frac{\int_a^b H'(t) f(t)^2 \;dt}
   {\int_a^b f(t)^2 \;dt} ,
\end{equation}
where $a,b$ are two zeros of $f$ (for example, two consecutive
zeros).
\end{theorem}

We comment that if we crudely approximate $|H'(t)| \approx
(10^{10}\mbox{ light years})^{-2}$ and $\Upsilon \approx (2\mbox{
light years})^{-1}$, then $|\epsilon| \approx 4 \cdot 10^{-20}$ at
the current age of the universe. We leave it to others to find
better approximations for these values, but the point is that
$\epsilon$ is very small.  Hence, this scalar field dark matter
model automatically gives a cold dark matter model in the
homogeneous, isotropic case, sufficiently long after the Big Bang.

In contrast, in the WIMP model the particles are considered
independently of one another, and so their stress-energy tensors do
add, allowing for significant pressure terms to be present, even in
the homogeneous, isotropic universe case. Hence, unlike the WIMP
model, there is no hot version of the scalar field model of dark
matter in the homogeneous and isotropic universe case.  Since hot
dark matter is inconsistent with cosmological observations
\cite{WMAPobservations}, this is a nice property for a dark matter
theory to have.

Thus, scalar field dark matter is a serious candidate for cold dark
matter. Furthermore, since it appears that scalar field dark matter
and cold WIMP dark matter give similar predictions on the
cosmological scale, it makes sense to go to the galactic scale to
look for differences.

\section{The Galactic Scale}\label{GalacticScale}

We now begin our analysis which shows that scalar field dark matter
satisfying the Einstein-Klein-Gordon equations may be related to
spiral structure in disk galaxies and may predict plausible
brightness profiles for elliptical galaxies.   The cosmological
constant is sufficiently small to be well approximated by zero on
the galactic scale, so we need to study solutions to the
Einstein-Klein-Gordon equations
\begin{eqnarray}\label{eqn:EinsteinEquation}
   G &=& 8 \pi \mu_0 \; \left\{ 2 \frac{ df \otimes df}{\Upsilon^2}
   - \left(\frac{|df|^2}{\Upsilon^2} + f^2 \right)g \right\} \\
   \Box f &=& \Upsilon^2 f  .  \label{eqn:KleinGordon}
\end{eqnarray}

Studying these equations is particularly challenging because of
their nonlinearities. The trivial solution to these equations is the
Minkowski spacetime with zero scalar field.  Also, any vacuum
general relativity solution with zero scalar field is a solution. In
addition, when the scalar field is taken to be complex, there exist
spherically symmetric static solutions \cite{Kaup}.  As discussed in
\cite{MSBS}, one may think of these static solutions as solutions
where the dispersive characteristics of the scalar field, which are
mild for long wavelengths, are balanced with gravity. There are also
real scalar field solutions analogous to the complex ones, but these
are not quite static and are only known approximately \cite{RB1969}.

In 1992, \cite{Sin1992}, \cite{JiSin1992}, and then
\cite{LeeKoh1992}, \cite{LeeKoh1996} proposed scalar field wave
equations for dark matter as a way of explaining the observed
flatness of rotation curves \cite{RC1}, \cite{RC2}, \cite{RC3},
\cite{RC4} for most galaxies. As described in the survey article
\cite{JWLee}, this idea has been rediscovered many times (including
by the author at the beginning of this work). Translated into our
context, the exciting fact about the static solutions discussed
above is that they give qualitatively good predictions for the
rotation curves for galaxies. Since most of the matter in disk
galaxies is going in very circular orbits \cite{BT}, it makes sense
to define the rotation curve of a galaxy to be the velocity of the
matter of the galaxy as a function of radius.  It is a striking fact
that these curves are very flat. For example, most of the mass of
the Milky Way Galaxy is going approximately 220 km/s in circular
motion about its center \cite{BT}.  However, visible mass is not
massive enough to account for these rotation curves, which is one of
the motivations for the existence of dark matter. Hence, the
prediction of flat rotation curves is certainly intriguing, to say
the least.

However, most of the above solutions, in both the real and complex
case, are actually unstable \cite{MSBS}.  On the other hand,
combinations of the above solutions (with one being the stable
``ground state'') yield dynamic spherically symmetric solutions
which are stable according to numerical simulations \cite{MSBS} and
give somewhat flat rotation curves. Alternatively, it is suggested
in \cite{JWLee} that perhaps the dark matter halos, like most
everything else in the universe, are rotating, and that this
rotation provides additional stability. \cite{JWLee} also points out
that the bottom-up structure formation scenario of the $\Lambda CDM$
model could allow, or even require, that the dark matter have
angular momentum.

The answers to the questions implied in the previous paragraph are
not yet clear.  What is really needed are careful simulations of the
Einstein-Klein-Gordon equations in a perturbed cosmological setting
to see what typical ``blobs'' of scalar field dark matter look like.
These simulations could then address the question of what happens
when two blobs collide, whether they have enough dynamical friction
to combine, and how they carry angular momentum.  It may be
necessary to simulate the regular matter at the same time so that
energy from the dark matter can be transferred to the regular matter
and then dissipated through friction and radiation. Also, regular
matter could help stabilize the galactic potential which could help
to stabilize the dark matter at sufficiently small radii. There are
many important open problems in these areas.

On the other hand, it is very easy to write down solutions to the
Klein-Gordon equation if we fix the spacetime metric as the
Minkowski spacetime.  The Klein-Gordon equation then becomes
\begin{equation}
   \left( - \frac{\partial^2}{\partial t^2} + \Delta_x \right) f =
   \Upsilon^2 f
\end{equation}
where $\Delta_x$ is the standard Laplacian on $R^3$.  Solutions can
then be expanded in terms of spherical harmonics to get solutions
which are linear combinations of solutions of the form
\begin{equation}\label{basis}
   f = A \cos(\omega t) \cdot Y_n(\theta,\phi) \cdot r^n \cdot f_{\omega,n}(r)
\end{equation}
where
\begin{equation}\label{sphericalode}
   f_{\omega,n}''(r) + \frac{2(n+1)}{r} f_{\omega,n}'(r) =
   (\Upsilon^2 - \omega^2) f_{\omega,n}
\end{equation}
and $Y_n(\theta,\phi)$ is an $n$th degree spherical harmonic.  Note
that we require $f_{\omega,n}'(0) = 0$ but have not specified an
overall normalization.  Naturally, to get a complete basis of
solutions we also need to include solutions like the one above but
where $\cos(\omega t)$ is replaced by $\sin(\omega t)$. We will
study real solutions in this paper, but complex solutions are quite
analogous.

It is also easy to write down solutions to the Klein-Gordon equation
if we fix the spacetime metric to be a spherically symmetric static
spacetime.  For our purposes, suppose we approximate the spherically
symmetric static spacetime metric as
\begin{equation}
   ds^2 = -V(r)^2 dt^2 + V(r)^{-2} \left( dx^2 + dy^2 + dz^2 \right)
\end{equation}
as is standard.  The function $V(r)$ acts likes the gravitational
potential function from Newtonian physics.  In this case, the above
form in equation \ref{basis} is still fine, but equation
\ref{sphericalode} is modified to become
\begin{equation}\label{sphericalodemod}
   V(r)^2 \left( f_{\omega,n}''(r) + \frac{2(n+1)}{r} f_{\omega,n}'(r) \right) =
   \left(\Upsilon^2 - \frac{\omega^2}{V(r)^2}\right) f_{\omega,n}.
\end{equation}
Since dark matter, which makes up most of the mass of most galaxies,
is known to be mostly spherical \cite{BT}, even when the regular
matter is not, the spacetime metric of a typical galaxy can be
approximated reasonably well by one which is spherically symmetric
and static. Hence, we want to understand the spherically symmetric
static case with a potential well $V(r)$ coming from the mass
distribution of a galaxy as well as we can.

There is one main qualitative difference between the Minkowski
spacetime case and the spherically symmetric static case.  In the
Minkowski case, $r^n f_{\omega,n}(r)$ always decays sinusoidally
with amplitude decreasing like $1/r$.  Energy density decays like
the square of this, which means that solutions do not have finite
total energy.  In the spherically symmetric case, there are still
solutions like these, but there are also solutions which, if
$\omega$ is small enough, have an exponential behavior as $r$ goes
to infinity.  For most values of $\omega$, this exponential behavior
is actually increasing.  However, for a discrete set of values of
$\omega$, the coefficient in front of the increasing exponential
term is zero and all that is left is a decaying exponential
behavior. The corresponding energy density is also exponentially
decaying, and so solutions with finite total energy exist.

The key to understanding this behavior is the term $\left(\Upsilon^2
- \frac{\omega^2}{V(r)^2}\right)$.  When it is negative, the
solution $f_{\omega,n}(r)$ is in its oscillatory domain.  When this
term is positive, $f_{\omega,n}(r)$ is in its exponential domain.
The cases we are interested in are when a solution $f_{\omega,n}(r)$
begins in its oscillatory domain at $r=0$ but ultimately ends up in
its exponential domain as $r$ goes to infinity, with exponential
decay and hence finite total energy.  Hence, we must choose $\omega$
in the range
\begin{equation}\label{omegarange}
   \Upsilon \; V(0) < \omega < \Upsilon \lim_{r \rightarrow \infty}
   V(r).
\end{equation}
For such a solution $f_{\omega,n}(r)$, it is natural to define the
radius $\mbox{Rad}(f_{\omega,n}(r))$ to be the largest value of $r$
for which $\left(\Upsilon^2 - \frac{\omega^2}{V(r)^2}\right) = 0$.
In the spherically symmetric case with positive matter density,
$V(r)$ is an increasing function of $r$, so there is a unique value
for $r$ which makes this expression zero. The point to take away
from all of this is that for $r > \mbox{Rad}(f_{\omega,n}(r))$, we
get rapid decay for $f_{\omega,n}(r)$.  We also comment that for
$V(r)$ which are sufficiently asymptotic to $a - b/r$ at infinity,
for example (the standard scenario), and for each $n$, there are an
infinite number of discrete values of $\omega$ which give finite
energy solutions, with the values of these $\omega$ converging to
the upper limit in inequality \ref{omegarange} and the radii
$\mbox{Rad}(f_{\omega,n}(r))$ of these solutions converging to
infinity.  We refer the reader to \cite{Agmon} for more analysis of
these types of equations.

Furthermore, there is an analogy between Minkowski solutions and
solutions in a spherically symmetric potential with finite total
energy.  Without loss of generality (by rescaling the spacetime
coordinates), it is convenient to take $V(0) = 1$. Furthermore,
$V(r) \approx 1$ since we are in the low field limit of general
relativity. Hence, the range of allowable values for $\omega$ given
by equation \ref{omegarange} is quite narrow.  Furthermore, if we
compare equations \ref{sphericalode} and \ref{sphericalodemod}, we
see that solutions with the same $\omega$ will start out very nearly
the same for small $r$ since $V(0) = 1$. They will begin to differ
more and more as $r$ increases, until finally the solution in the
potential well will stop oscillating and decay rapidly to zero. From
these qualitative observations, we make the claim that we can get a
qualitative approximation for the solution in the spherically
symmetric potential well by taking the solution in Minkowski space
and modifying it by declaring it to be zero outside of some radius.
The purpose of this approximation is not to describe the nuances of
the solutions, but to get something that is qualitatively correct
and has qualitatively similar properties.  In this way we do not
need to know the specifics of the potential of a galaxy, just the
radius at which we wish to cutoff the matter density of the scalar
field dark matter solutions in the Minkowski spacetime.

At this point we keep our promise from the end of the first section
and clearly state that we have made three approximations. First, we
are approximating solutions in a spherically symmetric potential
well with Minkowski solutions which we will arbitrarily cutoff at
some radius.  Second, the exact solutions in a spherically symmetric
potential well would still be approximations in the cases we are
interested in, since the model for rotating dark matter we are using
does not give a spherically symmetric dark matter density or
potential, although one could approximate the potential as such.
Third, we have not shown and do not know for certain that there even
exist solutions to the full Einstein-Klein-Gordon equations which
are qualitatively similar to what we are doing here. Determining the
answer to this last issue is a very important open problem which
could benefit from simulations.

As a next step in the near future, it will be a good idea to
actually specify a spherically symmetric potential of a galaxy that
one is trying to model, and then solve equation
\ref{sphericalodemod} with that potential. It will then be desirable
to match up the resulting dark matter density (which may be
aspherical) with the closest spherically symmetric potential.  This
is a reasonable approach, but slightly more complicated than what we
do here. Hence, we have made the decision to keep things as simple
as possible for this preliminary analysis, which in this case
suggests using Minkowski spacetime solutions which are arbitrarily
cutoff at the radius of our choosing.  We justify letting ourselves
choose this cutoff radius since $\mbox{Rad}(f_{\omega,n}(r))$,
defined in the spherically symmetric potential case, takes on
arbitrarily large values for every value of $n$.

\begin{figure}\label{DMinMWG}
\includegraphics[height=39mm]{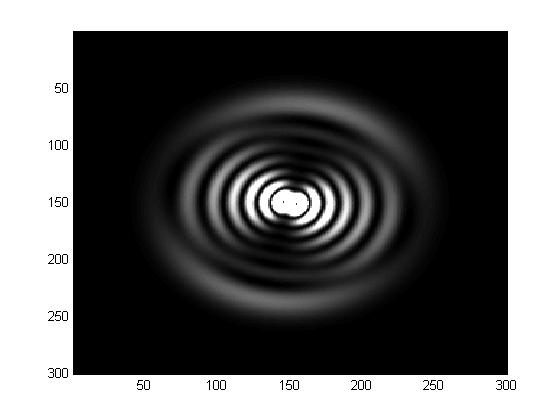}
\includegraphics[height=39mm]{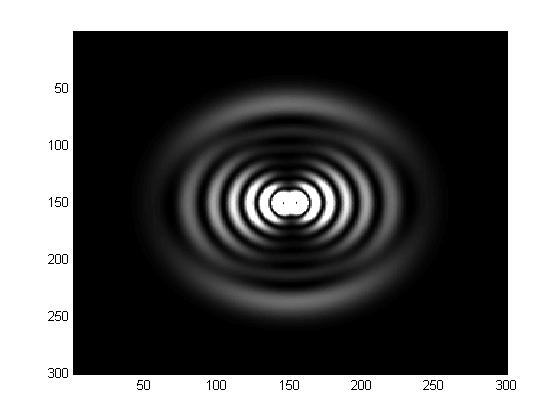}
\includegraphics[height=39mm]{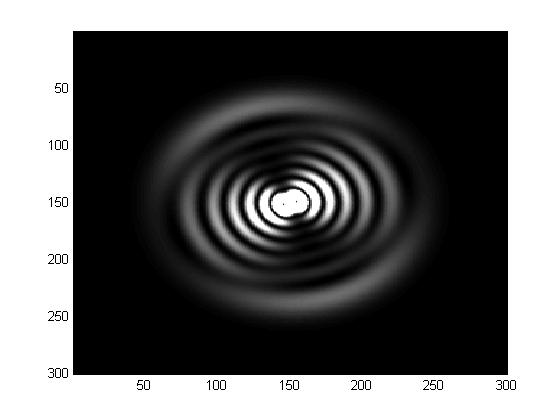}
\caption{Exact solution to the Klein-Gordon equation in a fixed
spherically symmetric potential well based on the Milky Way Galaxy
at $t = 0$, $t = 10$ million years, and $t = 20$ million years. The
pictures show the dark matter density (in white) in the $xy$ plane.
This solution, which one can see is rotating, has angular momentum.}
\end{figure}

The form of the Minkowski solution which we will study for the
remainder of this paper and on which our simulations are based is
\begin{equation}\label{angmomsol}
   f = A_0 \cos(\omega_0 t) f_{\omega_0,0}(r) + A_2 \cos(\omega_2 t -
   2\phi) \sin^2(\theta) r^2 f_{\omega_2,2}(r)
\end{equation}
where $f_{\omega_0,0}(r)$ and $f_{\omega_2,2}(r)$ satisfy equation
\ref{sphericalode}.  We note these solutions fall into the form of
equation \ref{basis} since both $\cos(2\phi)\sin^2(\theta)$ and
$\sin(2\phi)\sin^2(\theta)$ are second degree spherical harmonics.
Hence, the above Minkowski spacetime solution is the sum of a
spherically symmetric solution (degree $n=0$) and two degree two
solutions.  We have not included first degree spherical harmonics
since they would not preserve the center of mass at the origin.  We
would in general expect higher degree spherical harmonics in
galactic solutions as well, but for this paper we have chosen to
focus on the case when the second degree spherical harmonic terms
dominate.  As we will see, these solutions all have $180^\circ$
rotational symmetry with densities which rotate rigidly (which
naturally is not the case when more than two distinct $\omega$
frequencies are present). In the future, if one would want to try to
model spiral galaxies with something other than two arms, then these
other degree spherical harmonics should be studied.

Also, initial computations suggest that the approximation in
equation \ref{angmomsol} is a reasonable qualitative approximation
to a solution in the fixed spherically symmetric potential case. For
example, the pictures in figure 8 show the precise scalar field dark
matter densities in the $xy$ plane at $t = 0$, $t = 10$ million
years, and $t = 20$ million years in a fixed spherically symmetric
potential well based on the Milky Way Galaxy. Like the Minkowski
solution in equation \ref{angmomsol}, the solution depicted in
figure 8 is also a sum of a spherically symmetric solution and two
degree two solutions, but in a spherically symmetric potential.
Notice how the solutions appear to have compact support, although
actually the solution has simply decayed very rapidly outside a
finite radius. Otherwise, the interference pattern is very similar
to what we will see with the solutions in equation \ref{angmomsol},
with the most obvious difference being in the last ring or two,
which are a bit stretched out in the radial direction compared to
our model. Since all we ultimately care about are the qualitative
characteristics of the gravitational potential for the dark matter
that we ultimately produce, this minor difference is something we
are willing to tolerate for now.

The reason the dark matter density is rotating in figure 8 and in
our Minkowski model is most easily seen by making the substitution
\begin{equation}
   \alpha = \phi - \left(\frac{\omega_2 - \omega_0}{2}\right) t
\end{equation}
into our expression for $f$ in equations \ref{angmomsol} to get
\begin{equation}\label{angmomsol2}
   f = A_0 \cos(\omega_0 t) f_{\omega_0,0}(r) + A_2 \cos(\omega_0 t -
   2\alpha) \sin^2(\theta) r^2 f_{\omega_2,2}(r).
\end{equation}
Note that to the extent that $\alpha$ stays fixed in time, then the
above solution does not rotate and gives a fixed interference
pattern since both terms are oscillating in time with the same
frequency $\omega_0$.  Hence, since $|\omega_2 - \omega_0| <<
\omega_0$ because of inequality \ref{omegarange}, we see that we get
an dark matter interference pattern which is rotating according to
the formula
\begin{equation}\label{rotationformula}
   \phi_0 = \frac{\omega_2 - \omega_0}{2} t
\end{equation}
with period
\begin{equation}\label{periodformula}
   T_{DM} = \frac{4\pi}{\omega_2 - \omega_0},
\end{equation}
which may be related to the pattern period of the resulting barred
spiral patterns in the regular matter.

Our next goal is to approximate the energy density $\mu_{DM}$ due to
this scalar field dark matter solution and then to expand it in
terms of spherical harmonics. From equation \ref{angmomsol2}, we
have that
\begin{eqnarray}
   f &=& \cos(\omega_0 t) \left[A_0 f_{\omega_0,0}(r)
   + A_2 \cos(2\alpha) \sin^2(\theta) r^2 f_{\omega_2,2}(r)\right] \nonumber \\
   &&+ \sin(\omega_0 t) \left[A_2 \sin(2\alpha) \sin^2(\theta) r^2
   f_{\omega_2,2}(r)\right].
   \label{angmomsol3}
\end{eqnarray}
Hence, by equation \ref{eqn:EinsteinEquation}
\begin{eqnarray}
   \frac{\mu_{DM}}{\mu_0} &=& \frac{1}{8\pi \mu_0} G(\partial_t,
   \partial_t) \\
   &\approx& \frac{1}{\Upsilon^2}\left( f_t^2 + |\nabla_x f|^2 \right) +
   f^2 \\
   &\approx& \left( \frac{f_t}{\Upsilon} \right)^2 + f^2
\end{eqnarray}
where we have assumed a long wavelength solution in the
approximation. Again, with loss of generality, we will always assume
$V(r) \approx 1$, so that $\omega_0 \approx \Upsilon \approx
\omega_2$ by inequality \ref{omegarange}. This allows us to think of
$\alpha$ as being approximately fixed in time. Hence,
\begin{eqnarray}
   \frac{\mu_{DM}}{\mu_0} &\approx& \left[A_0 f_{\omega_0,0}(r)
   + A_2 \cos(2\alpha) \sin^2(\theta) r^2 f_{\omega_2,2}(r)\right]^2
   + \left[A_2 \sin(2\alpha) \sin^2(\theta) r^2
   f_{\omega_2,2}(r)\right]^2 \\
   &=& A_0^2 f_{\omega_0,0}(r)^2 + A_2^2 \sin^4(\theta) r^4
   f_{\omega_2,2}(r)^2 + 2A_0 A_2 \cos(2\alpha) \sin^2(\theta) r^2
   f_{\omega_0,0}(r) f_{\omega_2,2}(r) \label{expandedmess}
\end{eqnarray}

In order to expand the above result into spherical harmonics, we
need to recall that spherical harmonics defined on the unit sphere
are actually restrictions of homogeneous polynomials of the same
degree which are harmonic in $R^3$. For example,
$\cos(2\phi)\sin^2(\theta)$ is $x^2 - y^2$, a harmonic polynomial of
degree 2, restricted to the unit sphere and
$\sin(2\phi)\sin^2(\theta)$ is $2xy$ restricted to the unit sphere.
To summarize the quick facts that we need to know, here is a quick
list of homogeneous harmonic polynomials:  degree zero:  $1$, degree
one:  $x,y,z$, degree two:  $x^2 - y^2, 2xy, 3z^2 - r^2, 2xz, 2yz$,
and then one more of degree four:  $35z^4 - 30r^2z^2 + 3r^4$, where
$r^2 = x^2 + y^2 + z^2$.  It is easy to check that these are all
harmonic in $R^3$ and hence their restrictions to the unit sphere
are spherical harmonics of the same degree.

Since $\cos(2\alpha)\sin^2(\theta) = \cos(2\phi -
2\phi_0)\sin^2(\theta)$ is a rotated degree two spherical harmonic
and hence a spherical harmonic itself (taking $\phi_0$ to be fixed),
the last term of equation \ref{expandedmess} is already in the form
of a spherical harmonic times a function of $r$, as we desire. The
first term is as well, since it is already a function of $r$ and the
zeroth spherical harmonic is the constant function one. Hence, it is
only the middle term that we need to put into the desired form. To
do this, we note that
\begin{eqnarray*}
   r^4 \sin^4\theta &=& (r^2 - z^2)^2 = z^4 - 2r^2z^2 + r^4 \\
   &=& \frac{3}{105}(35z^4 - 30r^2z^2 + 3r^4) -\frac{40}{105}r^2 (3z^2 - r^2) + \frac{56}{105}r^4
   (1),
\end{eqnarray*}
which successfully expresses this term as the sum of three terms,
each of which is a function of $r$ times a spherical harmonic.
Hence,
\begin{eqnarray}\label{dmdensitysphericalharmonics}
   \frac{\mu_{DM}}{\mu_0} &\approx& U_0(r) + U_2(r) (3z^2 - r^2) +
   U_4(r) (35z^4 - 30r^2z^2 + 3r^4) \nonumber \\
   && + \tilde{U}_2(r) (r^2 \cos(2\alpha)
   \sin^2(\theta))
\end{eqnarray}
where
\begin{eqnarray}
   U_0(r) &=& A_0^2 f_{\omega_0,0}(r)^2 + \frac{56}{105}A_2^2 r^4 f_{\omega_2,2}(r)^2 \nonumber \\
   U_2(r) &=& -\frac{40}{105} A_2^2 r^2 f_{\omega_2,2}(r)^2 \nonumber \\
   U_4(r) &=& \frac{3}{105} A_2^2 f_{\omega_2,2}(r)^2 \nonumber \\
   \tilde{U}_2(r) &=& 2 A_0 A_2 f_{\omega_0,0}(r) f_{\omega_2,2}(r).
   \label{sphericalharmoniccomponents}
\end{eqnarray}

Our next step is to compute the gravitational potential $V$ of this
dark matter density $\mu_{DM}$.  Notice that the dark matter density
is the sum of a spherically symmetric term, two axially symmetric
terms, and a term with neither of those symmetries which is
rotating.  As it turns out, the potential will be of this same form.

We will solve for the potential by solving the Newtonian equation
\begin{equation}
   \Delta_x V = 4\pi \mu_{DM}
\end{equation}
for each value of $t$, where $\Delta_x$ is the Laplacian in $R^3$.
Since we are interested in solutions where the group velocities of
our dark matter scalar field is much less than the speed of light,
this Newtonian approximation will be fine.  We will use the fact
that it is easy to take the Laplacian of a function expressed in
terms of spherical harmonics since
\begin{equation}
   \Delta_x \left( \sum_\alpha W_\alpha(r) \cdot r^{n_\alpha}
   Y_\alpha(\theta,\phi) \right)
   =\sum_\alpha \left( W_\alpha''(r) + \frac{2(n_\alpha + 1)}{r} W_\alpha'(r) \right) r^{n_\alpha}
   Y_\alpha(\theta,\phi)
\end{equation}
where $\alpha$ is a general index and $Y_\alpha$ is a spherical
harmonic of degree $n_\alpha$.  This identity follows from the facts
that $r^{n_\alpha} Y_\alpha(\theta,\phi)$ is a harmonic function in
$R^3$, the radial part of the Laplacian is $\partial_r^2 +
(2/r)\partial_r$, and $\Delta_x (ab) = a \Delta_x b + 2 \langle
\nabla_x a, \nabla_x b \rangle + b \Delta_x a$.  It follows that
\begin{eqnarray}\label{potentialsphericalharmonics}
   \frac{V}{4\pi\mu_0} &\approx& W_0(r) + W_2(r) (3z^2 - r^2) +
   W_4(r) (35z^4 - 30r^2z^2 + 3r^4) \nonumber \\
   && + \tilde{W}_2(r) (r^2 \cos(2\alpha)
   \sin^2(\theta))
\end{eqnarray}
where
\begin{equation}\label{potentialcomponents}
   W_n''(r) + \frac{2(n + 1)}{r} W_n'(r) = U_n(r)
\end{equation}
with the boundary conditions
\begin{equation}
   W_n'(0) = 0 \;\;\;\mbox{ and }\;\;\; \lim_{r \rightarrow \infty} W_n(r) = 0.
\end{equation}
Note that we mean to apply the above equation and boundary
conditions to $\tilde{W}_2(r)$ as well.  These boundary conditions
produce a potential function which goes to zero at infinity.
Naturally any constant may be added to the potential if one desires.

Since we are going to cutoff the scalar field dark matter density
outside of a radius $r_{max}$ of our choice, for $r > r_{max}$ we
will have
\begin{equation}
   W_n''(r) + \frac{2(n + 1)}{r} W_n'(r) = 0,
\end{equation}
which, to satisfy our boundary condition at infinity, has solution
\begin{equation}
   W_n(r) = \frac{k}{r^{2n+1}}.
\end{equation}
Hence, for $r> r_{max}$,
\begin{equation}\label{effectivebdycondition}
   W_n(r) = - \frac{r}{2n+1} W_n'(r).
\end{equation}
In our computer simulations, we will not actually take $r$ to
infinity, but instead, after starting with $W_n'(0) = 0$ and solving
the o.d.e. out to $r=r_{max}$, add whatever constant is necessary to
$W_n(r)$ to satisfy equation \ref{effectivebdycondition} outside the
dark matter radius $r_{max}$.

Thus, in summary, given any $A_0, A_2, \omega_0, \omega_2$ and
$\Upsilon$ (which is a fixed fundamental constant of the universe in
our theory but is not precisely known), as well as any $r_{max}$
(the radius at which we cutoff the dark matter wave function and
density), we have written down formulas in this section for
everything else.  The wave function $f$ is defined by equations
\ref{sphericalode} and \ref{angmomsol}.  The dark matter density
rotates at constant angular speed according to equation
\ref{rotationformula} with period given in equation
\ref{periodformula}. The dark matter density, defined by equations
\ref{dmdensitysphericalharmonics} and
\ref{sphericalharmoniccomponents}, has gravitational potential
defined by equations \ref{potentialsphericalharmonics} and
\ref{potentialcomponents}.  Furthermore, all of these quantities can
be computed since it is particularly straightforward to solve
ordinary differential equations as in equations \ref{sphericalode}
and \ref{potentialcomponents} on a computer.

\section{Spiral Galaxies}\label{SpiralGalaxies}

The standard text book on galactic dynamics is the book by the same
name by Binney and Tremaine \cite{BT}.  A great companion to this
book is ``Galactic Astronomy'' by Binney and Merrifield \cite{BM}.
We also refer the reader to Toomre's review article entitled
``Theories of Spiral Structure'' \cite{ToomreReview} which \cite{BT}
says ``is still worth careful reading, even after several decades.''
This review article begins with three interesting quotes, which we
repeat here:
\begin{quotation}
``Much as the discovery of these strange forms may be calculated to
excite our curiosity, and to awaken an intense desire to learn
something of the laws which give order to these wonderful systems,
as yet, I think, we have no fair ground even for plausible
conjecture.''

\vspace{.05in} \hfill Lord Rosse (1850)
\end{quotation}
\begin{quotation}
``A beginning has been made by Jeans and other mathematicians on the
dynamical problems involved in the structure of the spirals.''

\vspace{.05in} \hfill Curtis (1919)
\end{quotation}
\begin{quotation}
``Incidentally, if you are looking for a good problem ... ''

\vspace{.05in} \hfill Feynman (1963)
\end{quotation}
To these quotes we add the opening statements of Toomre's review
article as a description of where things stood in 1977:
\begin{quotation}
``The old puzzle of the spiral arms of galaxies continues to taunt
theorists.  The more they manage to unravel it, the more obstinate
seems the remaining dynamics.  Right now, this sense of frustration
seems greatest in just that part of the subject which advanced most
impressively during the past decade - the idea of Lindblad and Lin
that the grand bisymmetric spiral patterns, as in M51 and M81, are
basically compression waves felt most intensely by the gas in the
disks of those galaxies.  Recent observations leave little doubt
that such spiral ``density waves'' exist and indeed are fairly
common, but no one still seems to know why.

To confound matters, not even the $N$-body experiments conducted on
several large computers since the late 1960s have yet yielded any
decently \emph{long-lived} regular spirals.  ''

\vspace{.05in} \hfill Toomre (1977)
\end{quotation}

To this 160 year old ``puzzle of the spiral arms'' we offer evidence
of the possibility that scalar field dark matter density waves,
which arise naturally due to the wave nature of the Klein-Gordon
equation, are the primary driver of barred spiral density waves in
disk galaxies, at least in many cases.  Our proposed contribution to
understanding disk galaxies ends there - we do not simulate the
gravitational interactions of the stars, much less the complicated
dynamics of the interstellar medium composed of gas and dust, nor
star formation and supernova. The contribution we attempt to make
here is simply to qualitatively model the effect of scalar field
dark matter satisfying the Klein-Gordon equation on the regular
matter of a galaxy, which we demonstrate may possibly cause barred spiral
patterns in a wide range of shapes.

If indeed dark matter is responsible for the shapes of many
galaxies, this would explain why it has taken so long to understand
spiral structure.  This is especially the case if the dark matter
has a dynamics which is different than most commonly thought.  Since
the bulk of dark matter has only been detected so far by its
gravitational effects, it is reasonable to look to the details of
the gravitational predictions made by each dark matter theory. We
consider the possibility that scalar field dark matter satisfying
the Klein-Gordon equation may have a characteristic signature which
provides the beginnings of a unified theory for spiral and bar
patterns in disk galaxies.

Also, we comment that we do not mean to dismiss other possible
explanations of spiral structure but instead are simply adding
another possibility which seems to hold promise. We recommend the
book ``Spiral Structure in Galaxies: A Density Wave Theory'' by G.
Bertin and C.C. Lin \cite{BL}, published in 1996, as a fascinating
description of the progress of the Lin-Shu density wave theory of
spiral structure.  Our dark matter density wave theory of spiral
structure has much in common with this famous theory as both
theories try to explain observed density waves in the visible
matter.  In the Lin-Shu theory, these waves are induced by the
regular matter itself which plays the dominant role. In our dark
matter density wave theory, we suggest that the dark matter provides
the primary effect leading to barred spiral structure, for two
reasons:  there is more dark matter than regular matter, at least
for large radii, and scalar field dark matter may have a greater
tendency to produce density waves because of its wave nature.  Of
course in our theory regular matter still has gravity too, so this
effect must still be considered. Hence, we see an opportunity for a
combined theory of spiral structure coming out of our proposal which
accounts for the gravitational effects of both the dark matter and
the regular matter.  Of course for a complete picture, it seems very
likely that the interstellar medium, star formation, and supernova
will need to be modeled too.  Until all of these elements are put
together into a convincing model, it will not be clear what the
ultimate implications of the preliminary simulations we do here will
be.

We begin with a few qualitative conclusions.  By looking at figure
8, for example, we see that the dark matter density deviates from
spherical symmetry and rotates rigidly for these very special
solutions. Thus, the level sets of the resulting potential function
will not be spheres but instead will resemble triaxial ellipsoids.
Hence, the potential function for our dark matter can be thought of
as a rotating triaxial potential.

Another way to get a rotating triaxial galactic potential is to have
another galaxy pass nearby.   In \cite{ToomreSwingAmplification},
Toomre shows how quadrupole forces (modeled by positive masses at 0
and 180 degrees and negative masses at 90 and 270 degrees rotating
around the origin), which also produce something resembling a
rotating triaxial potential, can cause surprisingly strong spiral
patterns in test particles which were otherwise rotating in circular
motion around the origin. Hence, before we even begin our
simulations, Toomre's work suggests that we are likely to get spiral
patterns to emerge.

As described in \cite{BT}, one characteristic of disk galaxies is
that they have significant amounts of gas and dust.  Unlike stars
which effectively never collide and hence do not have friction, gas
and dust do have friction.  In fact, the presumption is that it is
precisely this friction which causes the collapse of gas and dust
clouds into disks in the first place.  Friction decreases total
energy but conserves angular momentum, and a disk configuration with
everything rotating circularly in the same direction allows for
relatively large total angular momentum for its total energy since
the angular momentum vectors of the individual masses are aligned.
And sure enough, most of the visible mass of disk galaxies are
indeed moving in roughly circular orbits.  In addition, the surface
brightness of disk galaxies is approximately modeled by
\begin{equation}\label{dsl}
   I(R) = I_d e^{-R/R_d},
\end{equation}
where $R_d$ is called the disk scale length of the galaxy. Hence,
our simulations begin with a large number of point masses
representing the mass of the galaxy in circular motion with area
density modeled by the above equation.  Of course since we are not
modeling the gravitational effects of the regular matter, its
initial density profile is not critically important.  In fact, each
of these point masses should be thought of as test particles which,
for now, act as if they have zero mass in our simulations.

The command line for our spiral galaxy simulations, all done in
Matlab, is
\begin{eqnarray}
\mbox{spiralgalaxy}( dr, r_{max}, A_0, A_2, \lambda_0, \lambda_2,
T_{DM}, \mu_0, R_d, n_{particles}, T_{total}, dt).
\end{eqnarray}
This Matlab .m file may be downloaded at
http://www.math.duke.edu/faculty/bray/darkmatter/darkmatter.html which
also contains the author's Matlab .m file, ellipticalgalaxy.m, for
doing elliptical galaxy simulations.  The input variables $r_{max}$,
$A_0$, $A_2$, $T_{DM}$, $\mu_0$, and $R_d$ have already been
defined. The input variable $dr$ is the step size with which the
o.d.e.s in equations \ref{sphericalode} and
\ref{potentialcomponents} are approximately solved.  The user is
allowed to enter either a positive or negative value for $\mu_0$ as
only the absolute value is used for the actual value of $\mu_0$,
where a positive value denotes counterclockwise rotation for the
regular matter and a negative value denotes clockwise rotation for
the regular matter.  Similarly, a positive value for $T_{DM}$ gives
counterclockwise rotation for the dark matter density whereas a
negative value gives clockwise rotation.  The input variable
$n_{particles}$ is the number of test particles used in the
simulation, $T_{total}$ is the total time of the simulation, and
$dt$ is the step size in time used to compute the paths of the test
particles.

As is clear from equation \ref{sphericalode}, let us define
\begin{equation}
   \lambda_k = \frac{2\pi}{\sqrt{\omega_k^2 - \Upsilon^2}}
\end{equation}
for $k=0,2$ to be the spatial wavelengths of our two terms.  Then
using the above two equations and equation \ref{periodformula}, we
can solve for $\omega_0$, $\omega_2$, and $\Upsilon$ in terms of
$\lambda_0$, $\lambda_2$, and $T_{DM}$ to get
\begin{eqnarray}
   \omega_0 &=& \frac{\pi}{2}\left(\frac{1}{\lambda_2^2} - \frac{1}{\lambda_0^2}\right) T_{DM} -
   \frac{2\pi}{T_{DM}}\\
   \omega_2 &=& \frac{\pi}{2}\left(\frac{1}{\lambda_2^2} - \frac{1}{\lambda_0^2}\right) T_{DM} +
   \frac{2\pi}{T_{DM}}\\
   \Upsilon^2 &=& \left(\frac{2\pi}{T_{DM}}\right)^2 +
   \left(\frac{\pi T_{DM}}{2}\right)^2 \left(\frac{1}{\lambda_2^2} - \frac{1}{\lambda_0^2}\right)^2
   -2\pi^2 \left(\frac{1}{\lambda_2^2} +
   \frac{1}{\lambda_0^2}\right)\label{UpsilonEquation}
\end{eqnarray}
which is routine to derive.  In this way we get to choose the dark
matter pattern period, presumably roughly equal to the regular
matter pattern period, and the two wavelengths which manifests
themselves somewhat in the rotation curve data.  In theory, one
could try to find best matches to known galaxies using our
simulation and then use this to estimate $\Upsilon$, although we
have not made a careful attempt at this yet.

\subsection{Spiral Galaxy Simulation \# 1}\label{SGS1}

As a first example, consider
\begin{equation}
   \mbox{spiralgalaxy}(1,75000,1,-1,2000,1990,25000000,8.7e-13,7500,5000,50000000,10000)
\end{equation}
which produces the simulated image in figure 1 as the $t = 25$
million years image of figure 16, rotated $90$ degrees
counterclockwise. In figure 1, we compare this image to NGC 1300, a
barred spiral galaxy of type SBbc.  Notice that in this example we
have $|A_2| = |A_0|$.  The Matlab code normalizes $f_{\omega_0,
0}(r)$ and $r^2 f_{\omega_2,2}(r)$ to have the same magnitudes for
large $r$, so this choice maximizes both the constructive and
destructive interferences of the two terms in equation
\ref{angmomsol} and hence roughly maximizes the ellipticity of the
level sets of the potential function.  We have observed in other
simulations that this has the effect of making the bar in the
simulation more like a bar and less like an oval (which we will
demonstrate in our next example).

Figure 9 shows the graphs of $f_{\omega_0, 0}(r)$ and $r^2
f_{\omega_2,2}(r)$ which, as one can see, have been normalized to
have roughly the same magnitudes for large $r$.  Naturally $r^2
f_{\omega_2,2}(r)$ is the one which equals zero at the origin.  The
input parameters $\lambda_0$ and $\lambda_2$ are the wavelengths of
these functions in the limit as $r$ goes to infinity.

Figure 10 shows planar cross sections of the dark matter density
produced by our choice of input parameters in the $xy$, $xz$, and
$yz$ planes at $t = 0$.  These densities rotate rigidly in time with
period equal to $T_{DM}$, which is $25$ million years in this
example. The densities in figure 10, which decrease roughly like
$1/r^2$, have been multiplied by $r^2$ in the plots to be visible
for large radii. Note how the dark matter densities have been cutoff
at $r = 75,000$ light years.  We point out that the discontinuity at
that radius is not as extreme as it appears because of the $r^2$
factor. Also, there is dark matter density at the origin, but the
$r^2$ factor suppresses this fact in figure 10.  However, this
density is smooth and bounded at the origin, as can be seen from
direct calculation. These images may be compared to figure 8 which
is very similar, except that figure 8 shows the densities without a
factor of $r^2$.

\begin{figure}
   \begin{center}
   \includegraphics[height=59mm]{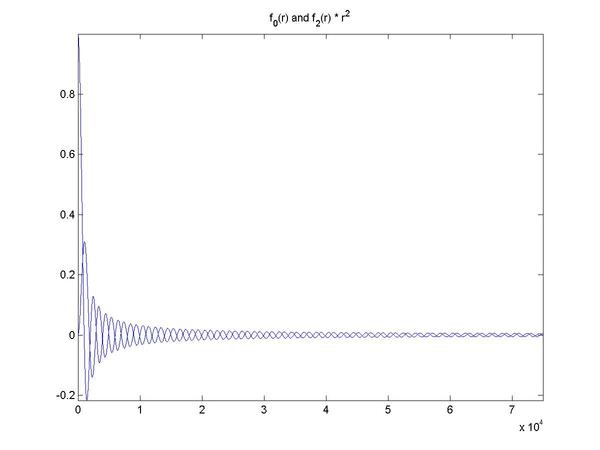}
   \includegraphics[height=59mm]{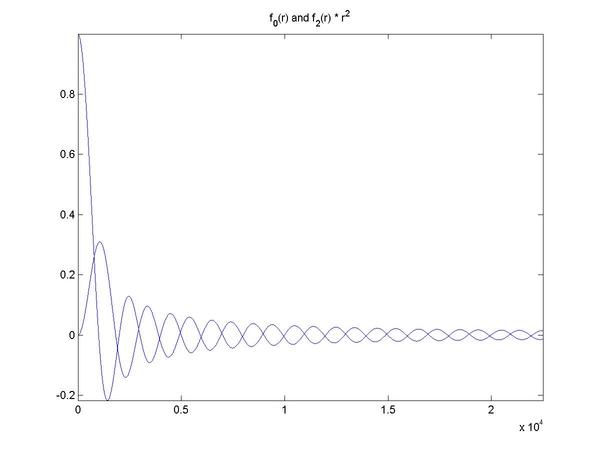}
   \end{center}
   \caption{Graphs of $f_{\omega_0,0}(r)$ and $r^2 f_{\omega_2,2}(r)$ for $r$ up to $75,000$ light years (left) and
   $22,500$ light years (right) in Spiral Galaxy Simulation \# 1.  Notice how the functions begin out
   of phase but are gradually becoming more in phase since they have slightly different wavelengths.}
\end{figure}

One way to think of the densities in figure 10 is as two ``blobs''
of scalar field dark matter rotating around each other like a binary
star.  If we conjecture that the dark matter, in some situations,
will choose a configuration which maximizes angular momentum
(perhaps because of many collisions and mergers with other blobs of
dark matter), then the configuration shown here suggests what these
configurations could look like, at least qualitatively.

Figure 11 shows the plots of the radial functions $W_0(r)$, $-r^2
W_2(r)$, $3r^4 W_4(r)$, and $r^2 \tilde{W}_2(r)$ which define the
potential function $V$ given by equation
\ref{potentialsphericalharmonics}.  The factors in front of the $W$
functions are those relevant for $z=0$, as can be seen in equation
\ref{potentialsphericalharmonics}, so that the relative
contributions of the terms can be judged.  In this case we see that
the spherically symmetric term is most dominant, followed by the
rotating term defined by $\tilde{W}_2(r)$.

Figure 12 shows the end result of all of these computations, the
potential function $V$, graphed in planar cross sections.  At first
glance, this result is amazingly boring.  After all, these images
look very much like a generic perturbation of a spherically
symmetric potential with a second degree spherical harmonic term,
which of course they basically are.  In the next two spiral galaxy
simulation examples, it is even harder to see the perturbation, as
the example shown here roughly maximizes the ellipticities of the
level sets of the potential.

\begin{figure}
   \begin{center}
   \includegraphics[height=59mm]{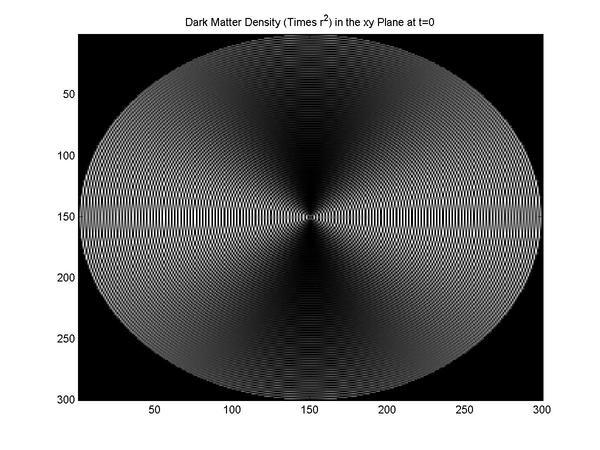}
   \includegraphics[height=59mm]{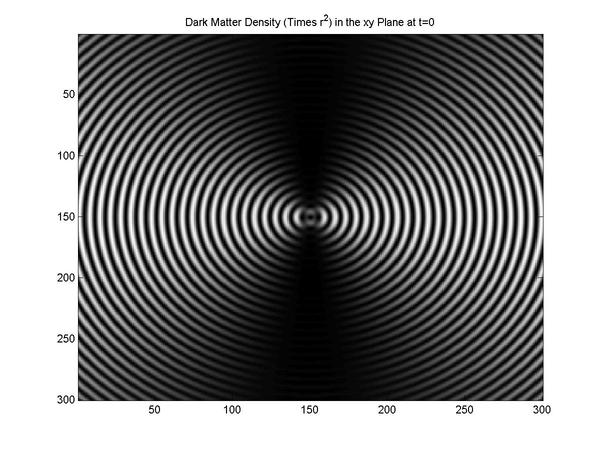}
   \includegraphics[height=59mm]{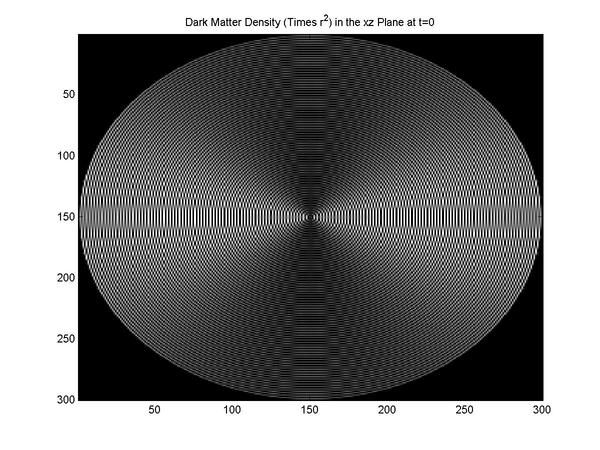}
   \includegraphics[height=59mm]{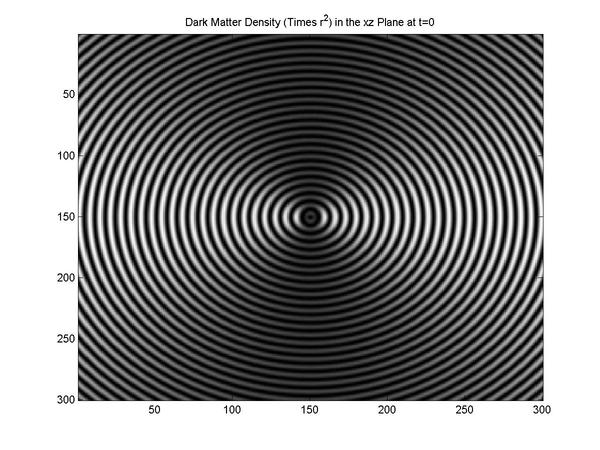}
   \includegraphics[height=59mm]{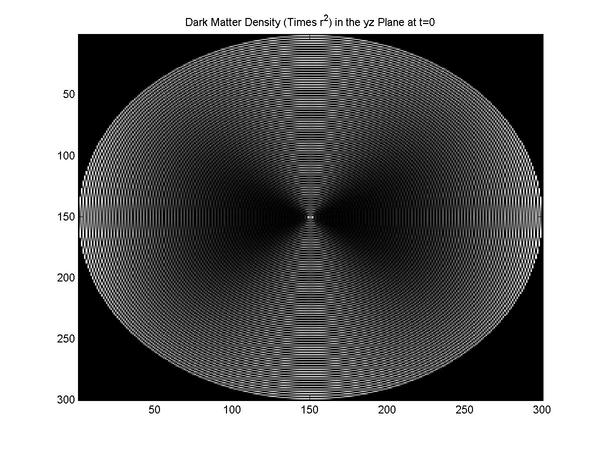}
   \includegraphics[height=59mm]{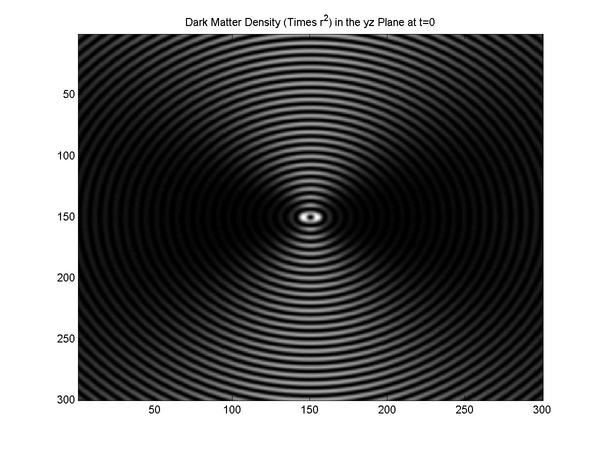}
   \end{center}
   \caption{Plot of densities (in white) of dark matter times $r^2$ for Spiral Galaxy Simulation \#1.  The
   left column has a radius of $75,000$ light years and the right column
   is zoomed in at a radius of $22,500$ light years.
   The top row is the density in the $xy$ plane, the middle row is the density in the $xz$
   plane, and the bottom row is the density in the $yz$ plane.  The densities, which roughly decrease
   like $1/r^2$, have been multiplied by $r^2$ to be more easily visible.  Hence, the cutoff of the dark
   matter density at $r = 75,000$ light years is not as severe as it appears.}
\end{figure}

\begin{figure}
   \begin{center}
   \includegraphics[height=50mm]{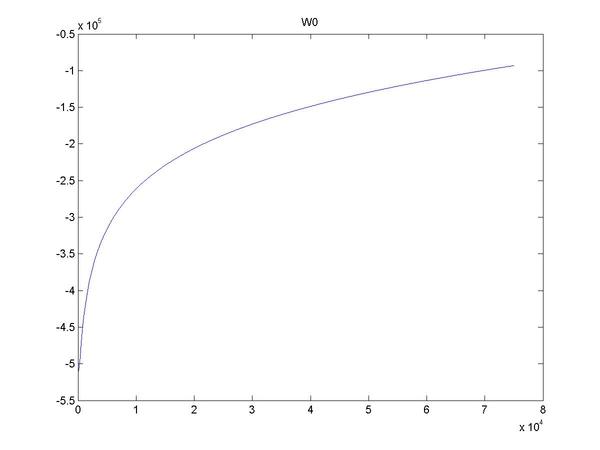}
   \includegraphics[height=50mm]{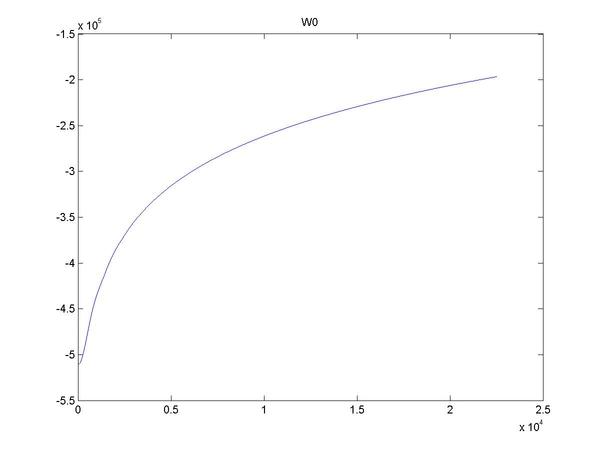}
   \includegraphics[height=50mm]{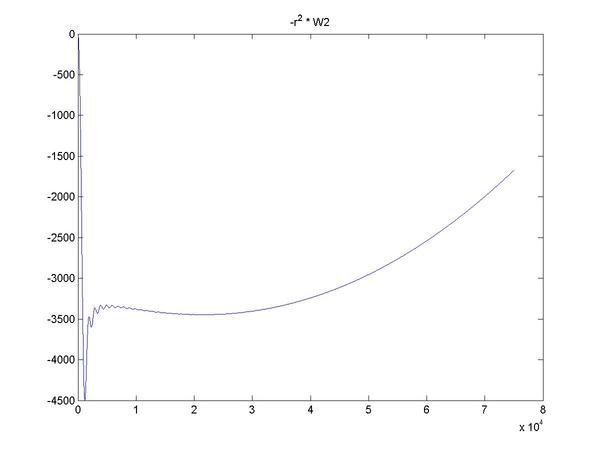}
   \includegraphics[height=50mm]{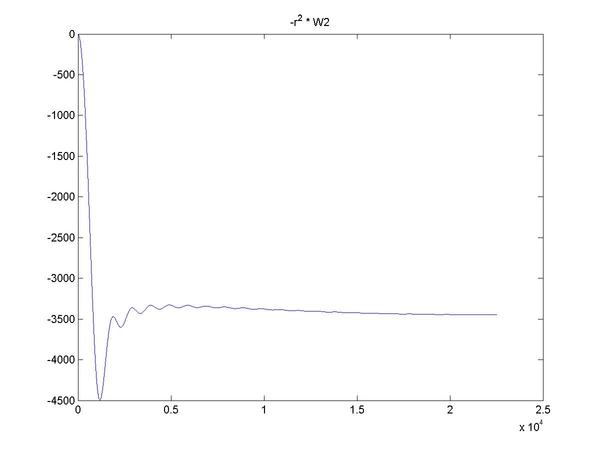}
   \includegraphics[height=50mm]{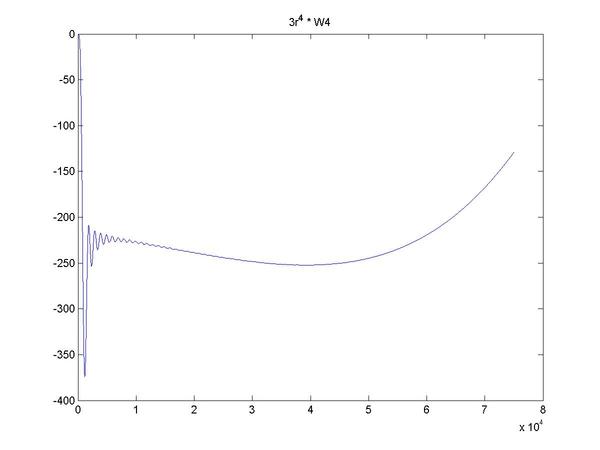}
   \includegraphics[height=50mm]{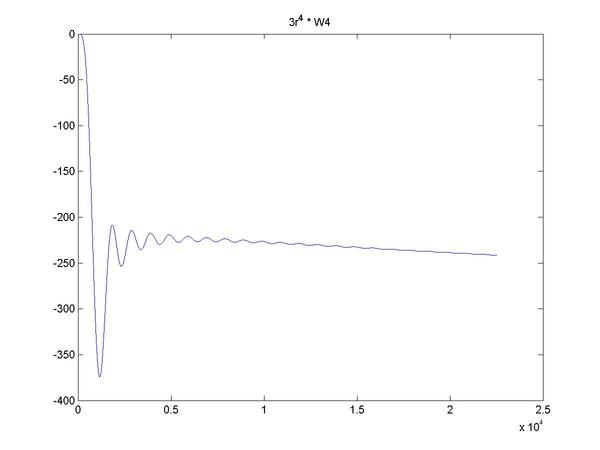}
   \includegraphics[height=50mm]{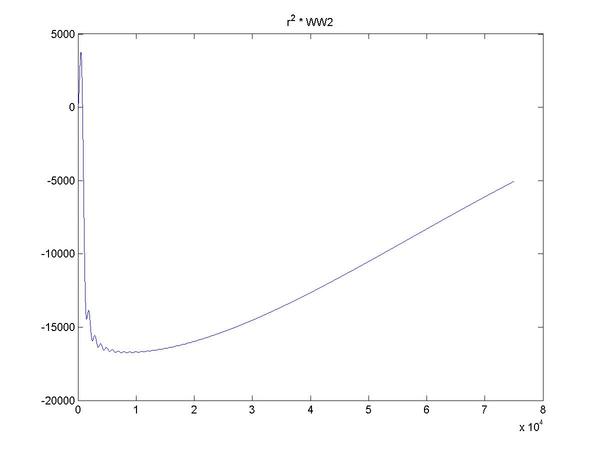}
   \includegraphics[height=50mm]{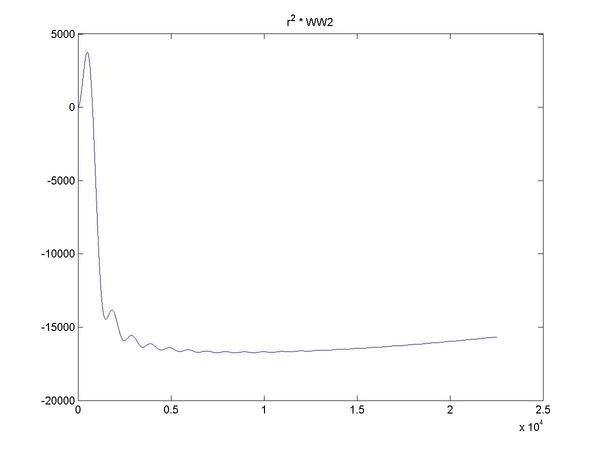}
   \end{center}
   \caption{Plots of the radial functions composing the potential function for Spiral Galaxy Simulation
   \#1. The left column has a radius of $75,000$ light years and the right
   column is zoomed in at a radius of $22,500$ light years.  The first row is $W_0(r)$, the second row is
   $-r^2 W_2(r)$, the third row is $3r^4 W_4(r)$, and the fourth row is $r^2 \tilde{W}_2(r)$.}
\end{figure}

\begin{figure}
   \begin{center}
   \includegraphics[height=59mm]{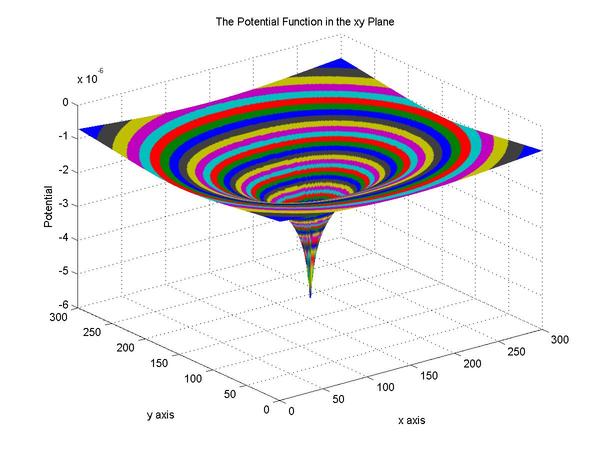}
   \includegraphics[height=59mm]{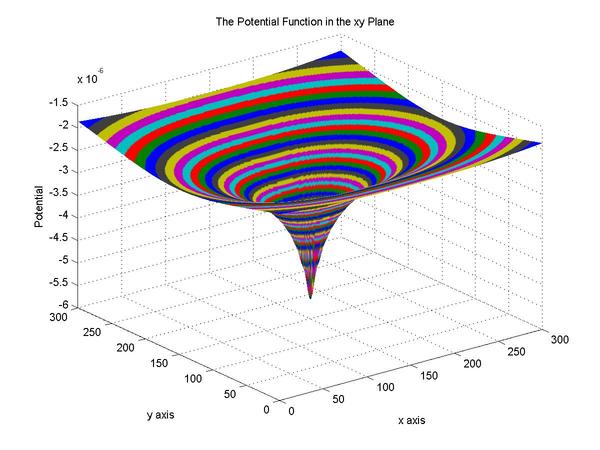}
   \includegraphics[height=59mm]{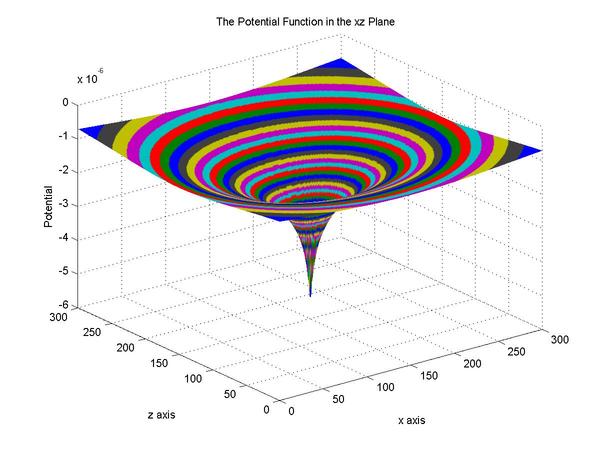}
   \includegraphics[height=59mm]{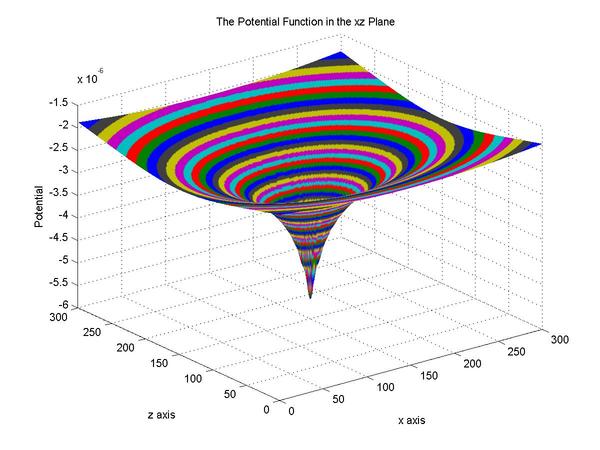}
   \includegraphics[height=59mm]{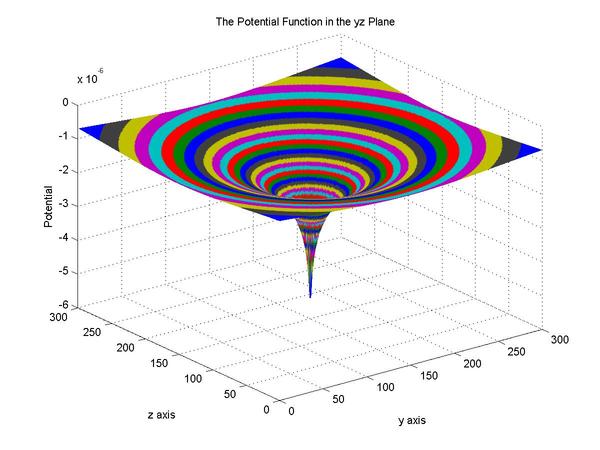}
   \includegraphics[height=59mm]{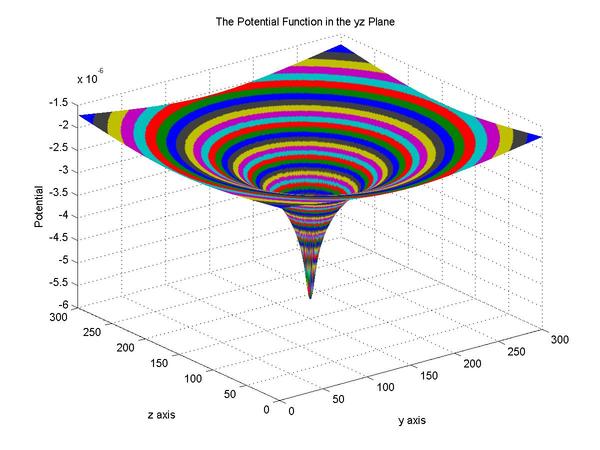}
   \end{center}
   \caption{Graphs of the potential function in the $xy$ plane (top row), the $xz$ plane (middle row),
   and the $yz$ plane (bottom row) for Spiral Galaxy Simulation \#1.
   The first column graphs have a radius of $75,000$ light years
   whereas the second column graphs have a radius of $22,500$ light years.  Notice how the top order
   appearance is like $V(r)$ = $log(r)$, but rounded off at the origin and perturbed everywhere
   else to give triaxial ellipsoidal level sets instead of spherical ones.}
\end{figure}

\begin{figure}
   \begin{center}
   \includegraphics[height=59mm]{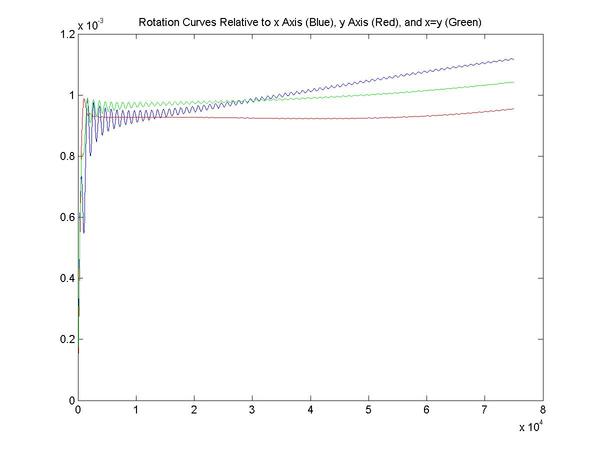}
   \includegraphics[height=59mm]{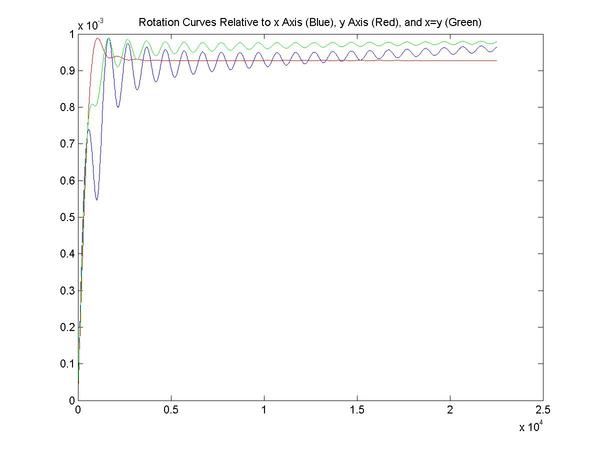}
   \end{center}
   \caption{Approximate rotation curves for Spiral Galaxy Simulation \#1
   out to a radius of $75,000$ light years (left)
   and $22,500$ light years (right).  We have approximated the rotation curves with
   graphs of $\sqrt{r |\nabla V|}$ (which is exactly correct in the spherically symmetric case)
   along the $x$ axis (in blue), along the $y$ axis (in red), and along $y=x$ (in green).}
\end{figure}

We are led to believe (although we have not studied this carefully)
that generic perturbations of the potential of this type (defined
appropriately) may lead to spiral patterns emerging. We refer the
reader to \cite{ToomreSwingAmplification} for more related
discussion. Of course, it is only reasonable to consider potentials
which come from physically plausible and common scenarios.  Hence, a
reasonable thought is that the main advantage that scalar field dark
matter offers is a physically plausible and common way of achieving
these triaxial potentials which cause spiral patterns.

Figure 13 shows three approximate rotation curves for this
simulation.  We have not graphed the actual velocities of the test
particles, although this is a good idea for a later version of this
simulation.  Instead, we have graphed $v(r) = \sqrt{r |\nabla V|}$
which, in the spherically symmetric case, gives the velocity of test
particles in circular motion as a function of $r$.  The three graphs
show the value of this function along the $x$ axis, $y axis$, and
the line $y=x$.  We comment that we use the $y=x$ velocity function
as the initial velocity (in the $xy$ plane, with zero radial
component) of our test particles which start out in roughly circular
motion.

The interesting feature of the approximate rotation curves, of
course, is how much they resemble the rotation curves of many
galaxies, especially the way they are somewhat flat.  As far as the
author is aware, rotation curve data for NGC 1300, the galaxy in
figure 1 to which we compare the results of this simulation, is not
available. This is a common problem because the galaxies with the
best pictures are usually ones which face us directly, whereas the
ones for which the best rotation curve data can be found are ones at
an angle to us, which allows for redshift data to be used to
determine the velocities of the stars and gas and dust in the
galaxy.  Hence, we are left to speculate that, based on other
galaxies for which there is rotation curve data \cite{RC1},
\cite{RC2}, \cite{RC3}, \cite{RC4}, that the estimates of the
rotation curve in figure 13 have a very plausible general form. Of
course our simulated rotation curves only account for the dark
matter, so this must be taken into account.

Another general comment worth making is that we have not tried to
match the exact radius of our simulation to the radius of NGC 1300.
The reason is that one can always scale everything in our simulation
to any scale that one desires.  Of course, the fundamental constant
$\Upsilon$ would have to scale as well, but since we do not know
what it is yet anyway, this is okay for now.  Hence, in these first
simulations we are simply trying to establish that there is the
potential for making good fits of simulations to actual galaxies.

\begin{figure}
   \begin{center}

   \includegraphics[height=52mm]{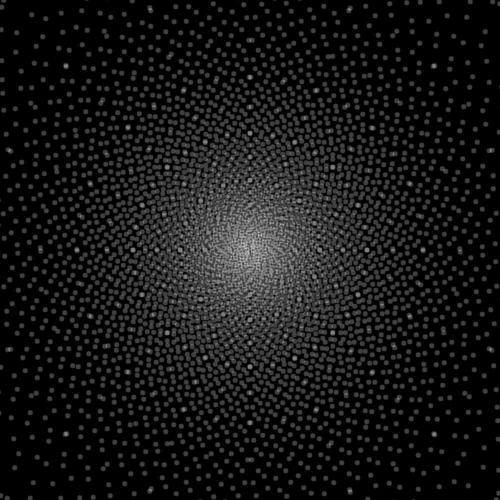}
   \includegraphics[height=52mm]{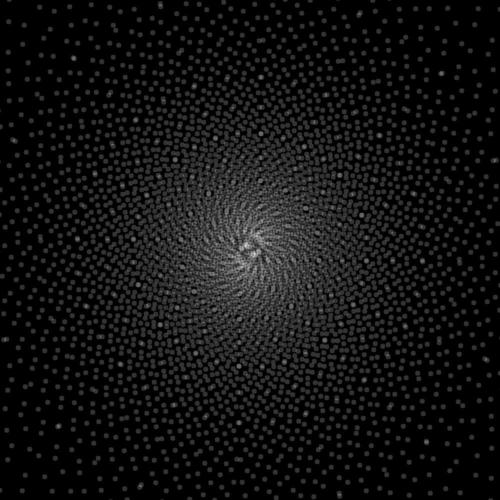}
   \includegraphics[height=52mm]{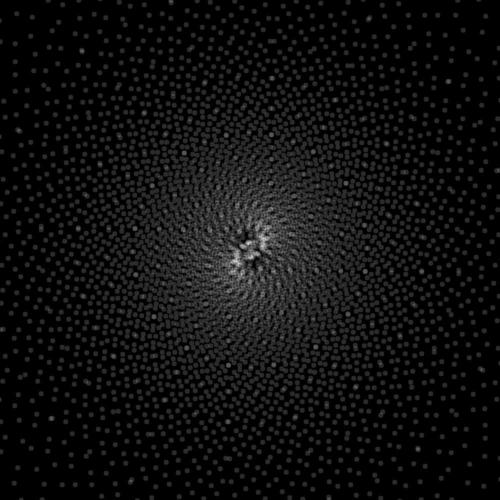}
   \vspace{.03in}

   \includegraphics[height=52mm]{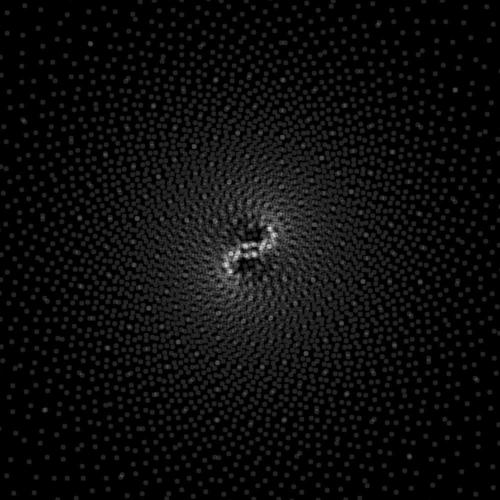}
   \includegraphics[height=52mm]{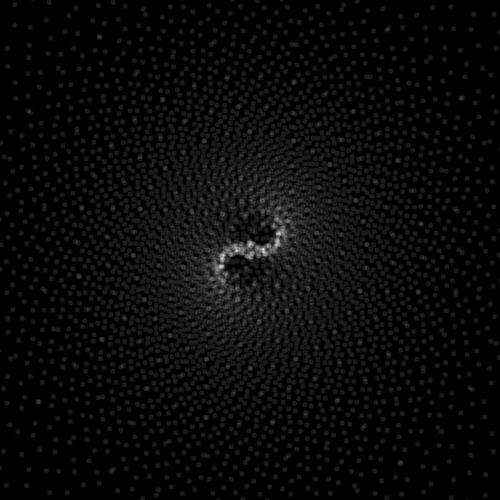}
   \includegraphics[height=52mm]{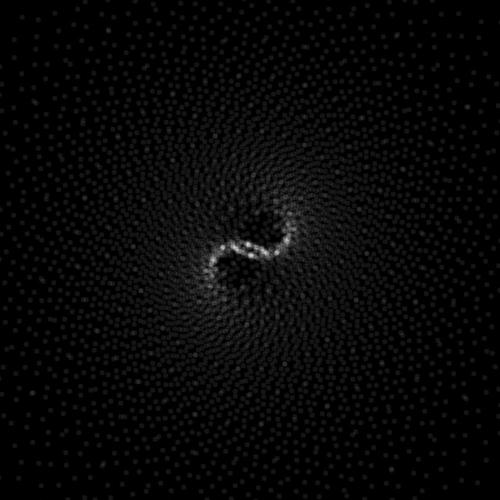}
   \vspace{.03in}

   \includegraphics[height=52mm]{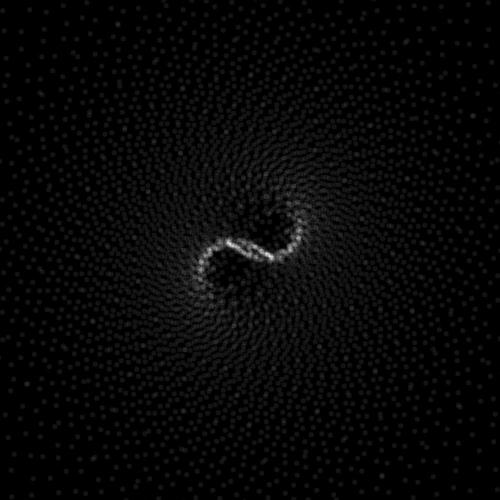}
   \includegraphics[height=52mm]{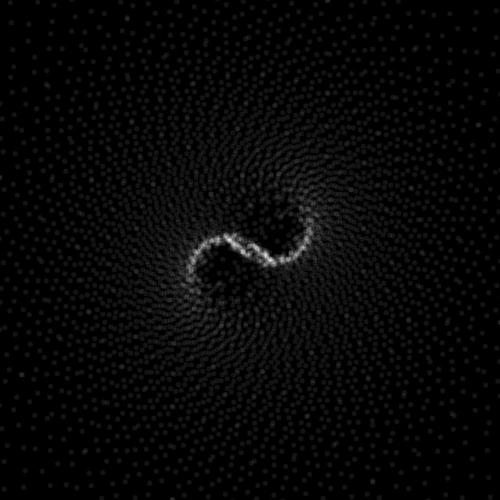}
   \includegraphics[height=52mm]{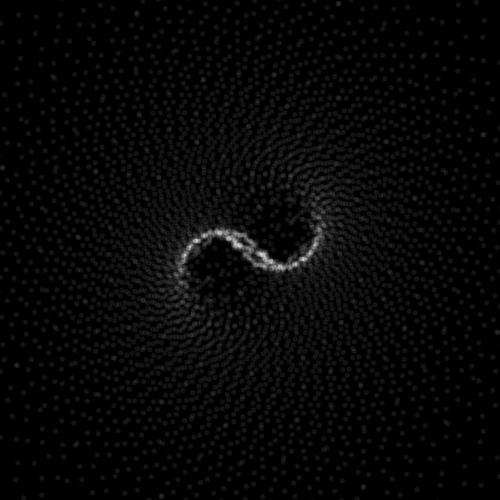}
   \vspace{.03in}

   \includegraphics[height=52mm]{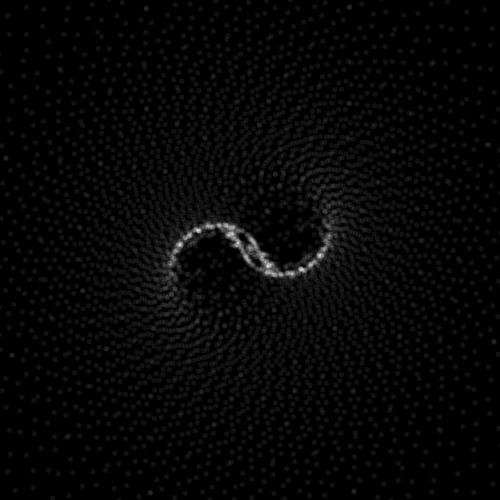}
   \includegraphics[height=52mm]{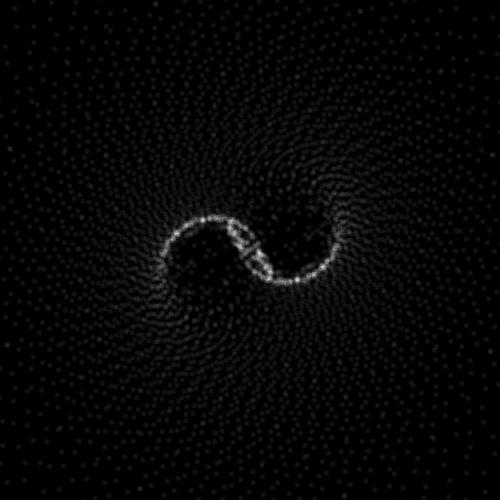}
   \includegraphics[height=52mm]{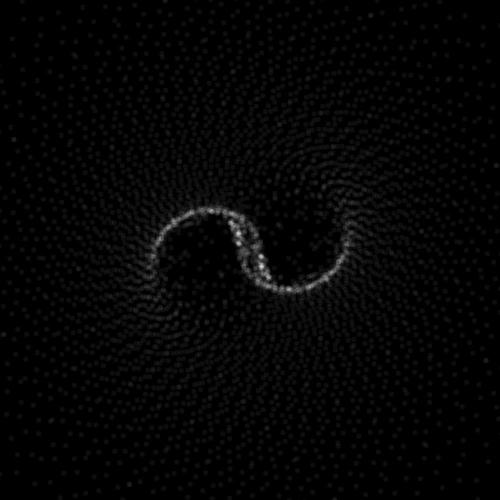}
   \vspace{.03in}

   \end{center}
   \caption{$t=0$ to $t = 11$ million years for Spiral Galaxy Simulation \#1.}
\end{figure}

\begin{figure}
   \begin{center}

   \includegraphics[height=52mm]{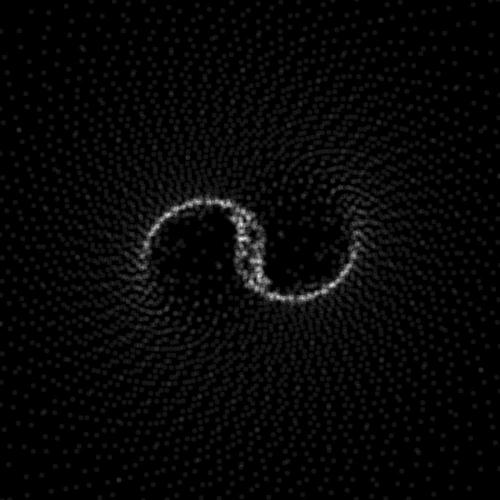}
   \includegraphics[height=52mm]{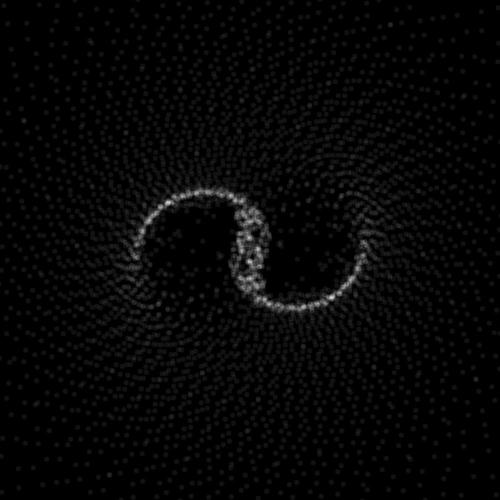}
   \includegraphics[height=52mm]{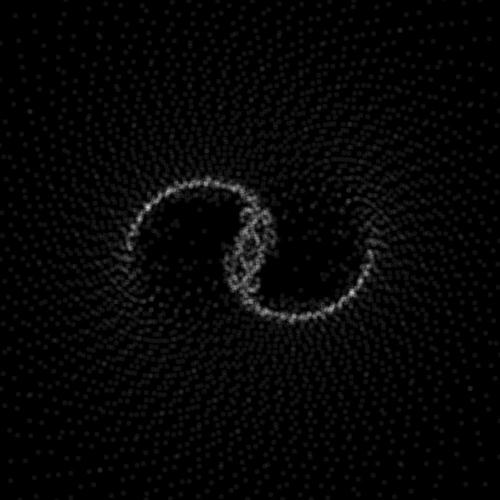}
   \vspace{.03in}

   \includegraphics[height=52mm]{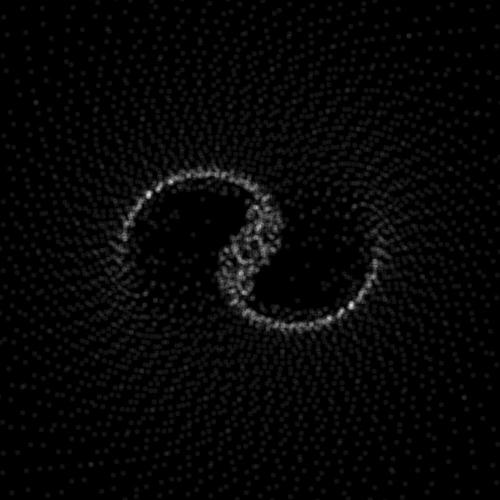}
   \includegraphics[height=52mm]{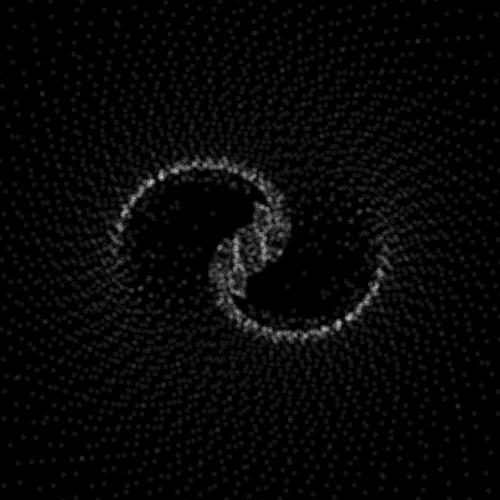}
   \includegraphics[height=52mm]{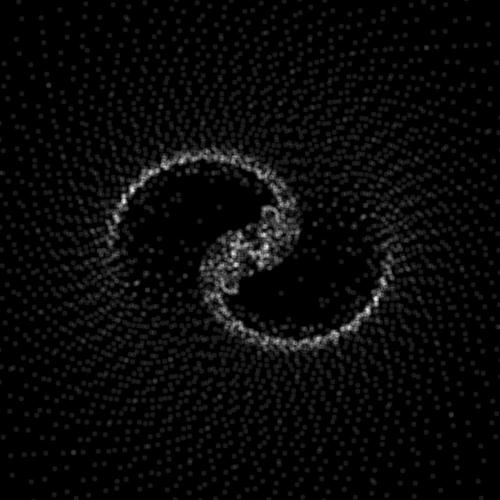}
   \vspace{.03in}

   \includegraphics[height=52mm]{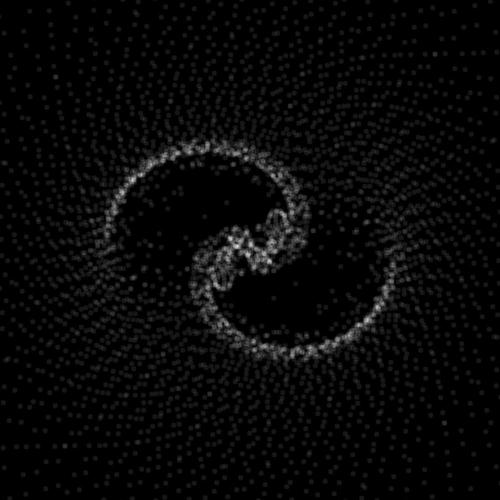}
   \includegraphics[height=52mm]{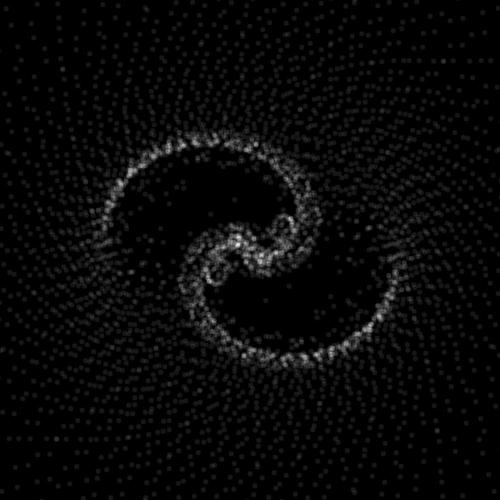}
   \includegraphics[height=52mm]{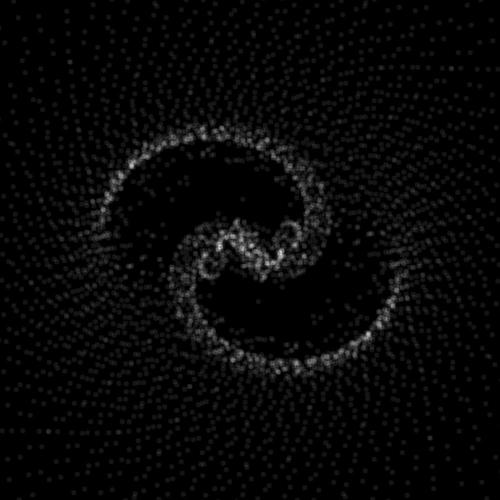}
   \vspace{.03in}

   \includegraphics[height=52mm]{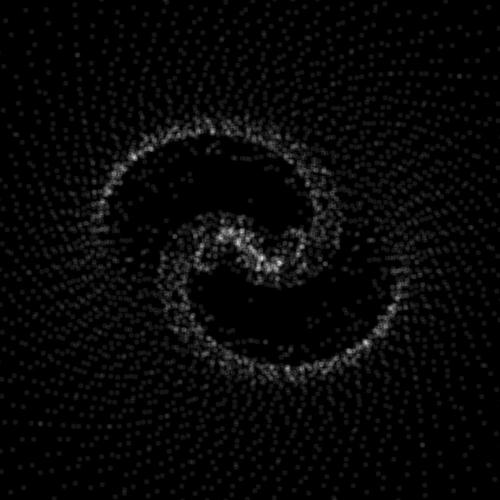}
   \includegraphics[height=52mm]{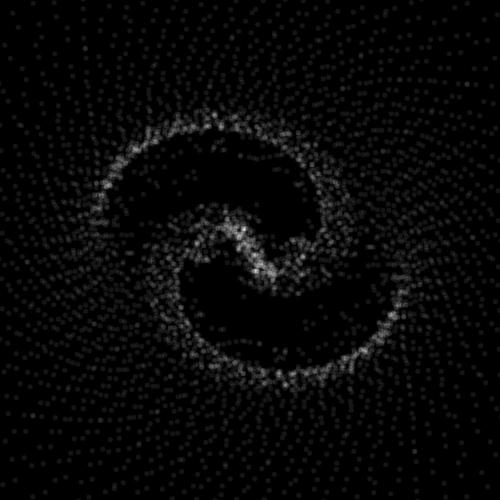}
   \includegraphics[height=52mm]{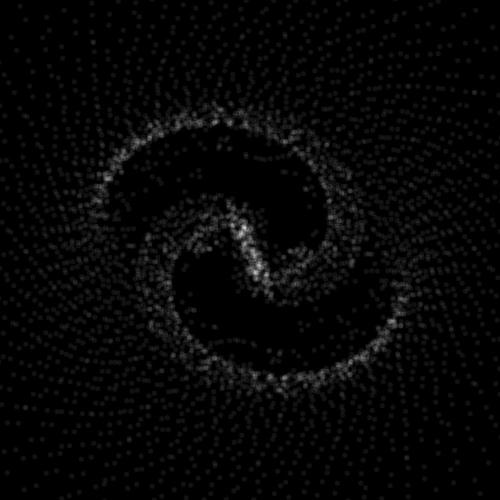}
   \vspace{.03in}

   \end{center}
   \caption{$t=12$ million years to $t = 23$ million years for Spiral Galaxy Simulation \#1.}
\end{figure}

\begin{figure}
   \begin{center}

   \includegraphics[height=52mm]{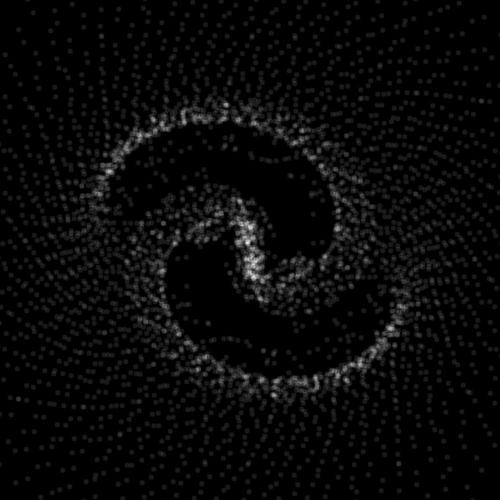}
   \includegraphics[height=52mm]{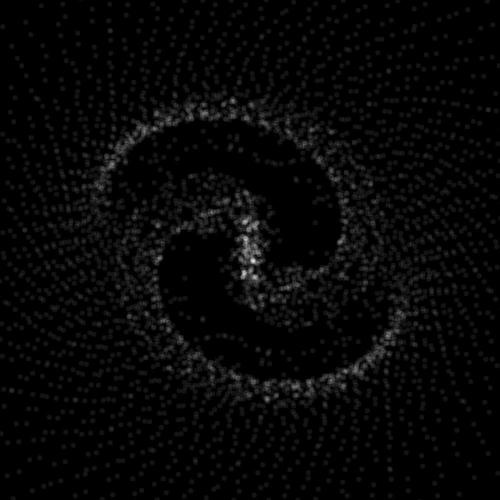}
   \includegraphics[height=52mm]{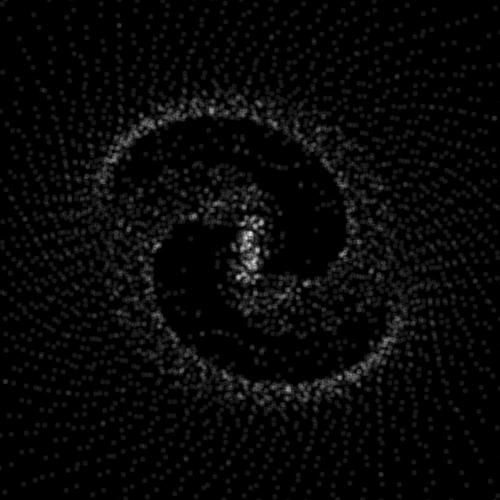}
   \vspace{.03in}

   \includegraphics[height=52mm]{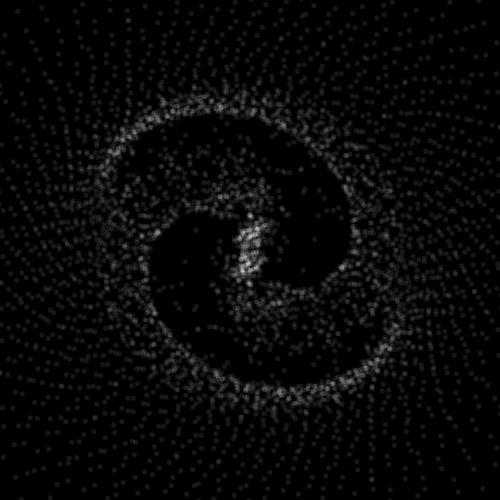}
   \includegraphics[height=52mm]{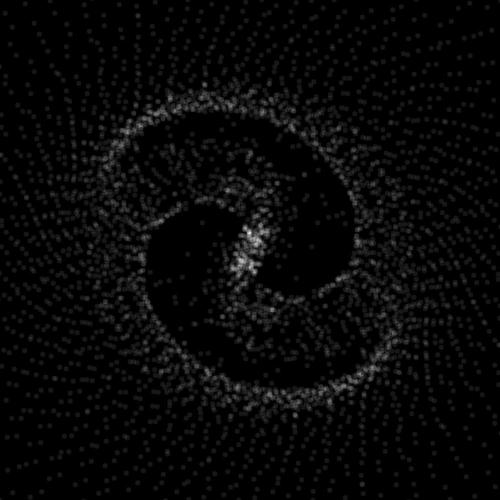}
   \includegraphics[height=52mm]{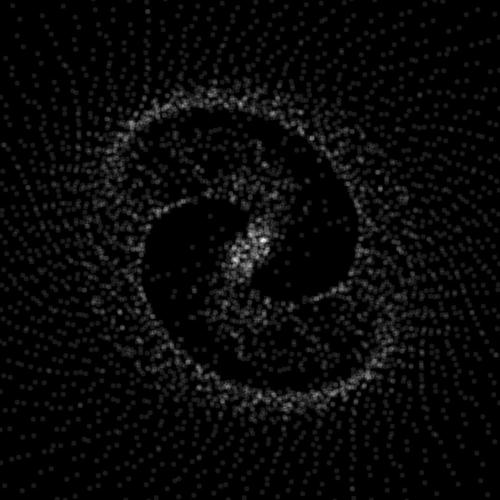}
   \vspace{.03in}

   \includegraphics[height=52mm]{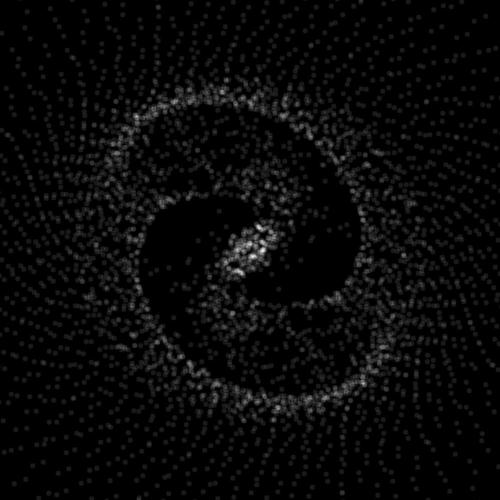}
   \includegraphics[height=52mm]{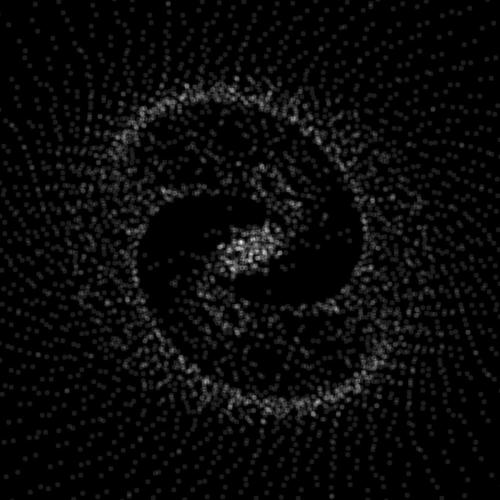}
   \includegraphics[height=52mm]{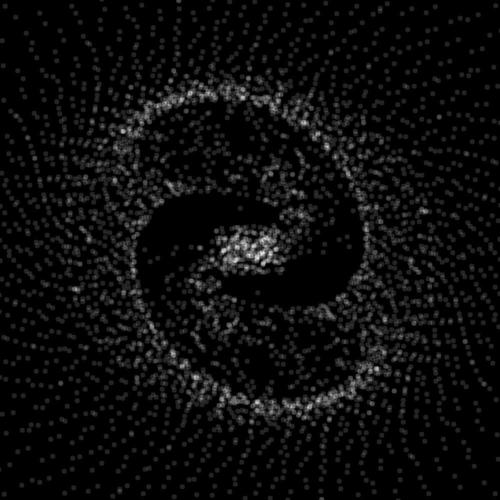}
   \vspace{.03in}

   \includegraphics[height=52mm]{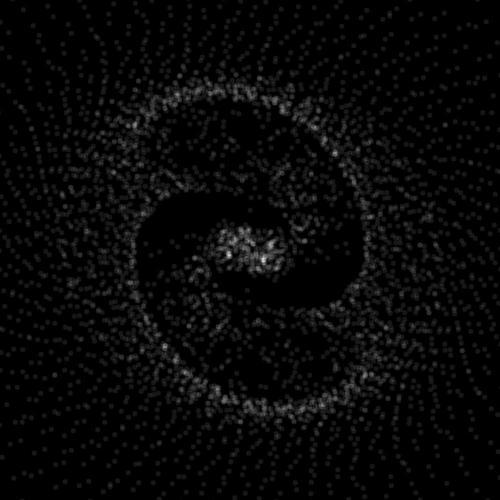}
   \includegraphics[height=52mm]{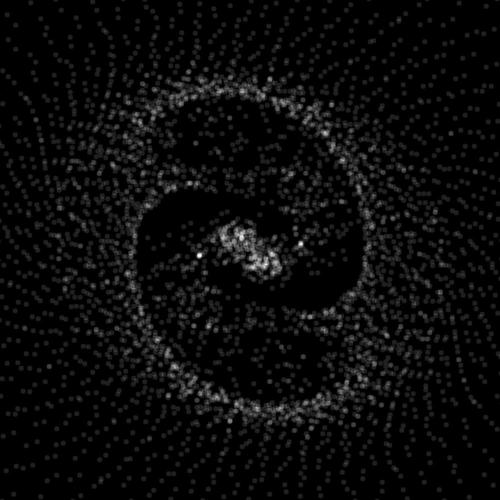}
   \includegraphics[height=52mm]{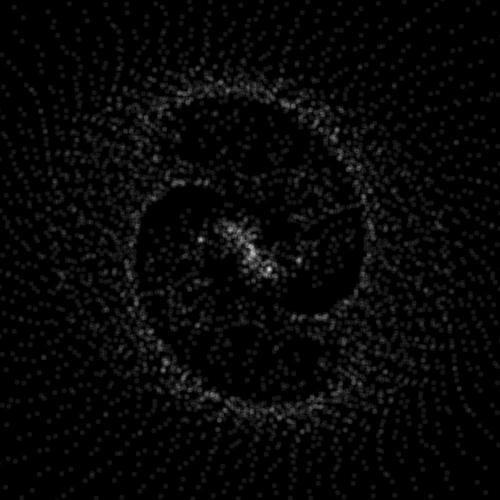}
   \vspace{.03in}

   \end{center}
   \caption{$t=24$ million years to $t = 35$ million years for Spiral Galaxy Simulation \#1.}
\end{figure}

\begin{figure}
   \begin{center}

   \includegraphics[height=52mm]{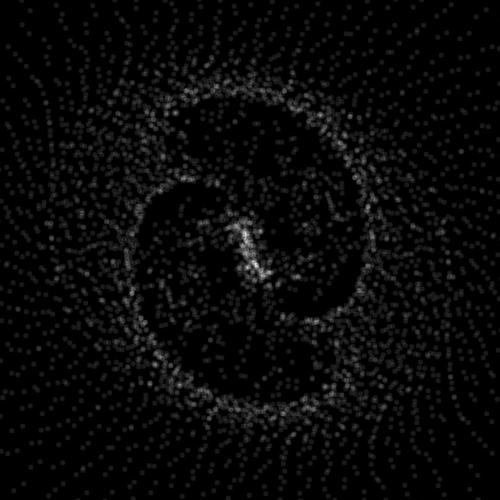}
   \includegraphics[height=52mm]{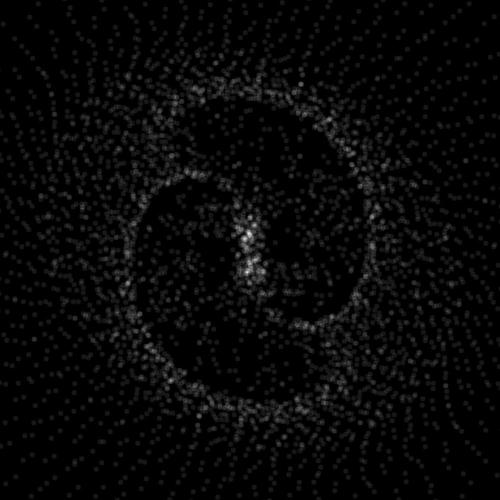}
   \includegraphics[height=52mm]{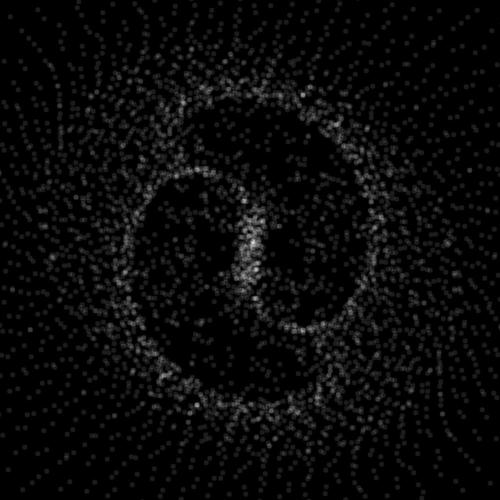}
   \vspace{.03in}

   \includegraphics[height=52mm]{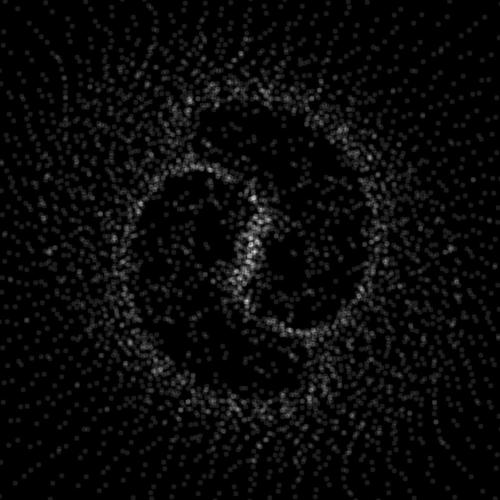}
   \includegraphics[height=52mm]{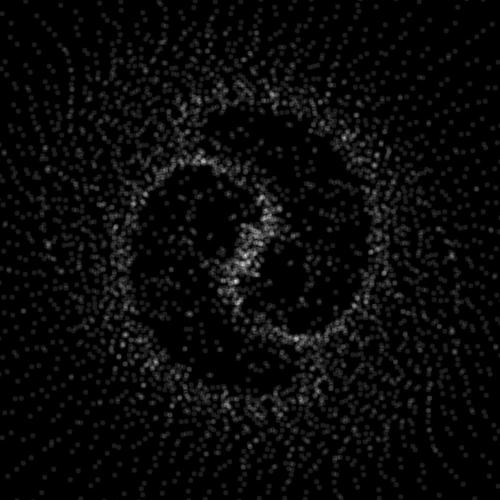}
   \includegraphics[height=52mm]{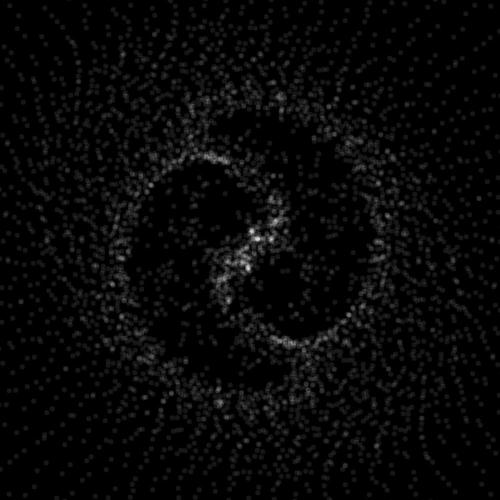}
   \vspace{.03in}

   \includegraphics[height=52mm]{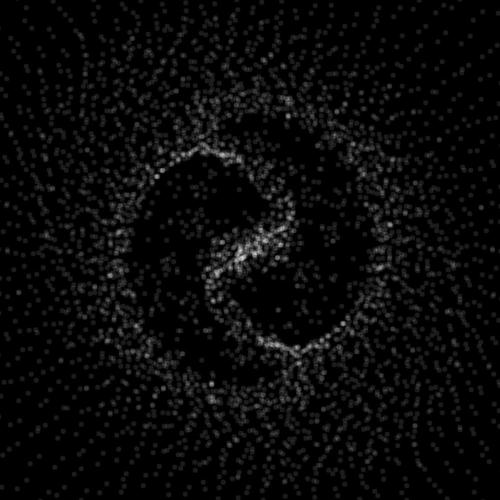}
   \includegraphics[height=52mm]{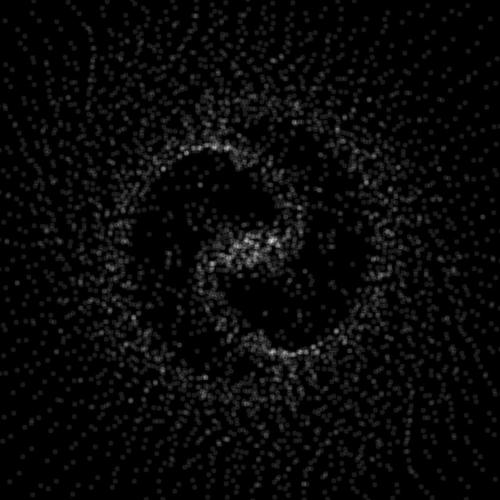}
   \includegraphics[height=52mm]{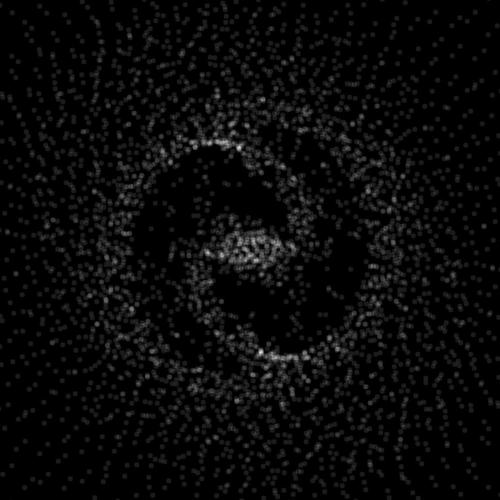}
   \vspace{.03in}

   \includegraphics[height=52mm]{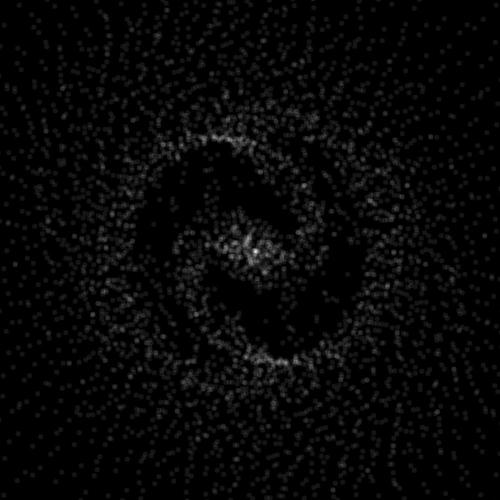}
   \includegraphics[height=52mm]{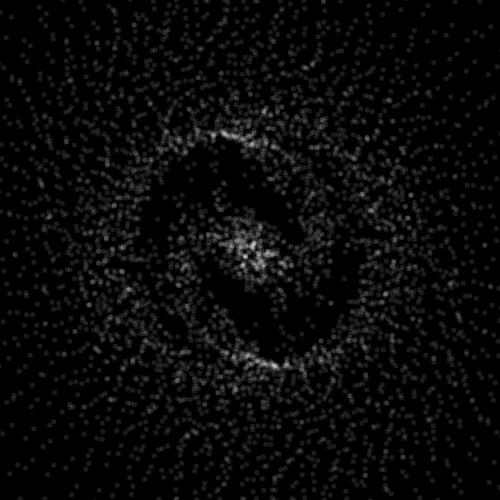}
   \includegraphics[height=52mm]{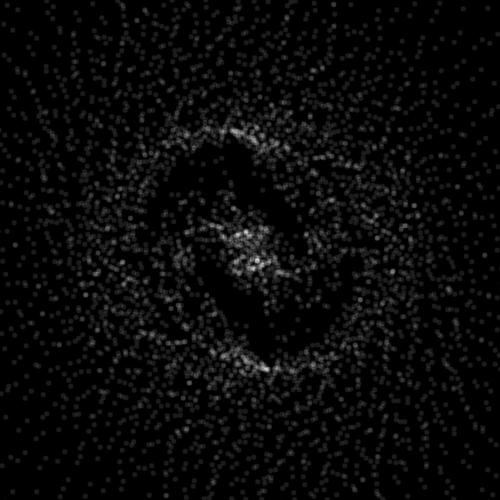}
   \vspace{.03in}

   \end{center}
   \caption{$t=36$ million years to $t = 47$ million years for Spiral Galaxy Simulation \#1.}
\end{figure}

\begin{figure}
   \begin{center}
   \includegraphics[height=52mm]{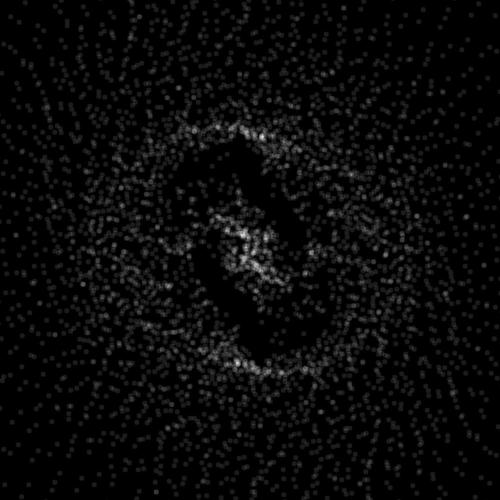}
   \includegraphics[height=52mm]{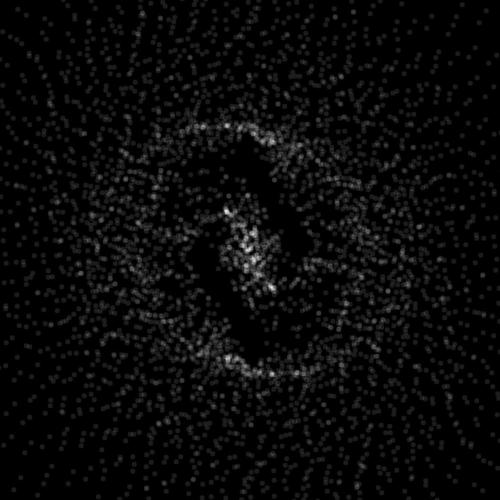}
   \includegraphics[height=52mm]{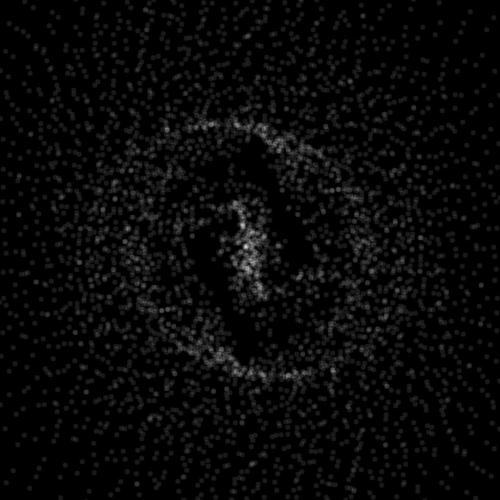}
   \end{center}
   \caption{$t=48$ million years to $t = 50$ million years for Spiral Galaxy Simulation \#1.}
\end{figure}

In figures 14 - 18, we show the results of our simulations.  In this
example, $5000$ test particles begin in circular motion with an
initial exponentially decaying density as described by equation
\ref{dsl}, which can be seen in the top left picture of figure 14.
If the potential function where spherically symmetric (corresponding
to $A_2 = 0$), these test particles would continue in perfectly
circular motion which would preserve the original density and image.
However, since our potential is not spherically symmetric, the
particles deviate from perfectly circular paths and respond to the
aspherical gravitational influence of the scalar field dark matter's
gravitational potential. This results in density waves in the test
particles, as can be seen in the figures. Each frame represents $1$
million years of elapse time.

The image visualization process that we have used to generate the
images in figures 14-18 is very important.  When plotting many data
points in the standard way, it is hard to see what the true density
of data points is in regions where the data points overlap each
other a lot. Instead, we have used the following method: Think of
each data point as a disk of a small radius with uniform brightness.
Define the overall brightness function to be the sum of the
brightnesses of all of the test disks, and then graph this overall
brightness function.  In this way, the overall brightness of each
frame in the figures should be the same. However, we have normalized
each frame so that the maximum value of the brightness in each frame
corresponds to $100\%$ white and zero brightness corresponds to
black, so the overall apparent brightness of each frame varies
slightly.

\subsection{Star Formation in Spiral Galaxies}\label{StarFormation}

Probably the most striking feature of the images in figures 14 - 18
is the development of the barred spiral pattern which, after roughly
$25$ million years, somewhat resembles the appearance of NGC 1300.
Even more generally, it is interesting that somewhat stable patterns
develop at all.  At a mini-conference at the Petters Research
Institute in Dangriga, Belize, Arlie Petters recognized the
existence of folds in these images, and pointed out that folds
naturally have a short time stability because of their topological
nature.  Furthermore, the density goes to infinity along a fold. To
be explicit, consider the mapping $\Phi_t: R^2 \rightarrow R^2$
represented by mapping the position of the particles at time zero to
their positions at time $t$. Naturally the density at $\Phi_t(x)$ at
time $t$ will increase by a factor of $1/|D\Phi_t(x)|$, which goes
to infinity along a fold in the mapping.  Hence, when a generic fold
in our simulation develops, it will cause a curve of infinite
density for as long as the generic fold remains.  Of course one
could expand this analysis to all of phase space for other more
general situations, but there is no need to do this here since the
velocity of our particles is a function of position at $t=0$.

Hence, this ``fold argument'' predicts density waves with short time
stability.  Moreover, along the fold curve the density goes to
infinity.  Naturally, the theoretical prediction of regions of
infinite density could be very important for understanding the large
scale process of star formation, especially since there is
dramatically increased star formation in the arms of spiral galaxies
\cite{BT}.  We note that this same fold argument can be generalized
to apply to three dimensional space as well. Hence, we put forth the
possibility that dark matter enforcing a ``fold dynamics'' upon gas
and dust clouds could be a major driver, possibly even the primary
driver, of star formation. We leave this as a very interesting
question to study.

\subsection{Long Time Behavior}

Before moving to the next simulation, we point out that it is not
entirely clear what all of the implications of the first simulation
are. For example, is NGC 1300 in a steady-state configuration with a
precise pattern speed, or will it evolve at some rate similar to the
pictures in figures 14-18?  At this point we can not answer this
question because the simulation just described does not account for
anything other than the gravity of the dark matter.  Hence, we
really can not say anything definitive on this question.

One possibility, which we discuss here for the sake of argument, is
that spiral galaxies result from at least two important effects, the
first being the gravity of dark matter, and the second, for example,
being friction of gas and dust and supernovae which replenish the
supply of gas and dust.  This second effect, whatever it is exactly,
would be responsible for keeping most of the matter of the galaxy in
the disk with roughly circular motion (and hence would probably not
be time reversible). Ideally, this second effect would partially
justify our choice of initial conditions in our simulations in some
general sense. In this manner, if such a second effect like this
exists, we could view a galaxy like NGC 1300 as being the result of
two effects fighting against each other, one pushing the state
toward something similar to the $t=0$ initial conditions, and the
other pushing distributions of regular matter into spiral patterns.
This could conceivably lead to something close to a steady state
equilibrium of a trailing spiral pattern rotating at a constant
pattern speed.  If this second effect is not time reversible (like
friction), then we are immune from the anti-spiral theorem
\cite{antispiral}.  Hence, the existence of steady state trailing
spiral patterns would not necessarily imply the existence of steady
state leading spiral patterns, which is good since they are
typically not observed \cite{BT}.  Clearly very careful simulations
are needed to shed light on these questions, about which at this
point we can only speculate.

The other issue to consider is the behavior of old stars.  Old stars
also are mostly in the disk of the galaxy going in roughly circular
motion, but with a higher variance than the gas and dust and younger
stars \cite{BT}.  Does dynamical friction of these older stars with
the gas, dust, and other stars or some other process ``cool'' the
velocities of these older stars and keep them in the galactic disk
with roughly circular orbits?  If so, then the simulations presented
here might be relevant for producing spiral patterns in the older
stars as well.  An answer to the above question is needed, at a
minimum, though, before one can make definitive conclusions.

\subsection{Spiral Galaxy Simulation \#2}\label{SGS2}

As a second example, consider
\begin{equation}
   \mbox{spiralgalaxy}(1,75000,1,-0.15,2000,1990,25000000,8.7e-13,7500,5000,50000000,50000)
\end{equation}
which produces the simulated image in figure 2 as the $t = 30$
million years image of figure 23. In figure 2, we compare this image
to NGC 4314, a barred spiral galaxy of type SBa.  The main
difference between this example and the previous simulation is that
$A_2$, which controls the size of the interference pattern which
rotates, is now only $0.15$ instead of $1$.  Hence, the resulting
potential is much more spherically symmetric.  (The $dt$ step size
is also $50,000$ now up from $10,000$, but this makes little
difference.)  Interestingly, we notice that this smaller value for
$A_2$ results in a more oval bar than in the previous example.
Hence, the shape of the bar, whether narrow like a bar as in the
previous example or thick like an oval as in this example, is a
characteristic which can be modeled by our simulations.

\begin{figure}
   \begin{center}
   \includegraphics[height=59mm]{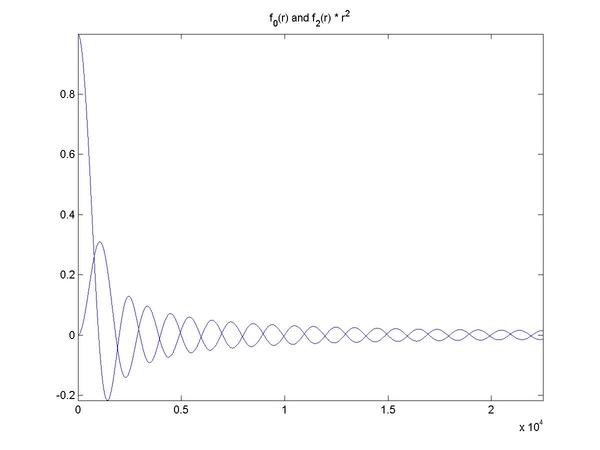}
   \includegraphics[height=59mm]{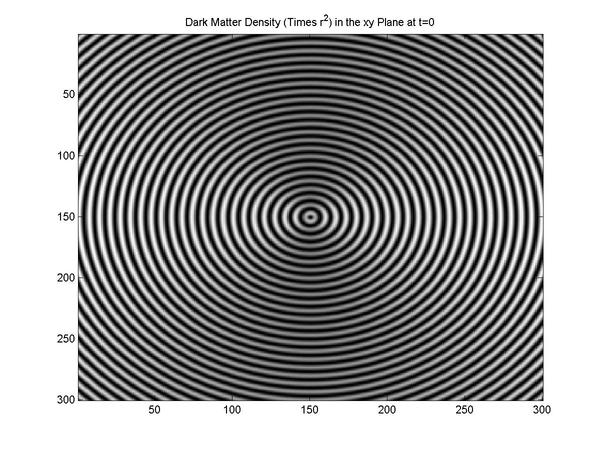}
   \includegraphics[height=59mm]{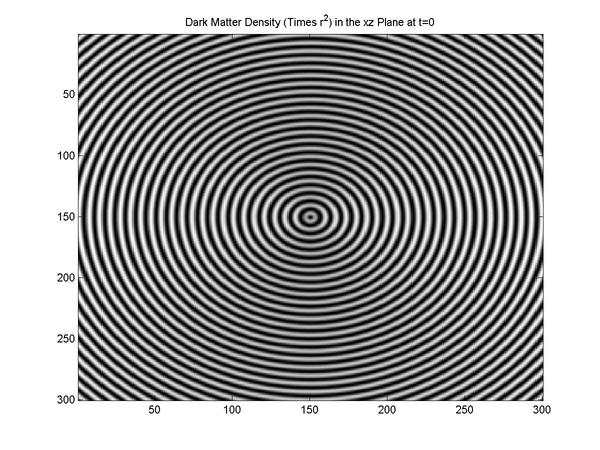}
   \includegraphics[height=59mm]{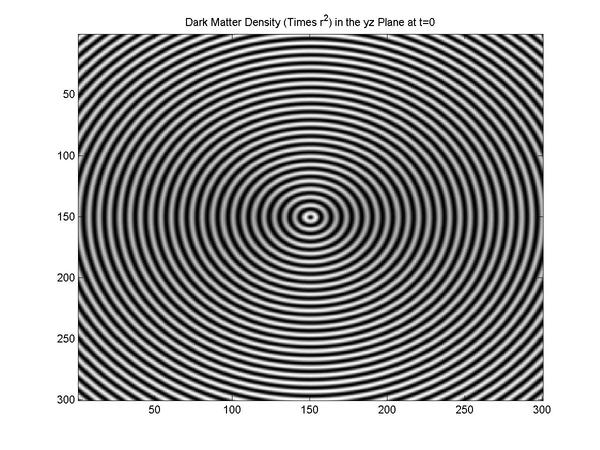}
   \end{center}
   \caption{Spiral Galaxy Simulation \# 2:  Graphs of $f_{\omega_0,0}(r)$ and $r^2 f_{\omega_2,2}(r)$ for
   $r$ up to $22,500$ light years (top left).  The other three images, each with a radius
   of $22,500$ light years, are plots of the dark matter
   density (in white) times $r^2$ in the $xy$ plane (top right), in the $xz$ plane (bottom left),
   and in the $yz$ plane (bottom
   right).  The densities, which roughly decrease like $1/r^2$, have been multiplied by $r^2$
   to be more easily visible.  Notice how the interference pattern is much subtler
   since $|A_2 / A_0| = 0.15$ now as opposed to $1$ in the previous simulation.}
\end{figure}

\begin{figure}
   \begin{center}
   \includegraphics[height=59mm]{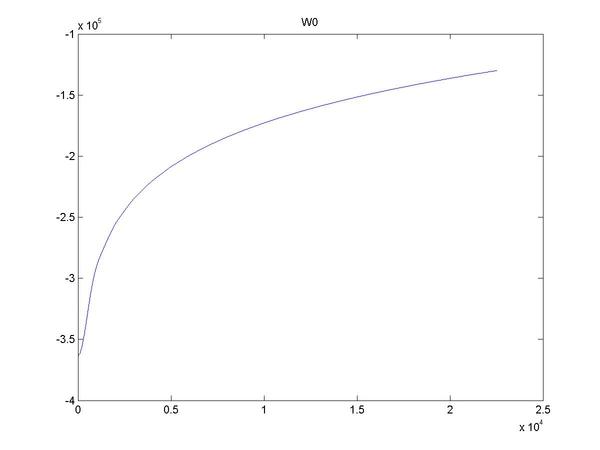}
   \includegraphics[height=59mm]{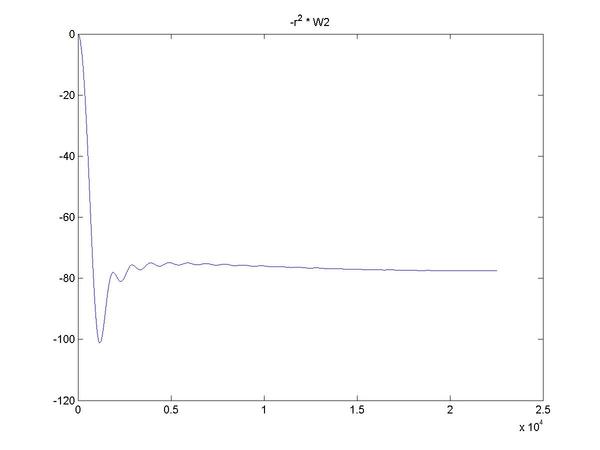}
   \includegraphics[height=59mm]{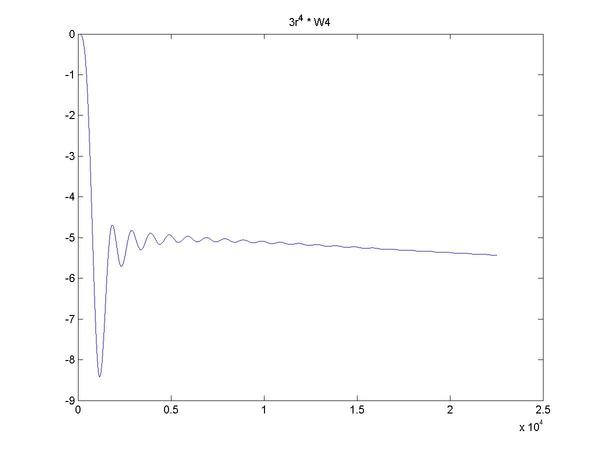}
   \includegraphics[height=59mm]{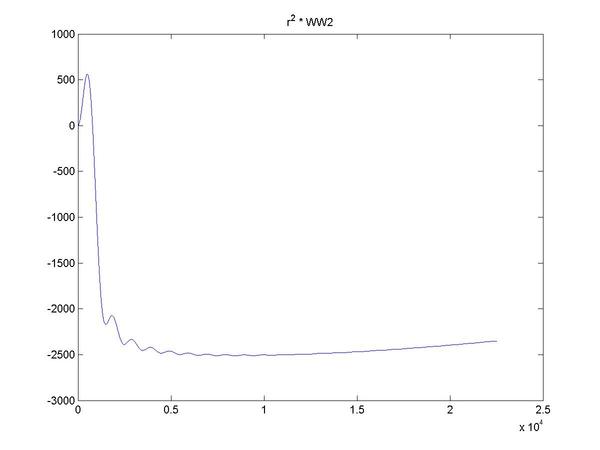}
   \end{center}
   \caption{Plots of the radial functions $W_0(r)$ (top left), $-r^2 W_2(r)$ (top right), $3r^4 W_4(r)$ (bottom left),
   and $r^2 \tilde{W}_2(r)$ (bottom right) composing the potential function for Spiral Galaxy Simulation \#2.
   Notice how
   the spherically symmetric contribution given by $W_0(r)$ dominates the rotating component described by
   $r^2 \tilde{W}_2(r)$ which in turn dominates the remaining two terms.}
\end{figure}

\begin{figure}
   \begin{center}
   \includegraphics[height=59mm]{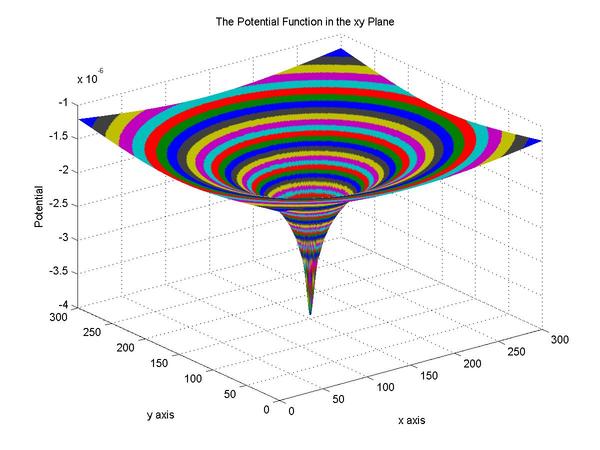}
   \includegraphics[height=59mm]{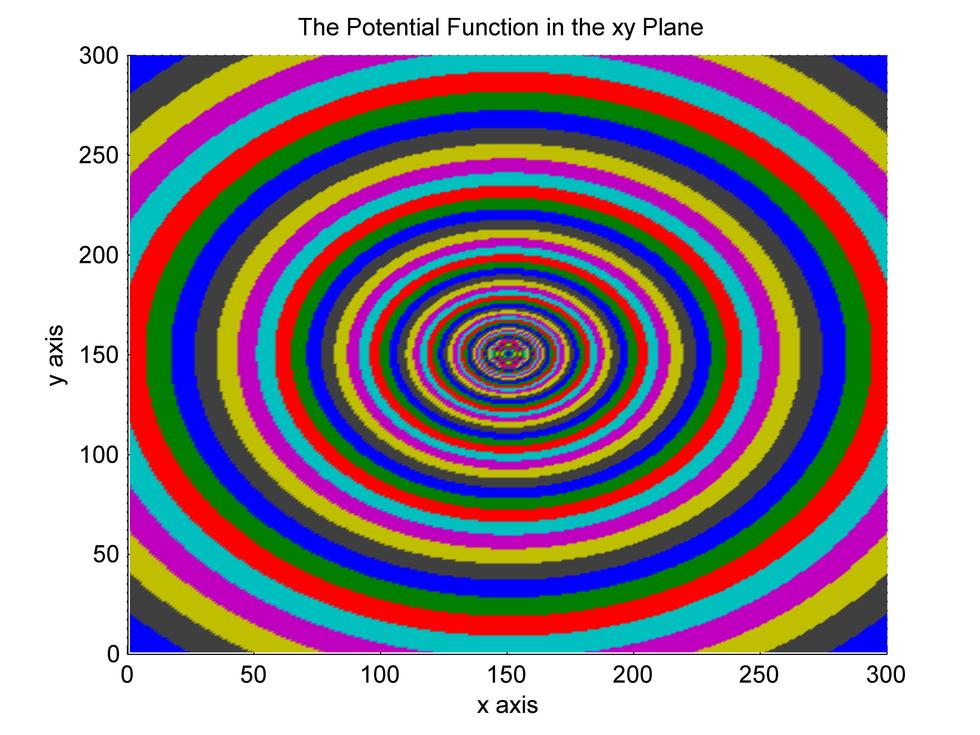}
   \includegraphics[height=59mm]{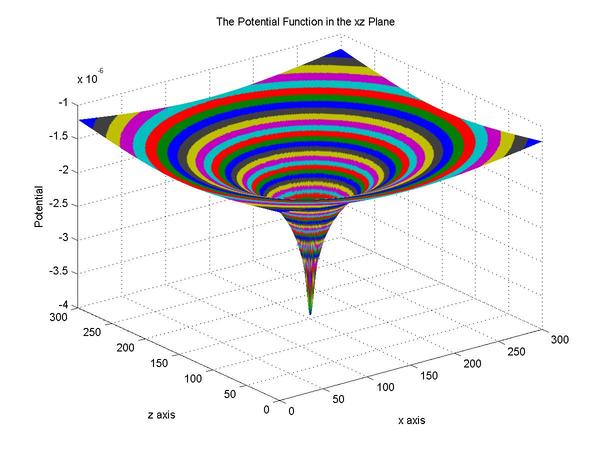}
   \includegraphics[height=59mm]{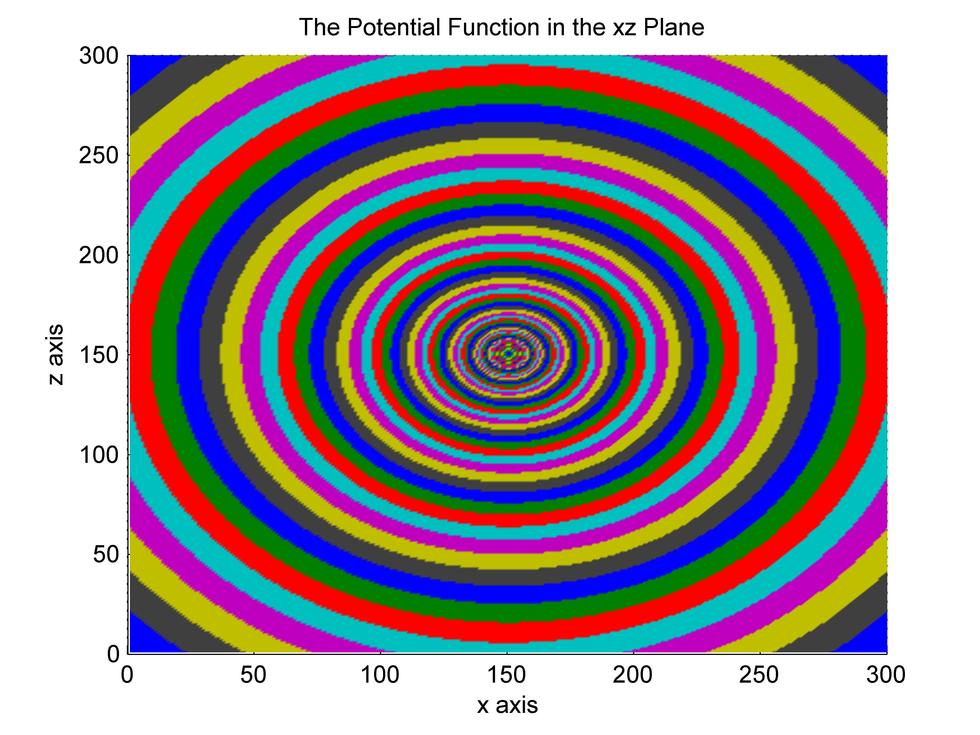}
   \includegraphics[height=59mm]{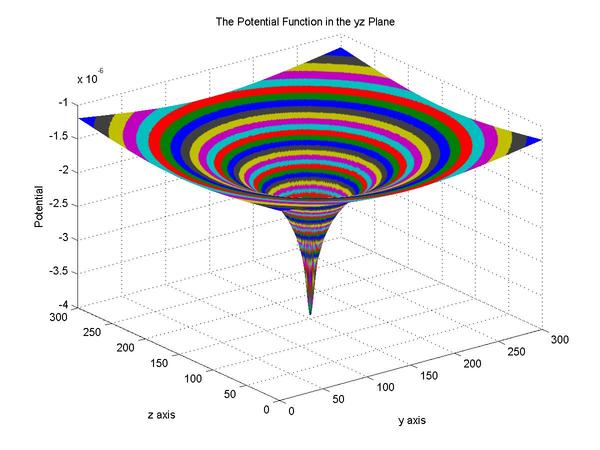}
   \includegraphics[height=59mm]{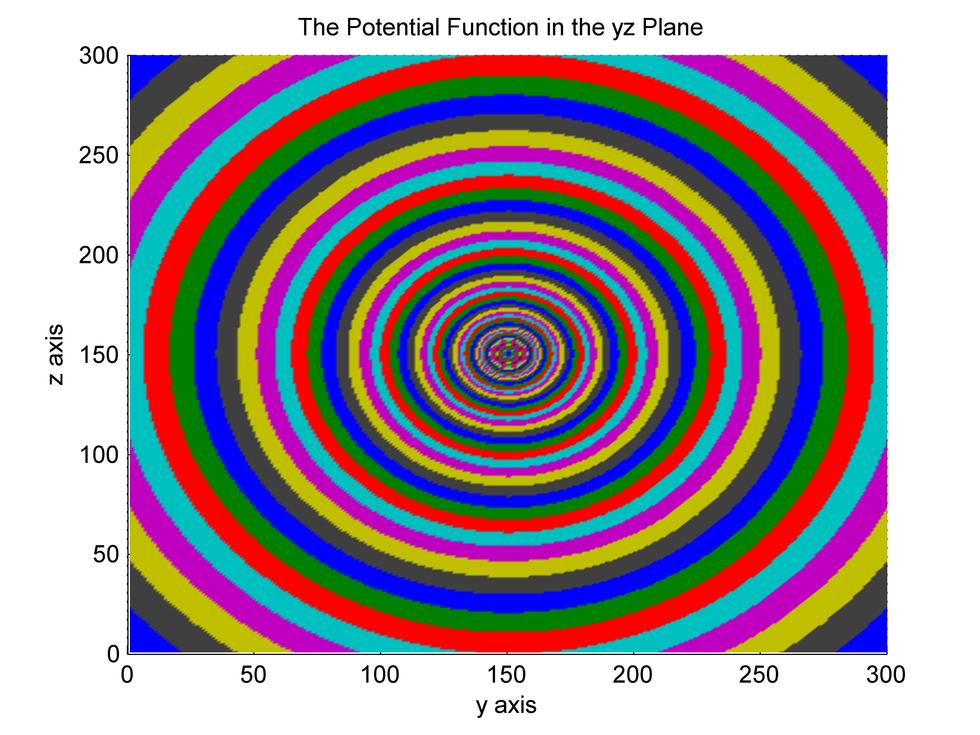}
   \end{center}
   \caption{Graphs of the potential function in the $xy$ plane (top row), the $xz$ plane (middle row),
   and the $yz$ plane (bottom row) out to a radius of $22,500$ light years for Spiral Galaxy Simulation \#2.
   The second column is the same as the first column except that the point of view is looking straight down so
   that we can see the level sets of the potential function in each plane.  There is some distortion in the image by the
   Matlab graphics algorithm in that the domains on the right are actually perfect squares, not rectangles.  From this
   we can deduce that the level sets are slightly ellipsoidal.}
\end{figure}

\begin{figure}
   \begin{center}
   \includegraphics[height=120mm]{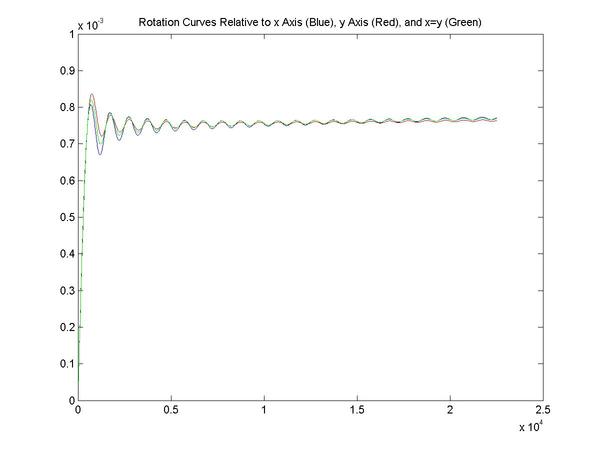}
   \end{center}
   \caption{Approximate rotation curves for Spiral Galaxy Simulation
   \#2
   out to a radius of $22,500$ light years.  We have approximated the rotation curves with
   graphs of $\sqrt{r |\nabla V|}$ (which is exactly correct in the spherically symmetric case)
   along the $x$ axis (in blue), along the $y$ axis (in red), and along $y=x$ (in green).  Notice how the three
   curves are much more similar than in the previous simulation since the potential is closer to being spherically
   symmetric.}
\end{figure}

\begin{figure}
   \begin{center}

   \includegraphics[height=52mm]{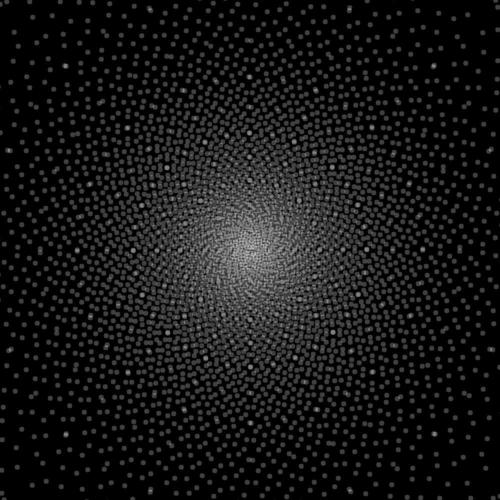}
   \includegraphics[height=52mm]{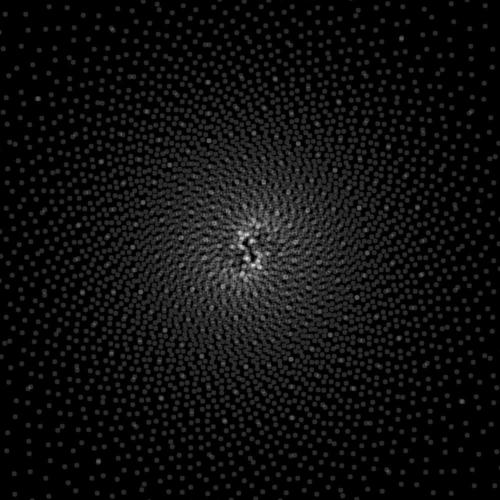}
   \includegraphics[height=52mm]{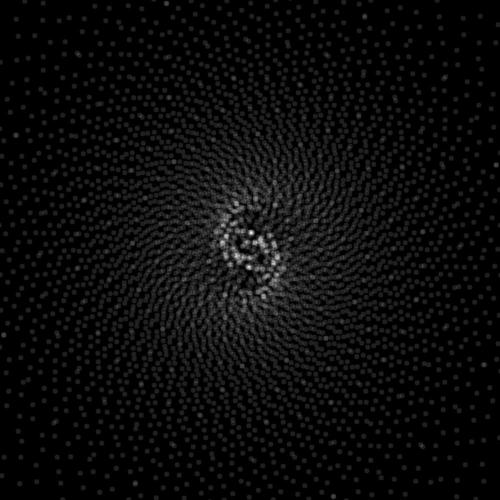}
   \vspace{.03in}

   \includegraphics[height=52mm]{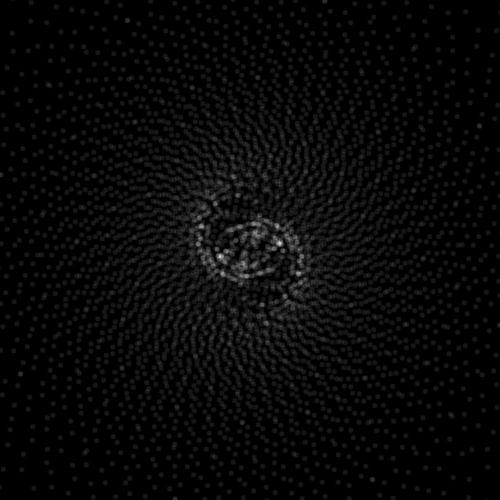}
   \includegraphics[height=52mm]{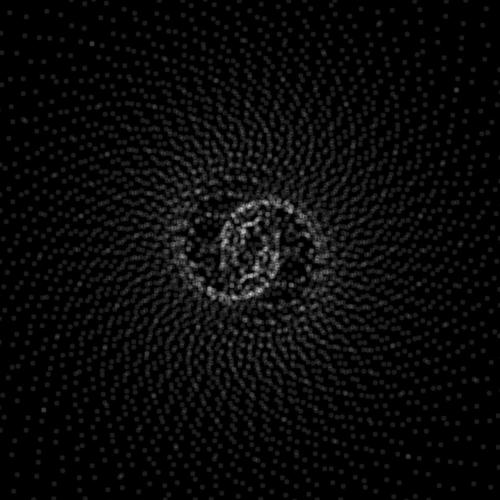}
   \includegraphics[height=52mm]{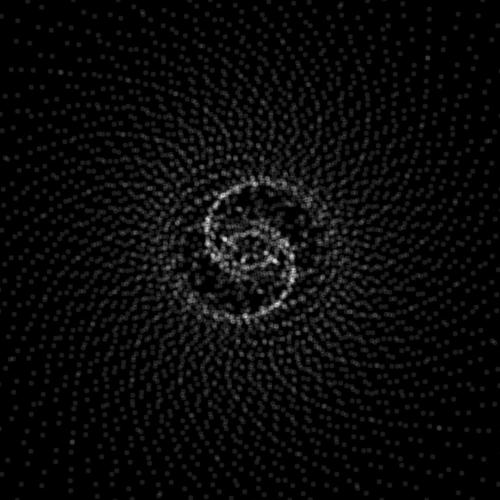}
   \vspace{.03in}

   \includegraphics[height=52mm]{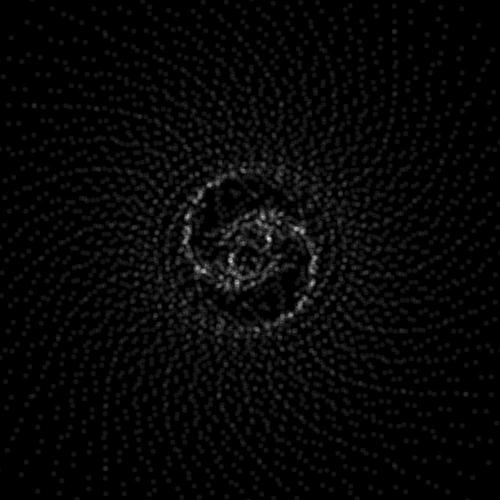}
   \includegraphics[height=52mm]{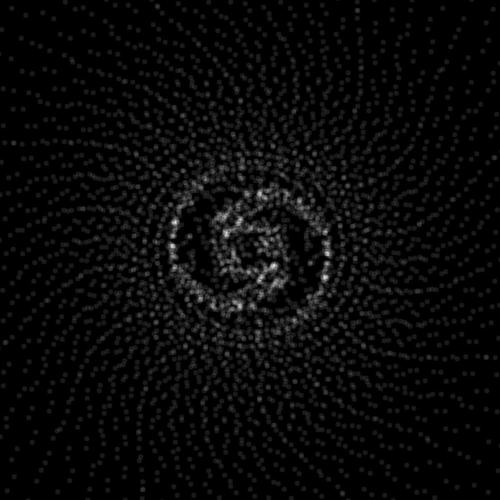}
   \includegraphics[height=52mm]{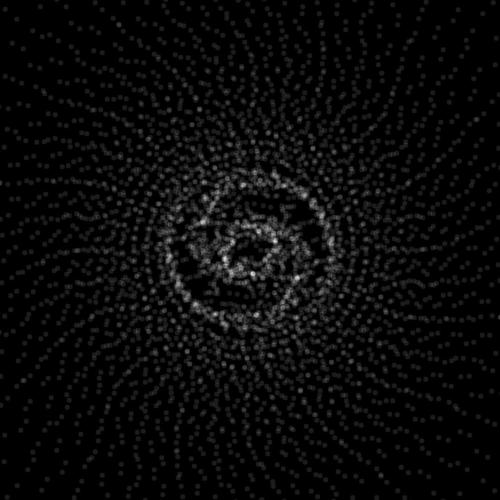}
   \vspace{.03in}

   \includegraphics[height=52mm]{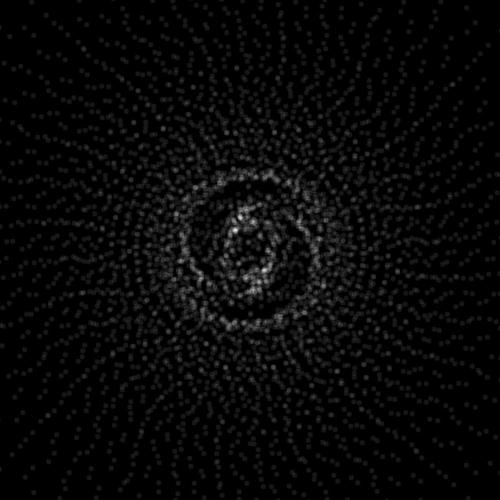}
   \includegraphics[height=52mm]{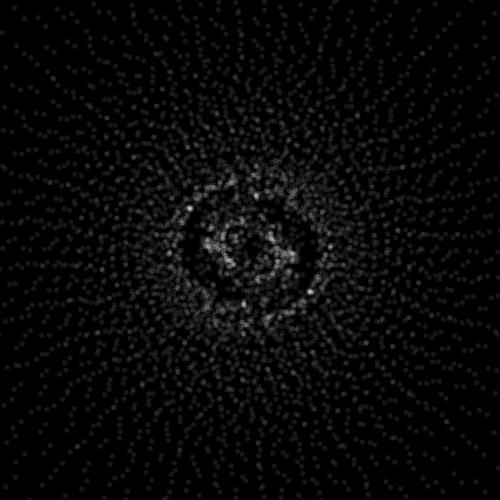}
   \vspace{.03in}

   \end{center}
   \caption{$t=0$ to $t = 50$ million years (in steps of $5$ million years) for Spiral Galaxy Simulation \#2.}
\end{figure}

The effect of lowering $A_2$ to $0.15$ can be seen in figures 19 -
22.  In figure 19, notice how the interference pattern, which still
rotates around once every $25$ million years as in the previous
example, is much more subtle now.  Also, in figure 20, the
spherically symmetric term represented by $W_0(r)$ dominates much
more than before, which is seen by looking at the values on the $y$
axes of these graphs.  In figure 21, the triaxiality of the
potential function is so mild now that it is hard to even be sure
that the level sets are not spheres from first inspection.  Finally,
in figure 22, the three approximate rotation curves, defined as
before along the $x$ axis, the $y$ axis, and the line $y = x$, are
very nearly equal, another consequence of being close to spherical
symmetry.

Figure 23 shows the results of the test particle simulations with
the computed dark matter potential (figure 21) rotating rigidly with
a period of $25$ million years.  As before, the test particles begin
in roughly circular motion but, over time, form the patterns shown.
As before, there appear to be folds which produce relatively dense
spiral arms as well as a general concentration in the central oval
bar region.

We use this example to make three more points:  First, in our spiral
galaxy simulations (unlike the elliptical galaxy simulations to
come), we assume an exponential initial density in the test
particles.  We do not derive this density, nor do we explain it with
our model.  We take this initial density as given, presumably from
other important effects like friction present in spiral galaxies
which we do not model.  Hence, the fact that the brightness goes to
zero for large radii is not something that we have attempted to
model for spiral galaxies (although we will attempt this for
elliptical galaxies).

Second, our initial distribution is not smooth but instead is
composed of $5,000$ tiny bright disks. While this visualization
method works well for us, this technique also produces artificial
patterns in the pictures which are not physical but instead reflect
the initial regular pattern in which these tiny bright disks were
originally placed.

Third, what is real, we believe, is that the background material not
in the spiral pattern does appear to be a larger percentage of the
test particles than in the previous example.  Hence, this also
appears to be a feature which is controlled by this model.

\newpage
\subsection{Spiral Galaxy Simulation \#3}\label{SGS3}

As a third example, consider
\begin{equation}
    \mbox{spiralgalaxy}(1,75000,1,-0.15,2000,1990,100000000,8.7e-13,7500,5000,95000000,50000)
\end{equation}
which produces the simulated image in figure 3 as the $t = 45$
million years image of figure 24, rotated $90$ degrees clockwise. In
figure 3, we compare this image to NGC 3310, a ``starburst'' spiral
galaxy of type Sbc which is forming stars at a rate much higher than
most other galaxies.  The dark matter density and the dark matter
potential in this example are exactly the same as in the previous
example.  In fact, the only difference between this third simulation
and the previous one is that the rate at which the dark matter
rotates has been slowed to a period of $100$ million years. This
would appear to have the effect of unwinding the spiral arms
compared to the previous example.

We also wonder if more slowly rotating dark matter density waves
could be correlated with starburst activity in galaxies where
greater numbers of stars are being formed, as in the case of NGC
3310.  This idea is consistent with the earlier ``fold dynamics''
suggestion that dark matter plays an important role in star
formation.  The slower the dark matter density waves move, the more
time they have to act on the regular matter before the dark matter
density wave passes through. This idea could conceivably be tested
in part by seeing if there are certain morphology types of galaxies
more commonly associated with starburst activity, although it is not
yet definitively clear what the prediction, if any, might be. Of
course, since there may be more than one cause of starburst activity
and multiple conditions required (such as sufficient gas and dust),
this is a delicate, but very interesting, question to study.

We also point out that changing the period of the dark matter
rotation changes the value of $\Upsilon$, as seen in equation
\ref{UpsilonEquation}.  In the first two simulations, $\Upsilon^{-1}
\approx 10.1$ light years.  In this third simulation, $\Upsilon^{-1}
\approx 2.53$ light years.  On the other hand, we have not paid much
attention to the overall scale of these galaxies. That is, we have
noticed that these simulations match the shapes of certain galaxies,
up to scale.  To actually estimate the value of $\Upsilon$ would
both require finding the range of values of our input parameters
which give best fits for these galaxies, up to scale, and then
scaling the resulting output morphologies to fit the actual
galaxies.  As we said before, this may be a good project, but we do
not attempt this in this paper.

Finally, one thing that we must say about these simulations in
general is that every run, for wide ranges of values for the input
variables, produces simulated images which look a lot like galaxies.
That is, we have not ``cherry picked'' the images shown in this
paper, other than to pick the ones which look as much as possible to
example galaxies for which we happen to have good pictures.  The
images we have not picked look a lot like galaxies too.  Of course,
since galaxies come in such a wide variety of shapes, this does not
prove anything.

On the other hand, not every galactic morphology is realized by
these simulations, nor was that expected.  For example, we
effectively built in a $180$ degree symmetry into our model by only
looking at a spherically symmetric component plus a second degree
spherical harmonic term. The next logical step is to widen the class
of solutions to the Klein-Gordon equation that are considered so
that additional morphologies of galaxies can be attempted to be
modeled.

\begin{figure}
   \begin{center}

   \includegraphics[height=52mm]{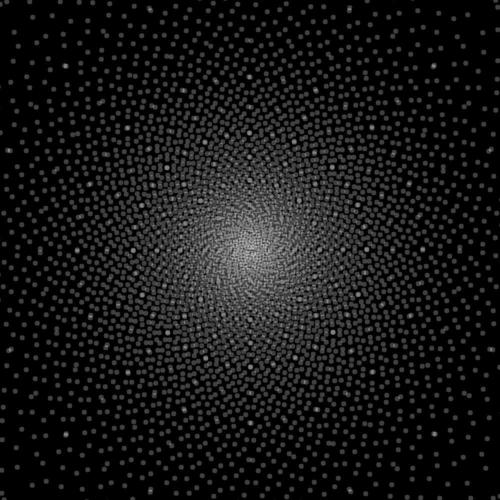}
   \includegraphics[height=52mm]{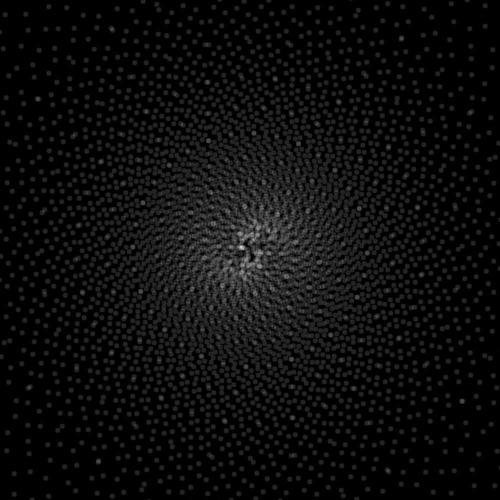}
   \includegraphics[height=52mm]{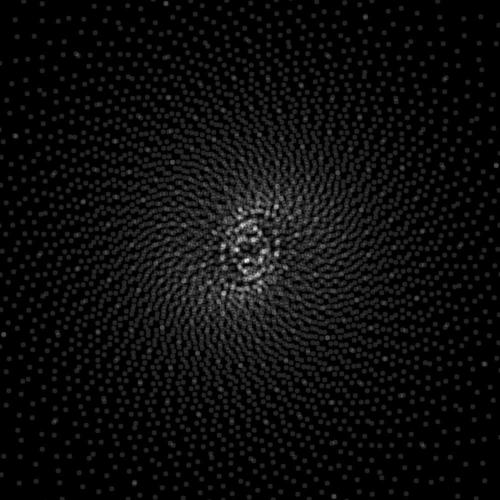}
   \vspace{.03in}

   \includegraphics[height=52mm]{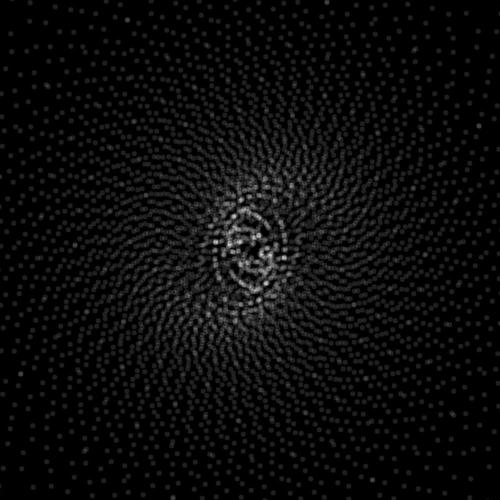}
   \includegraphics[height=52mm]{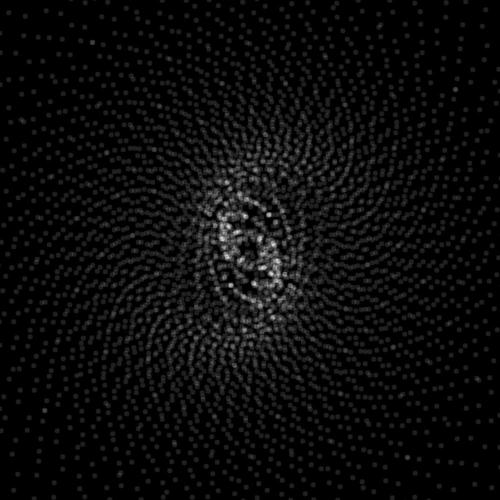}
   \includegraphics[height=52mm]{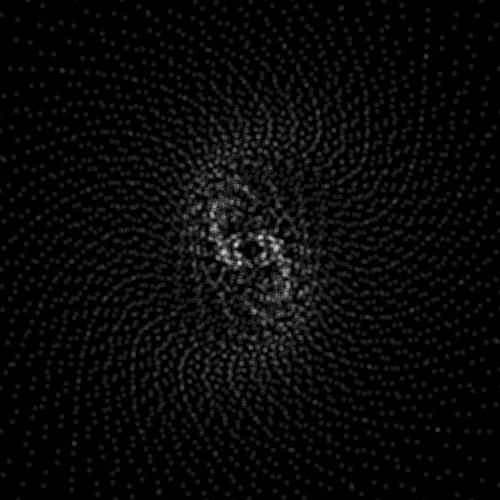}
   \vspace{.03in}

   \includegraphics[height=52mm]{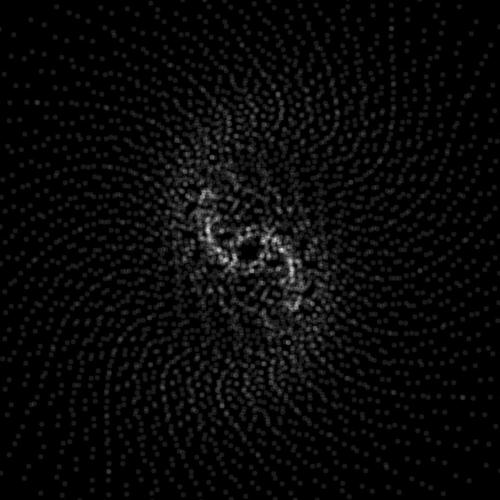}
   \includegraphics[height=52mm]{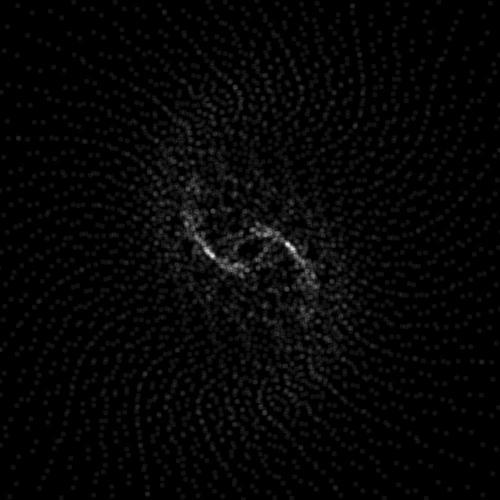}
   \includegraphics[height=52mm]{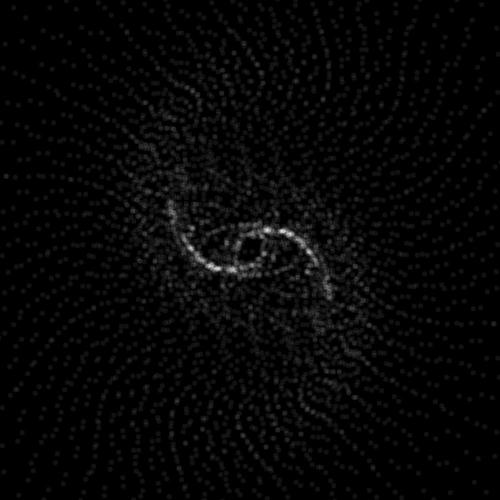}
   \vspace{.03in}

   \includegraphics[height=52mm]{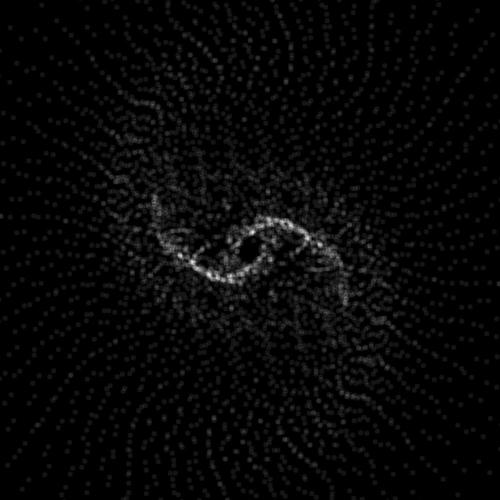}
   \includegraphics[height=52mm]{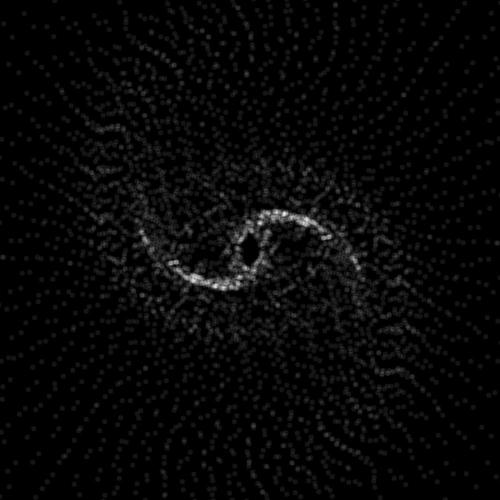}
   \includegraphics[height=52mm]{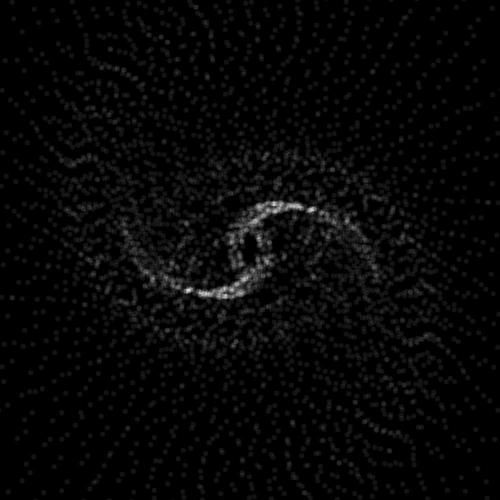}
   \vspace{.03in}

   \end{center}
   \caption{$t=0$ to $t = 55$ million years (in steps of $5$ million years) for Spiral Galaxy Simulation \#3.}
\end{figure}

\begin{figure}
   \begin{center}

   \includegraphics[height=52mm]{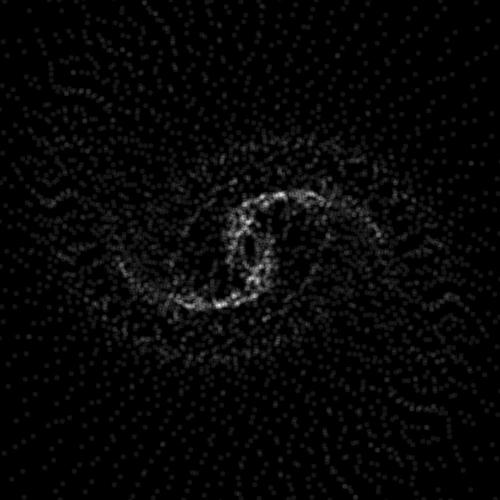}
   \includegraphics[height=52mm]{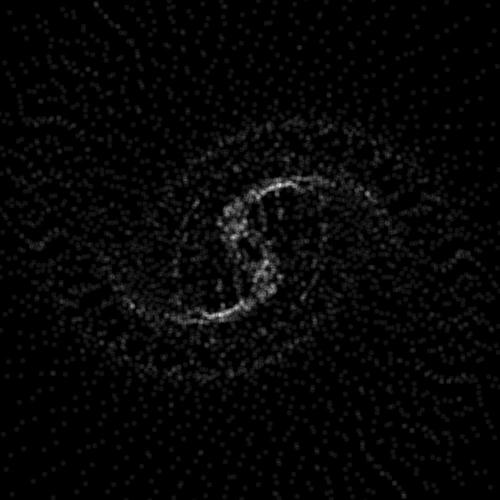}
   \includegraphics[height=52mm]{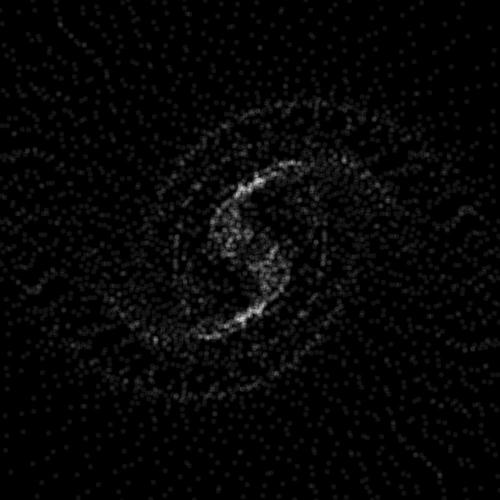}
   \vspace{.03in}

   \includegraphics[height=52mm]{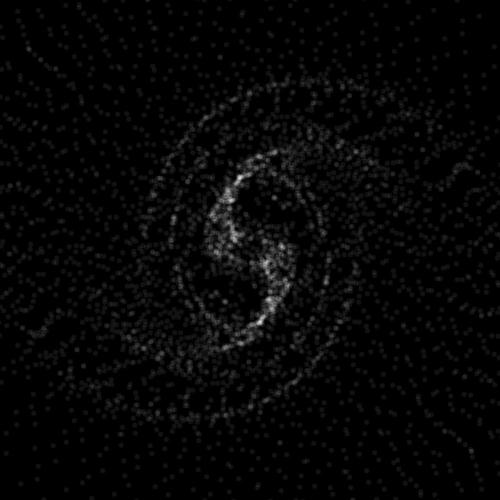}
   \includegraphics[height=52mm]{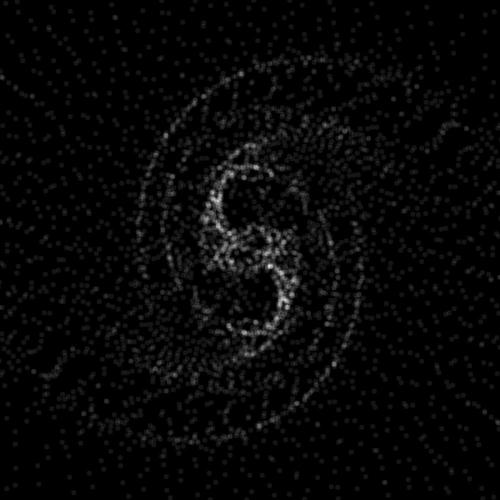}
   \includegraphics[height=52mm]{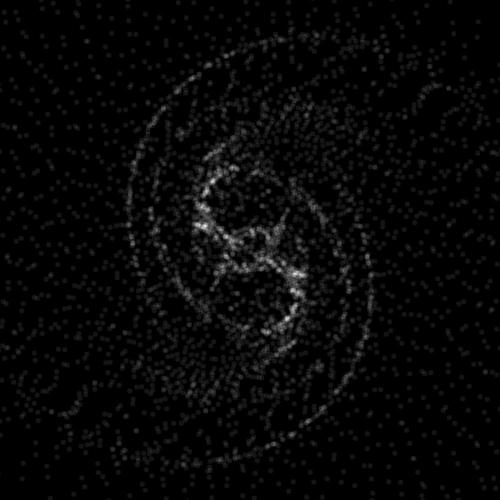}
   \vspace{.03in}

   \includegraphics[height=52mm]{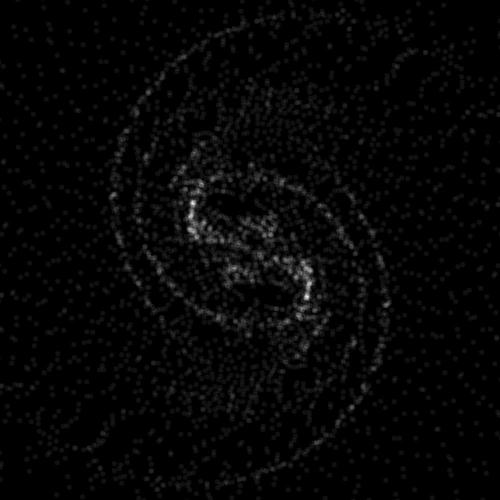}
   \includegraphics[height=52mm]{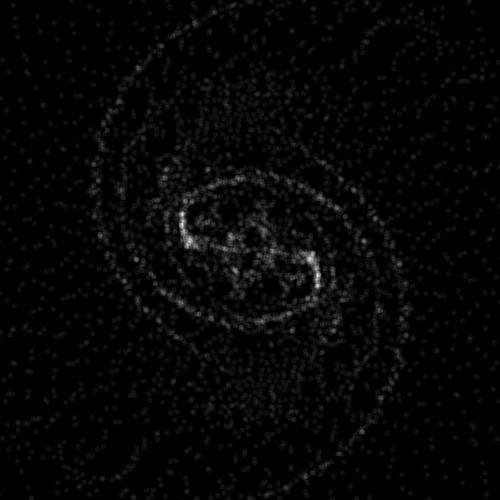}
   \vspace{.03in}

   \end{center}
   \caption{$t=60$ million years to $t = 95$ million years (in steps of $5$ million years) for Spiral Galaxy Simulation \#3.}
\end{figure}

\subsection{Spiral Galaxy Simulation \#4}\label{SGS4}

\begin{figure}
   \begin{center}
   \includegraphics[height=59mm]{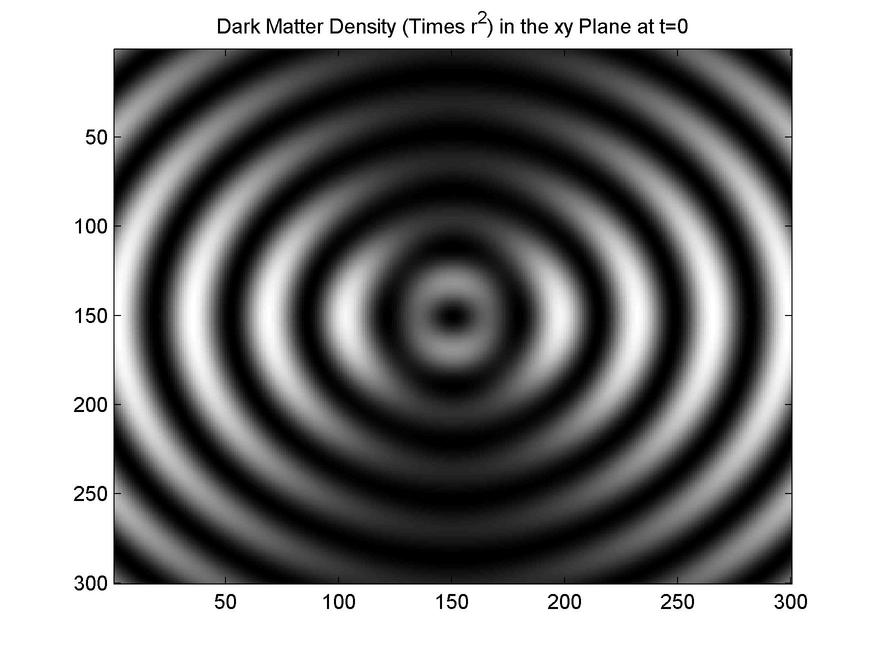}
   \includegraphics[height=59mm]{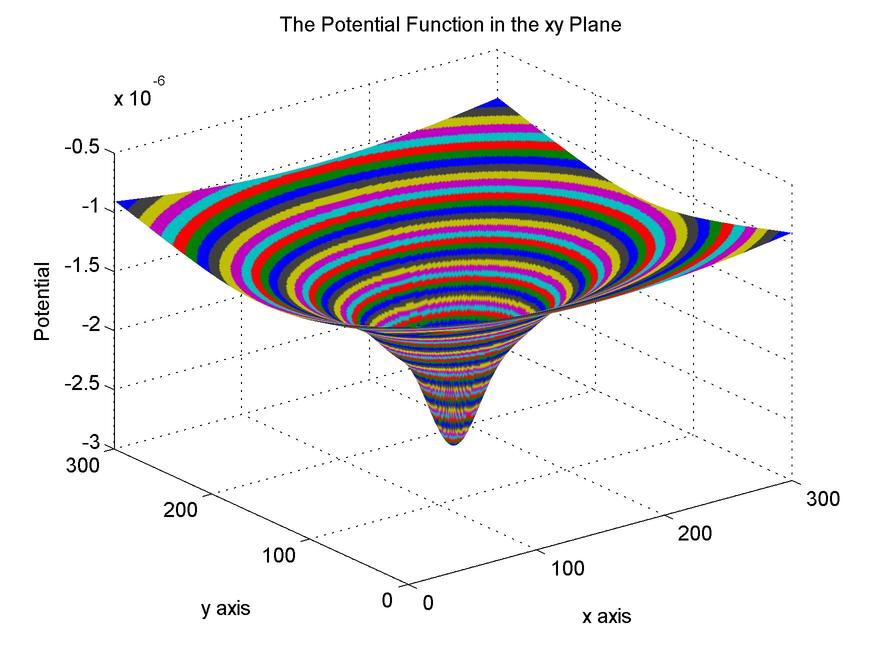}
   \includegraphics[height=59mm]{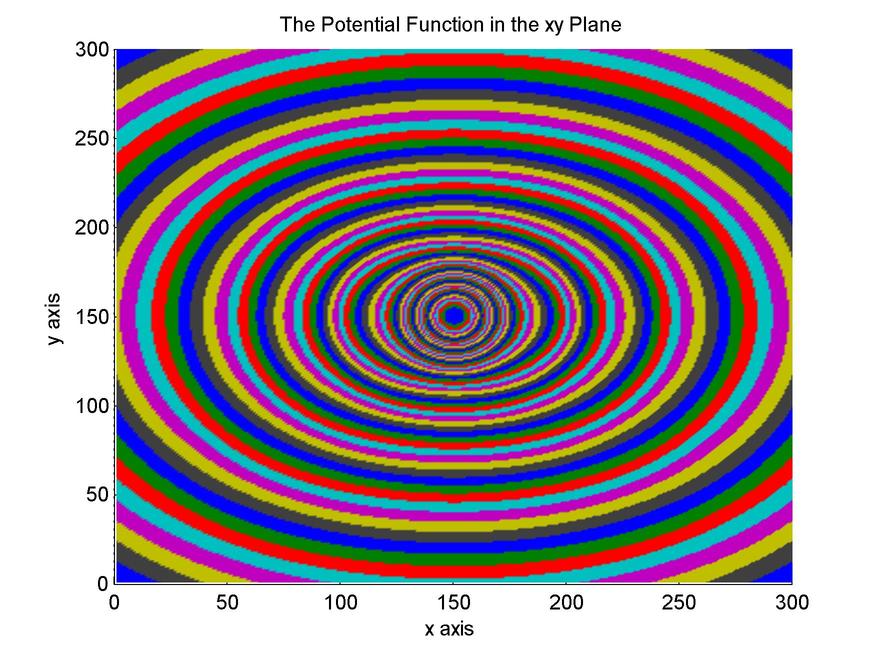}
   \includegraphics[height=59mm]{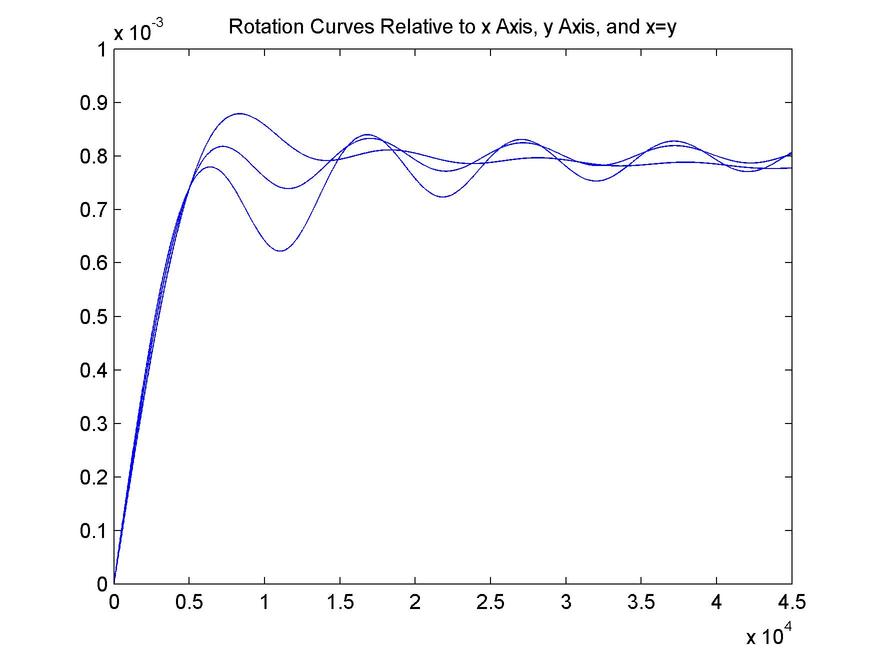}
   \end{center}
   \caption{Spiral Galaxy Simulation \#4:  The dark matter density times $r^2$ in the $xy$ plane (top left),
   the potential function in the $xy$ plane (top right), the level sets of the potential function
   in the $xy$ plane (bottom left), and the rotation curve (bottom right), all to a radius of $45,000$ light years.}
\end{figure}

As a final example, consider
\begin{equation}
\mbox{spiralgalaxy}(0.1, 100000, 1, -0.5, 20000, 19900, 50000000,
8.7e-15, 15000, 5000, 82000000, 20000)
\end{equation}
which produces the simulated image which we compare to NGC 488 in
figure 4. We only include one other figure for this example, namely
figure 26. As can be seen in that figure, the main difference in
this example from the previous ones is that we have increased the
wavelengths $\lambda_1$ and $\lambda_2$ by a factor of $10$.  As a
result, the scale of the wavelengths is approaching the scale of the
galaxy itself.  We have not done enough test runs of our simulations
to draw any careful conclusions about exactly what differences this
implies, but our first impression, based on figure 4, is that this
is one way to get tightly wound spirals.

We note that with these choices of wavelengths $\lambda_1,
\lambda_2$ and dark matter period $T_{DM}$ that $\Upsilon^{-1}
\approx 512$ light years, which is much different from our first
three examples.  However, we can always tweak the values of
$\lambda_1, \lambda_2,$ and $T_{DM}$ later.  At this stage we are
simply trying to understand the full range of the possible outcomes
of these simulations.

Finally, our earlier point about not optimizing our input parameters
to match the photos of actual spiral galaxies in figures 1-4 is
particularly relevant in this last example.  There was a resemblance
to NGC 488, so we put the simulated image next to this galaxy for
the sake of argument.  We only tried this one simulation in this
range of values, so it is very likely that the input parameters can
be adjusted, possibly by a lot, to create a better match.

Each run of these simulations, when done on the author's PC at home,
takes all night typically, so we have by no means exhausted all of
the possible scenarios to try to model. Of course, converting this
code to a faster programming language would speed things up greatly.
Using multiple processors, as the author has recently begun doing at
Duke University, is also a big help.

\section{Elliptical Galaxies}\label{EllipticalGalaxies}

In contrast to spiral galaxies, gas and dust are not thought to play
a major role in the overall dynamics of elliptical galaxies
\cite{BT}.  Furthermore, to a good approximation in many cases,
elliptical galaxies are mostly transparent \cite{BM}.  Light from
the stars of an elliptical galaxy is mostly not absorbed into the
dust of the galaxy but instead mostly makes it out to be observed by
telescopes. Hence, the observed brightness of elliptical galaxies is
a good indiction of the true brightness.

In addition, since gas and dust play a greatly reduced role,
elliptical galaxies should be fairly well modeled by $n$ body
simulations.  In retrospect, there is a fairly strong argument for
studying elliptical galaxies first, before disk galaxies, as a way
of testing a dark matter theory because elliptical galaxies are
easier to model. While spiral galaxies seem more spectacular at
first glance, there is plenty to understand about elliptical
galaxies as well.

For example, section 4.3 ``Photometry of Elliptical Galaxies'' of
\cite{BM} is a treasure trove of data about elliptical galaxies. One
of the most striking facts is that the surface brightness profiles
of observed images of many elliptical galaxies, especially NGC 1700
for example, are well modeled by
\begin{equation}
   I(R) = a \exp\{-b R^{1/4}\}
\end{equation}
for constants $a,b$.  It turns out to be more convenient to redefine
the constants such that
\begin{equation}\label{BrightnessModel}
   I(R) = I_e \exp\{-7.67[(R/R_e)^{1/4} - 1]\},
\end{equation}
where $7.67$ has been chosen so that half of the total brightness is
contained inside radius $R_e$ and $I_e$ is the brightness at this
radius.  That is,
\begin{equation}
   2 \int_0^{R_e} I(R) \; 2\pi R \; dR =
   \int_0^{\infty} I(R) \; 2\pi R \; dR = 7.22\pi R_e^2 I_e,
\end{equation}
as explained in \cite{BM}.  Naturally this model works better for
some elliptical galaxies than others, and in some cases there are
significant deviations \cite{BM}.

We refer the reader to \cite{EG1} as an example of previous work
trying to explain these observed brightness profiles of elliptical
galaxies.  Most other models depend greatly on the process by which
the elliptical galaxy is formed, which translates into the initial
conditions of the simulation.  We are also faced with this question.
However, we find that our case is qualitatively different.  The dark
matter density waves, which formed spiral patterns in disk galaxies,
now effectively ``randomize'' the energies of the test particles in
elliptical galaxies.  Some of the test particles gain energy and are
slung out to larger radii, while some of the test particles lose
energy and spend their time at smaller radii.  That is, unlike the
traditional WIMP dark matter case where the dark matter is taken to
create a roughly static potential, our rotating dynamic dark matter
potential removes the conservation of energy constraint for each
test particle. Physically, this can be thought of as a transfer of
energy between the test particles and the dark matter.  While this
happens in any dark matter theory, the scale of this transfer
appears to be much larger for our rotating scalar field dark matter
model than for the WIMP model.

To see if this model for rotating scalar field dark matter could
reproduce something close to the observed brightness profiles, we
created a Matlab simulation whose command line is
\begin{eqnarray}
\mbox{ellipticalgalaxy}(dr, r_{max}, A_0, A_2, \lambda_0, \lambda_2,
T_{DM}, \mu_0, R_I, n_{particles}, T_{total}, dt, s_{bin}).
\end{eqnarray}
This Matlab .m file may be downloaded at
http://www.math.duke.edu/faculty/bray/darkmatter/darkmatter.html which
also contains the author's Matlab .m file, spiralgalaxy.m, for doing
spiral galaxy simulations.

Notice that almost all of the input variables are the same as
spiralgalaxy.m.  In fact, the model for the rotating dark matter and
the corresponding rotating dark matter potential are exactly the
same.  The only differences, really, are the initial conditions that
we choose for our point particles and how we display the results.

Notice that we no longer have $R_d$ as an input parameter, which was
the disk scale length of the spiral galaxy, which determined the
initial density of the test particles.  Instead we are trying to
determine equilibrium densities for test particles in mostly radial
motion, as is the case for elliptical galaxies \cite{BM}.

This raises the question of what the initial placement of the test
particles should be.  We chose something very simple - place all of
the test particles on a sphere of radius $R_I$, the initial radius,
but with zero velocity.  When the test particles collapse down
toward the origin, the result will be a kind of explosion as the
stars come back out.  Elliptical galaxies are thought to be the
result, at least in many cases, of other galaxies which have merged,
perhaps many times \cite{BT}.  These initial conditions are inspired
by this as well.

The other input variable is $s_{bin}$ which has nothing to do with
the actual simulation but only with how the results of the
simulation are displayed.  Since we want to understand the predicted
brightness profile, we will want to know the distribution of the
test particles in the viewing plane.  To do this, we determine the
radius of each test particle in the viewing plane and then create a
histogram of the resulting radii, where the bin size for each
interval of radii is $s_{bin}$.  We will refer to these histograms
as graphs of the ``radial frequency in the $xy$ plane.''  We will
then compare our measured radial frequencies with that which is
predicted by equation \ref{BrightnessModel}, which we will
approximate by
\begin{equation}\label{BrightnessModel2}
   \mbox{radial frequency} \approx I(R) \cdot 2\pi R \cdot s_{bin}.
\end{equation}
This is easily done by defining $R_e$ to be the median value of the
measured radii in the viewing plane and by choosing $I_e$ to give
the correct overall normalization so that the total brightnesses are
the same.

We remind the reader that the test particles in these simulations,
which are mostly supposed to represent stars in this case, are still
treated as though they have zero mass.  Hence, we are only measuring
the gravitational effect of the dark matter on the stars.  This is
instructive for a first simulation.  It will also be roughly
accurate in cases where the dark matter makes up most of the mass of
the galaxy.  However, simulations which do account for gravitational
interactions between stars is a logical next step.

\subsection{Elliptical Galaxy Simulation \#1}\label{EGS1}

We will focus on a single elliptical galaxy example in this paper
which was used to create the simulated images in figures 5 and 6
using the following command line in Matlab:
\begin{equation}
\mbox{ellipticalgalaxy}(.1,100000,1,1,2000,1990,50000000,8.7e-13,15000,1500,1e20,1000,2000).
\end{equation}
We refer the reader to Spiral Galaxy Simulation \#1 and figures 9,
10, 11, 12 for qualitatively correct pictures of the dark matter
density and its corresponding potential which will be almost exactly
the same as in this example (except that we made a minor change in
the dark matter radius from $75,000$ light years to $100,000$ light
years). The sign of $A_2$ being switched simply rotates the
potential by $90$ degrees in the $xy$ plane, which is also
irrelevant for our purposes.  The only other difference is that in
this case the dark matter potential will rotate slower with a dark
matter period of $50,000,000$ million years.

\begin{figure}
   \begin{center}
   \includegraphics[height=59mm]{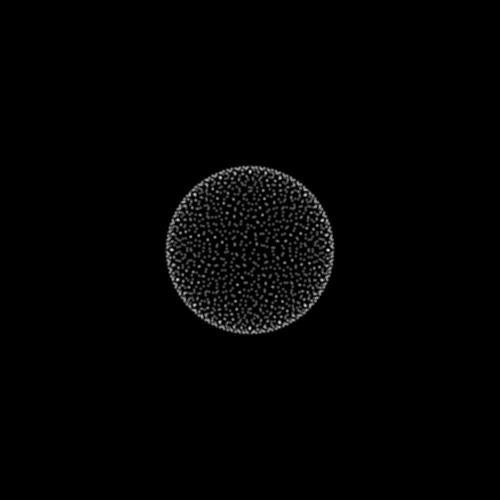}
   \includegraphics[height=59mm]{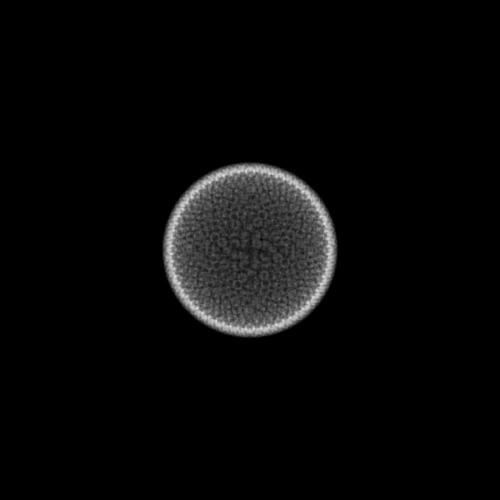}
   \end{center}
   \caption{The initial positions of the test particles for Elliptical Galaxy Simulation \#1 are on a sphere
   of radius $15,000$ light years.  The right
   image is simply a blurry version of the left image.  The initial velocities are all zero.}
\end{figure}

As previously mentioned, the initial placement of the test particles
is on spheres, which in this case will have a radius $15,000$ light
years as seen in figure 27. These spheres naturally collapse toward
the origin and then return back to something close to their original
positions, but modified because of the rotating dark matter
potential in the background. Over time, as previously described, a
wide spread in the energies of the test particles develops.  After a
billion years or so, the distribution of test particles settles down
to a kind of limiting distribution which looks qualitatively similar
to elliptical galaxies, as can be seen in figure 28.

\begin{figure}
   \begin{center}
   \includegraphics[height=59mm]{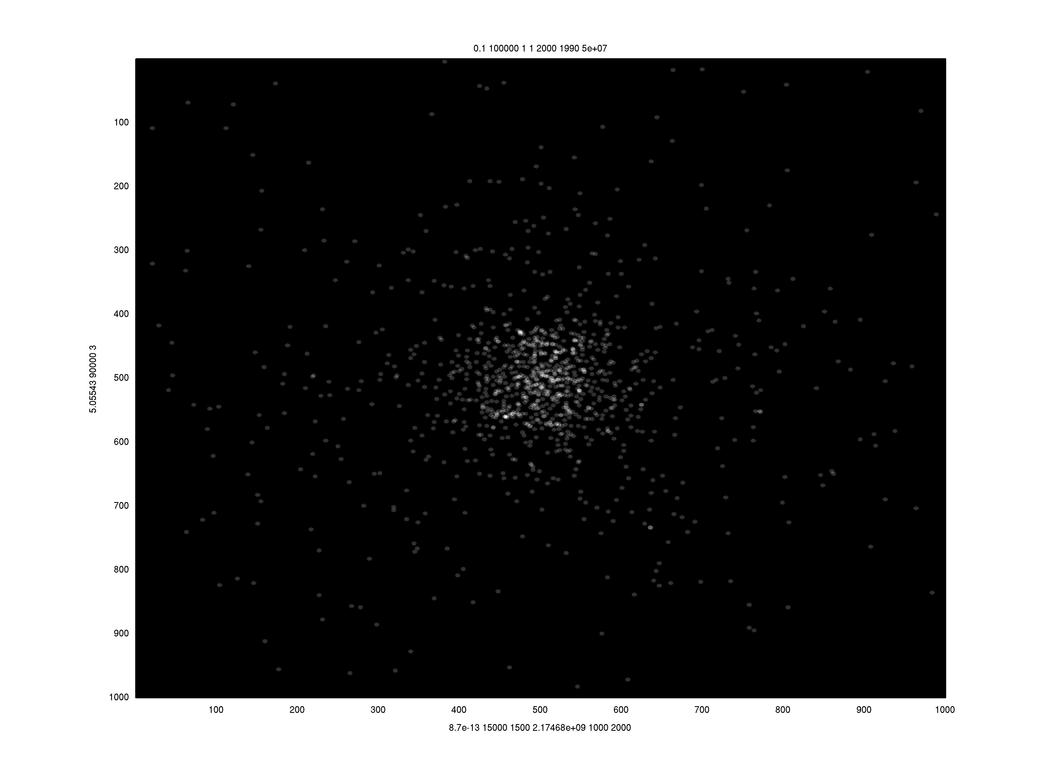}
   \includegraphics[height=59mm]{topview2-2174670000.jpg}
   \includegraphics[height=59mm]{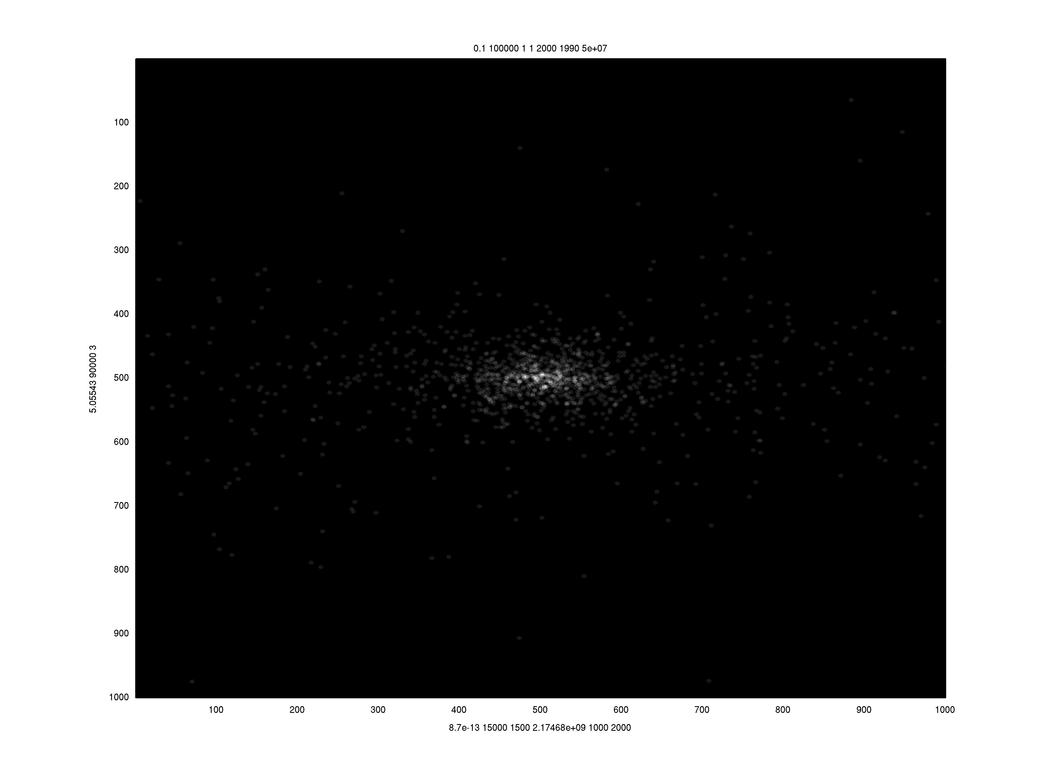}
   \includegraphics[height=59mm]{sideview2-2174670000.jpg}
   \end{center}
   \caption{The positions of the test particles for Elliptical Galaxy Simulation \#1 after $2.175$
   billion years.  The top row is a top view and the bottom row is a side view.  The right column is
   simply a blurry version of the left column.}
\end{figure}

We note that the images in figure 28, while very intriguing, are not
perfect. The ratio of the major axis to the minor axis of the top
view is very close to one, but this ratio is about 4:1 in the side
view (according to a formula used to calculate this in the
simulation which weights large radii more than small radii). Thus,
the side image appears to be too flat, since this ratio is generally
not observed to exceed 3:1.  Of course our simulation is very rough,
not even accounting for star - star gravitational interactions, so
we were only hoping to get something in the right ballpark. Also,
our initial conditions could be improved to be more reflective of
how elliptical galaxies are actually formed.  And finally, our model
for rotating scalar field dark matter is only an approximation.

Also, the side view in figure 28 is the view which maximizes the
apparent ellipticity.  A random point of view will rarely give the
side view, so a typical observed ellipticity will be much reduced.
We note that a histogram of the observed ellipticities of $2,135$
elliptical galaxies is found in figure 4.33 of \cite{BM}.  It is a
logical next step to study the predicted histogram of ellipticities
of any theory, so we list this as a good problem to study.

\begin{figure}
   \begin{center}
   \includegraphics[height=59mm]{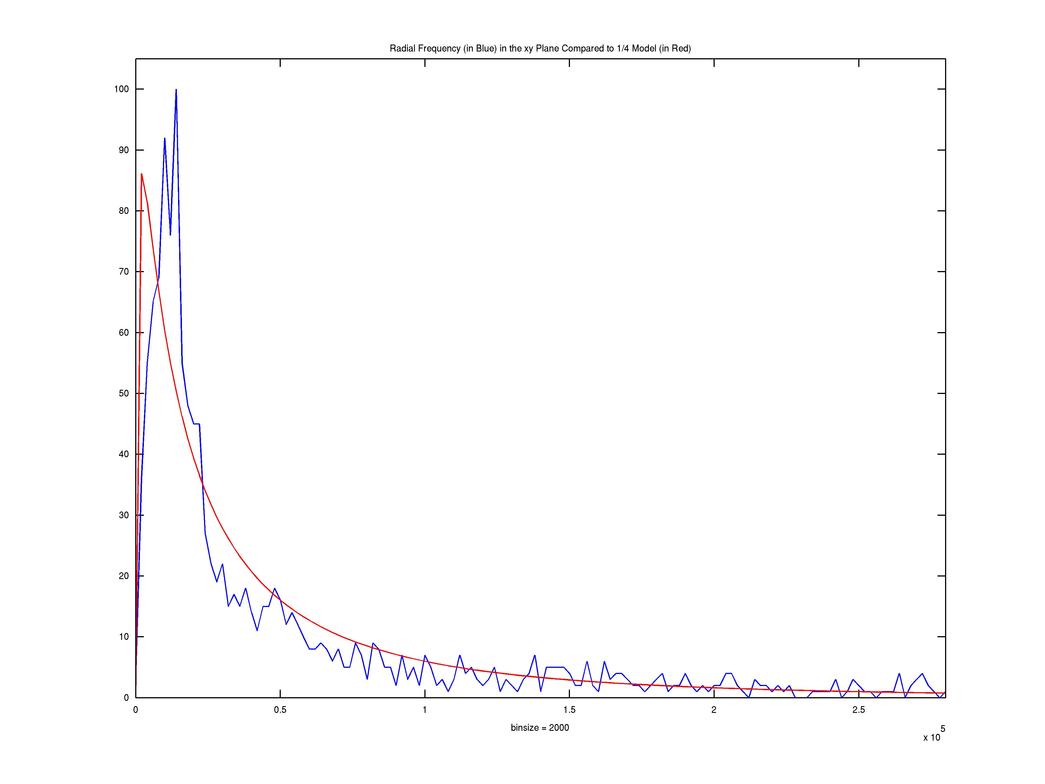}
   \includegraphics[height=59mm]{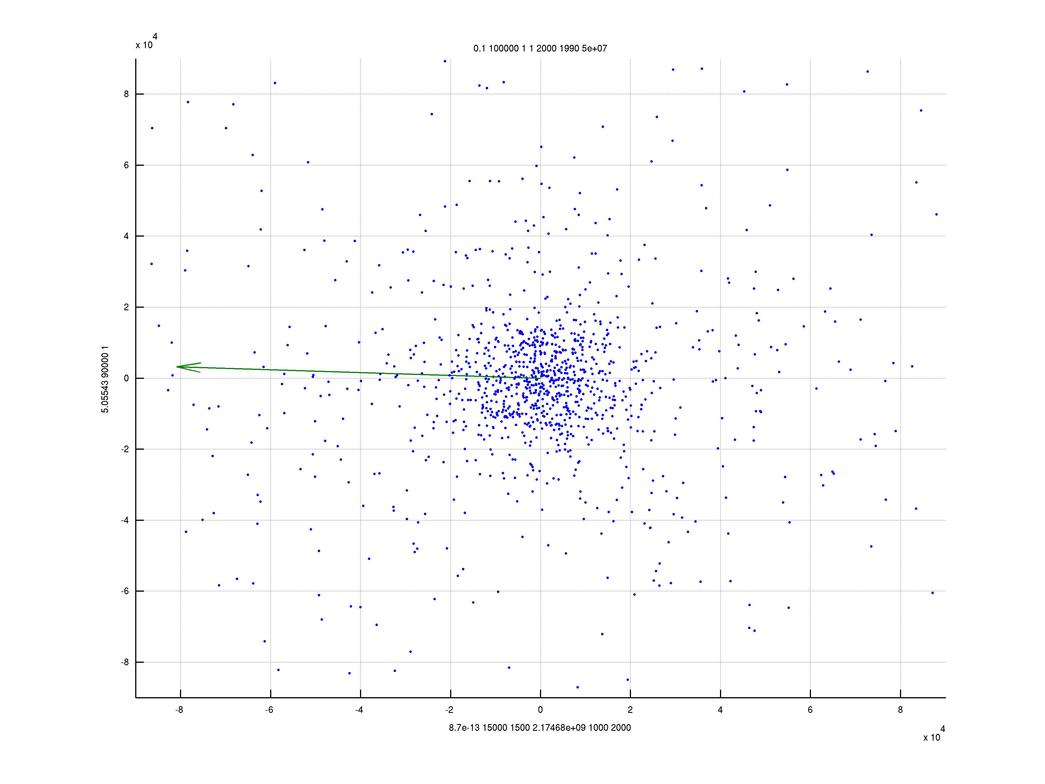}
   \end{center}
   \caption{Elliptical Galaxy Simulation \#1:
   The left image is the radial frequency in the viewing
   plane (in blue) compared to the standard model (in red) from equation
   \ref{BrightnessModel2}.  The right image is
   the location of test particles in the $xy$ viewing plane on which the
   the radial frequency is based.  The median radial value for the test particles
   is approximately $28,000$ light years.  The arrow in the right image is irrelevant and simply
   shows the direction of the minor axis of the potential in the $xy$ plane, which rotates.  This figure is a repeat of figure 7.}
\end{figure}

We can be even more quantitative about understanding these results
if we look at the radial frequency in the viewing plane, which is
related to the observed brightness profiles of elliptical galaxies.
In figure 29, we graph the radial frequency in the $xy$ viewing
plane correspond to the top view of the simulation in the top row of
figure 28.  We find such a close match, with such a generic initial
condition, to be very encouraging. Hence, we are very intrigued by
the possibility that scalar field dark matter with angular momentum
could be a major factor in determining the brightness profiles of
elliptical galaxies.

\subsection{More Brightness Profile Data}

We comment that we have found even better matches than the one in
figure 29 for the radial frequency of the test particles, which of
course is closely related to the brightness profile of a galaxy by
equation \ref{BrightnessModel2}.  We do not go into detail about
these simulations because of space considerations and because, in
retrospect, they use an initial condition which is suspect. Namely,
the test particles begin on a very small sphere of radius only
$5,000$ light years.  As a result, the test particles apparently
pick up a lot of angular momentum and rotate about the origin more
than desired.  This appears to result in eccentricities higher than
ideal. But we decided this data was worth sharing anyway, realizing
that it also is not quite perfect. What is interesting is that many
of the test particles get energized to radii many times their
initial $5,000$ light year radius well exceeding $100,000$ light
years or more, as long as the dark matter has enough angular
momentum.

\begin{figure}
   \begin{center}
   \includegraphics[height=59mm]{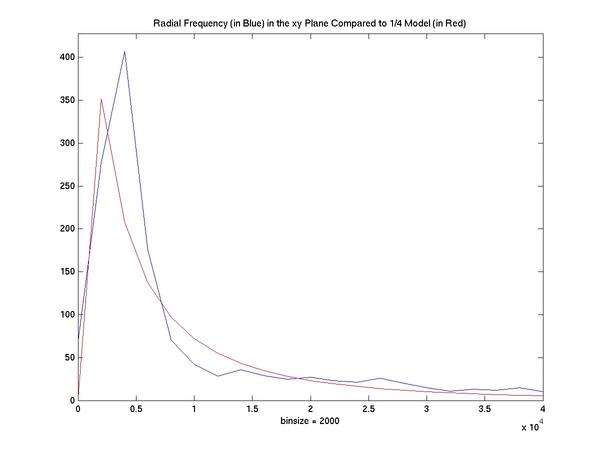}
   \includegraphics[height=59mm]{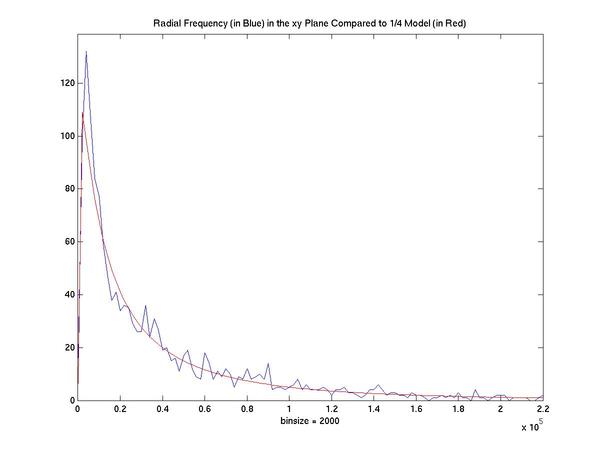}
   \includegraphics[height=59mm]{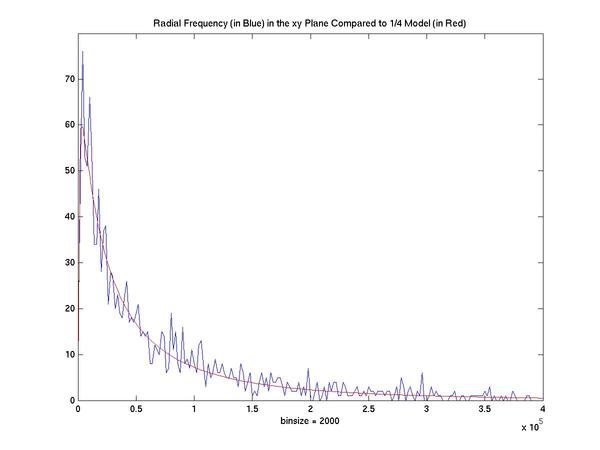}
   \includegraphics[height=59mm]{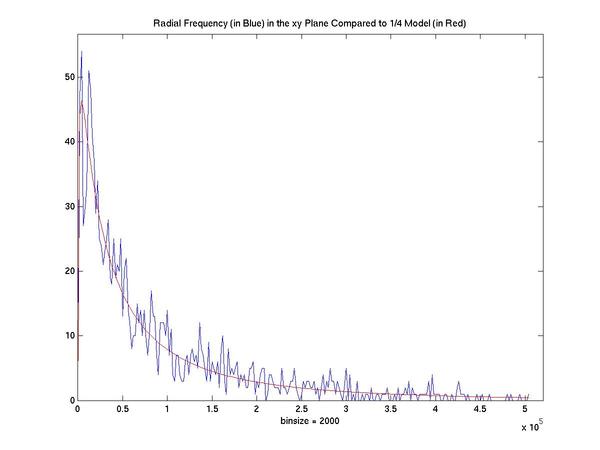}
   \includegraphics[height=59mm]{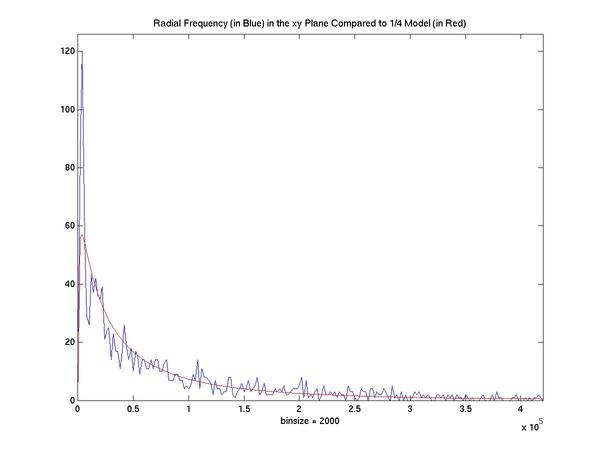}
   \includegraphics[height=59mm]{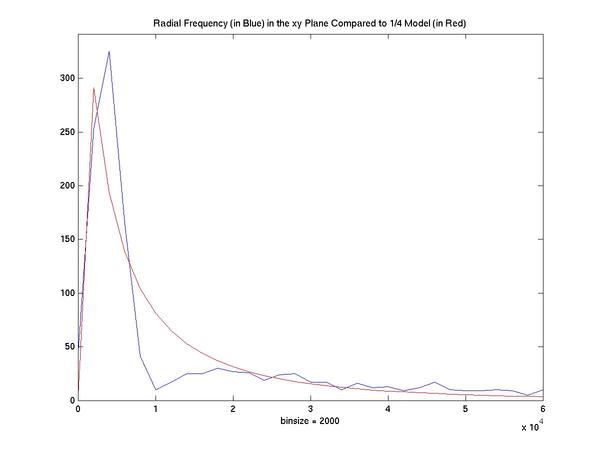}
   \end{center}
   \caption{These six radial frequency plots result from running the Matlab
   function described in the text where $A_2$ ranges over the
   values $0.5$ (top left), $0.75$ (top right), $1$ (middle left),
   $1.25$ (middle right), $1.5$ (bottom left), and $2$ (bottom right).  In that same order, the radial upper limits in these
   six plots are, respectively, $40000$, $220000$, $400000$, $500000$, $400000$, and $600000$ light years.}
\end{figure}

We begin by noting that if the dark matter is not rotating at all
($A_2 = 0$), then by conservation of energy in a fixed potential,
the (massless) test particles will never get more than $5,000$ light
years from the origin.  If the dark matter is only rotating a little
bit ($|A_2/A_1| << 1$), not much diffusion of energies occurs.

The six radial frequency plots in figure 30 result from running the
Matlab function
\begin{equation}
   \mbox{ellipticalgalaxy}(0.1,100000,1,A_2,2000,1990,50000000,8.7e-13,5000,1500,1000000000,1000,2000)
\end{equation}
where $A_2$ ranges over the values $\{0.5, 0.75, 1, 1.25, 1.5, 2\}$
from the top left to the bottom right.  Note that $A_1 = 1$.  The
greatest interference pattern occurs (as a percentage of the overall
pattern) when $|A_2| = |A_1|$, and interestingly, the corresponding
radial frequency plot in figure 30 is quite a good fit.

The fact that we get the best fit (in this imperfect example) when
the interference pattern is maximized suggests that elliptical
galaxies are best modeled in this theory by dark matter with large
angular momentum ($|A_2/A_1| \approx 1)$.  On the other hand, spiral
galaxies were successfully modeled with $|A_2/A_1|$ both equal to
$1.0$ (Spiral Galaxy Simulation \#1) and $0.15$ (Spiral Galaxy
Simulations \#2 and \#3).  It could be that practically all
elliptical galaxies have dark matter with large angular momentum due
to a major collision with another galaxy in their past, whereas
spiral galaxies are more likely to take on the whole range of
angular momenta.  More study is required to address this question.

\subsection{Ripples in Elliptical Galaxies}\label{ripples}

We could not end a section on elliptical galaxies without pointing
out some elliptical galaxies actually have visible ``ripples'' in
their images.  In fact, about $10$ to $20$ percent of early-type
galaxies have visible sharp steps in their luminosity profiles, as
can be seen in NGC 3923 in figure 4.40 of \cite{BM}.  This can not
be fully appreciated without looking at the image, so we highly
suggest the reader get a copy of \cite{BM}, look at the image, and
read the discussion.  The striking characteristic of these ripples
is that they tend to be smooth arcs of ellipses centered around the
center of the galaxy.  In NGC 3923, these arcs range in radii from 3
to 95 kpc \cite{BM}.  The ripples are on the order of $3\%$ to $5\%$
variations in the surface brightness of the galaxy.

The issue of these ripples is not settled.  Since our dark matter
density has ripples in it, as seen in figures 10 and 19 and
especially figure 26 where the wavelengths $\lambda_1$ and
$\lambda_2$ have been increased, it is only natural to wonder if
there is a connection.  We leave this as a very interesting open
question.

\section{Open Problems}

As exciting a project as this has been for the author, there are
vastly more questions left open than answers to problems solved.
Hence, it seems appropriate to end this paper with a list of open
problems.

\vspace{.1in} \noindent \underline{\textbf{Geometric Analysis
Problems}}

\begin{problem}
Prove or disprove conjecture \ref{ClassificationConjecture} in appendix \ref{geomcalcs}, that a
generic quadratic functional compatible with Axiom \ref{A1} leads to
the Einstein-Klein-Gordon equations.
\end{problem}

\begin{problem}
Prove or disprove that all actions compatible with Axiom \ref{A2}
are of the form in equation \ref{conjecture2}.
\end{problem}

\begin{problem}
Classify all actions of the form in equation \ref{conjecture2}, and
describe the resulting Euler-Lagrange equations.  What kinds of
phenomena exist for the solutions to these equations?
\end{problem}

\begin{problem}\label{stable}
What do generic stable solutions to the Einstein-Klein-Gordon
equations look like?  Are there solutions corresponding to a
gravitationally bound ``blob'' of scalar field with angular
momentum?
\end{problem}

\vspace{.1in} \noindent \underline{\textbf{Astrophysics Problems}}

\begin{problem}
Does dark matter enforcing a ``fold dynamics'' (discussed in section
\ref{StarFormation}) upon gas and dust clouds drive star formation?
\end{problem}

\begin{problem}
If dark matter is connected to star formation, what characteristics
of the dark matter are related to starburst activity in galaxies? Do
the magnitude and speed of the dark matter density wave affect the
star formation rate?  If the magnitude and speed of the dark matter
density wave can be associated with certain morphologies of spiral
galaxies, there could be a testable hypothesis.
\end{problem}

\begin{problem}
Is there a connection between the observed ripples in early-type
elliptical galaxies (discussed in section \ref{ripples}) and the
ripples in the scalar field dark matter density?
\end{problem}

\begin{problem}
What phenomena does this scalar field dark matter predict that can
be tested or detected by gravitational lensing?
\end{problem}

\begin{problem}
Study and explain the relationships between the input parameters of
the spiral galaxy simulation in this paper and the morphologies of
the disk galaxies created.
\end{problem}

\begin{problem}
Estimate the value of $\Upsilon$ (the constant in the Klein-Gordon equation) 
by making best fits of rotation
curves, brightness profiles, observed distribution of ellipticities,
and spiral patterns of galaxies.  Can the numbers and sizes of dwarf
galaxies be used to estimate $\Upsilon$?
\end{problem}

\begin{problem}
Compare the distribution of ellipticities predicted by simulations
of the scalar field dark matter theory (from random viewpoints) to
the observed distribution of ellipticities of $2,135$ elliptical
galaxies in figure 4.33 of \cite{BM}.  Be careful to use the same
procedure for defining ellipticity in the simulations as in the
observations.
\end{problem}

\vspace{.1in} \noindent \underline{\textbf{Computer Simulations}}

\noindent Note:  http://www.math.duke.edu/faculty/bray/darkmatter/darkmatter.html 
has the Matlab simulations described in this paper available for download.

\begin{problem}
Improve the elliptical galaxy simulation discussed in this paper by
adding in the gravitational influence of the stars on each other.
\end{problem}

\begin{problem}
Generalize the simulations in this paper by incorporating higher
degree spherical harmonics into the scalar field dark matter
solutions. As long as no more than two frequencies are used, the
solutions will also approximately rotate and produce potentials that
approximately rotate. See if spiral galaxies with more than two arms
can be simulated with these new rotating potentials.
\end{problem}

\begin{problem}
What are the dynamics of the stars and the gas and dust of a galaxy
with a typical stable solution (from problem \ref{stable}) to the
Einstein-Klein-Gordon equations in the background, and more
specifically, does spiral structure result for a wide range of
conditions?  How about observed elliptical galaxy characteristics?
\end{problem}

\begin{problem}
Does dynamical friction manifest itself in solutions to the
Einstein-Klein-Gordon equations, and if it does, how does it do it?
Is it necessary to have gas and dust around to dissipate energy to
allow a dynamical friction process to form larger and larger
gravitationally bound ``blobs'' of scalar field dark matter?
\end{problem}

\begin{problem}
Create careful simulations of the Einstein-Klein-Gordon equations in
a perturbed cosmological setting to see what typical ``blobs'' of
scalar field dark matter look like. These simulations could then
address the question of what happens when two blobs collide, whether
they have enough dynamical friction to combine, and how they carry
angular momentum.  It may be necessary to simulate the regular
matter at the same time so that energy from the dark matter can be
transferred to the regular matter and then dissipated through
friction and radiation. Also, regular matter could help stabilize
the galactic potential which could help to stabilize the dark matter
at sufficiently small radii.
\end{problem}

\vspace{.1in} \noindent \underline{\textbf{Physics Problems}}

\begin{problem}
Is there a way to modify Axiom \ref{A1} or Axiom \ref{A2} to include
quantum mechanics?
\end{problem}

\begin{problem}
What are the physical interpretations of the matter fields defined by Axiom \ref{A2} 
(discussed at the end of appendix \ref{BiggerTheory})?
\end{problem}

\begin{problem}
How does the connection on the tangent bundle of a spacetime
manifest itself physically?
\end{problem}

\begin{problem}
What is the
behavior of scalar field dark matter right after the Big Bang?
Note that $\bar{P}/\bar{\rho}$ (from theorem
\ref{cosmothm} in section \ref{cosmologicalpredictions}) is not necessarily close to zero 
in this case. 
\end{problem}

\begin{problem}
What are the physical consequences of the oscillating pressure of scalar field dark matter.  Note
that in the real scalar field case, the pressure necessarily oscillates between being positive and
negative, while averaging to something very close to zero, as described by 
theorem \ref{cosmothm} in section \ref{cosmologicalpredictions}.
\end{problem}

\begin{problem}
How does a scalar field model of dark matter affect the black holes at the centers of most 
galaxies?  At what rate does the scalar field dark matter fall into the black hole?  
\end{problem}

\section{Acknowledgments}

The author is indebted to Alar Toomre who generously and
enthusiastically made himself available on multiple occasions to
answer questions about spiral galaxies - explaining what was not
known, what was known, and how we knew it.  The author also wishes
to express his gratitude to Benoit Charbonneau, Felix Finster, Gary
Gibbons, Gary Horowitz, Gerhard Huisken, Claude LeBrun, Tonatiuh
Matos, Alan Parry, Arlie Petters, Ronen Plesser, Gerhard Rein, Alan
Rendall, Avy Soffer, Mark Stern, Scott Tremaine, and Marcus Werner
for suggestions, helpful insights, and interesting discussions on many aspects of
this paper, as well as to Andrew Schretter for help with the
computational aspects of this project.

The author also gratefully acknowledges support from the National
Science Foundation through grant \# DMS-0706794.  Also, some of the
work of this project was done at the Petters Research Institute in
Dangriga, Belize.

\newpage
\begin{appendix}

\section{A Brief Introduction to General Relativity}

In this appendix we provide a brief introduction to general
relativity relevant to the next section where we show how the
Einstein-Klein-Gordon equations with a cosmological constant result
from Axiom \ref{A1}.  The differential geometry discussed in these
two appendices is detailed but is also very standard.  Our
discussion will not require much knowledge beyond the first three
chapters of \cite{ONeill}, which we recommend as an excellent text
book for those who want to use differential geometry to study the
large scale structure of the universe.

Special relativity, Einstein's theory which unified space and time,
is really all about the study of the geometry of the flat Minkowski
spacetime \cite{ONeill}.  The defining idea leading to general
relativity is the idea that spacetime is not necessarily flat.
Philosophically, the question ``Why should spacetime be curved?''
has a great response:  ``Why should spacetime be flat?''  What is
remarkable is that removing the assumption that spacetime is flat
leads to general relativity as arguably the simplest way of defining
what the spacetime metric should be, as we will explain.  Moreover,
experiments have shown that general relativity describes gravity as
accurately as can be measured.

The central equation of general relativity is the Einstein equation
\[
   G = 8\pi T,
\]
where $G = Ric - \frac12 R g$ is the Einstein curvature tensor and
$T$, called the stress energy tensor, describes the local energy and
momentum density.  However, it is important to realize that this
equation by itself is only half a theory in that this equation
describes how the matter curves the metric but does not explain how
the matter evolves forward in time.  From this point on we will only
discuss complete theories where matter evolution equations are
specified, fully realizing that one can always add in an unspecified
stress energy tensor later if desired.  For example,
\[
   G = 0,
\]
the vacuum Einstein equation, is a complete theory in our
terminology because there is no matter term to evolve.

It is worth pointing out that vacuum general relativity is already
an intricate and very interesting theory.  For example, black holes
are solutions to the vacuum Einstein equation when all of the matter
has already fallen into the black holes.  This demonstrates an
interesting characteristic of general relativity - zero local energy
and momentum density does not imply that the system has zero total
mass. In fact, by the positive mass theorem \cite{SY3}, \cite{SY5},
an isolated asymptotically flat vacuum solution must have strictly
positive total mass, unless the spacetime is flat everywhere (and
hence is the Minkowski spacetime).

Also, matter can be approximated by black holes of various sizes.
For example, one could replace the Sun and the planets by black
holes of equal mass, and the corresponding n-body problem with black
holes is a solution to $G=0$.  For that matter, one could replace
every star, planet, meteor, and piece of dust in a galaxy with a
black hole of equal mass, and the resulting system would be a
solution to $G=0$.  The dynamics would be a good approximation to
the original system except when the objects got close to one
another.  All of the above statements are true assuming that an
existence theory for the vacuum Einstein equation $G = 0$
generically exists, which is currently an open problem.  The
difficulty in even proving an existence theory for $G = 0$ is
another indication of the highly nontrivial nature of the equation.

Einstein chose $G = Ric - \frac12 R g$ once he learned that the
divergence of $G$ equaled zero by the second Bianchi identity
\cite{ONeill}. This zero divergence property can be interpreted as a
local conservation property for energy and momentum density, a
understandably desirable property. Hilbert realized that the zero
divergence property of $G$ was a direct result of the fact that
metrics which are critical points of the functional
\begin{equation}\label{eqn:vacuumGR}
   H_U(g) = \int_{U} R \; dV,
\end{equation}
satisfy the corresponding Euler-Lagrange equation $G = 0$ on the
spacetime $N$.  The role of the open set $U$ in the above equation,
whose closure is required to be compact, is to keep the integral
finite. We define a metric $g$ to be a critical point of $H_U(g)$ if
this quantity does not change to first order for all smooth
variations of the metric with compact support in the interior of
$U$.  Note that we are not minimizing or maximizing anything but are
simply looking for critical points.

Moreover, Hilbert realized that the zero divergence property is true
for any Euler-Lagrange equation resulting from an action which is
invariant under reparameterizations of the metric, as the
Einstein-Hilbert action in equation \ref{eqn:vacuumGR} is.  Today we
think of this fact as a special case of Noether's theorem.  Hence,
the zero divergence property of $G$, while nice, is nothing
particularly special and is not the fundamental point. Instead,
vacuum general relativity is best understood as defining the metric
to be a critical point of the Einstein-Hilbert action in equation
\ref{eqn:vacuumGR}. This raises the truly relevant question:  What
is so special about the Einstein-Hilbert action?

We begin by noting that in some ways the Einstein-Hilbert action is
the simplest nontrivial scalar valued functional of a metric. The
scalar curvature of a spacetime metric is a scalar at each point, so
it is perfectly natural to integrate it.  However, there is another
characterizing property of this functional which is seen by
expressing it in a coordinate chart.

Given a coordinate chart mapping $\Phi: \Omega \rightarrow R^4 =
\{(x^0, x^1, x^2, x^3) \}$, define
\begin{equation}
   g_{ij} = g(\partial_i, \partial_j),\;\;\; \mbox{ where } \;\;\;
   \partial_i = D\Phi^{-1} \left( \frac{\partial}{\partial x^i} \right)
\end{equation}
to be the components of the metric in this coordinate chart.  Then
standard calculations show that the formula for the scalar curvature
in terms of the metric in a coordinate chart is
\begin{eqnarray}\nonumber
   R  &=& (g^{ik}g^{jl} - g^{ij}g^{kl}) g_{ij,kl} +
    g_{ij,k}g_{ab,c} \cdot \\&&
   \left(\frac34 g^{ia}g^{jb}g^{kc} - \frac12 g^{ia}g^{jc}g^{kb}
   -g^{ia}g^{jk}g^{bc} - \frac14 g^{ij}g^{ab}g^{kc} +
   g^{ij}g^{ac}g^{kb} \right)
\end{eqnarray}
where commas denote coordinate chart derivatives and the Einstein
summation convention of summing over indices which are both raised
and lowered is in effect.  As usual, we define $[g^{ij}] =
[g_{ij}]^{-1}$ as matrices. Also, the formula for the metric volume
form in terms of the metric in a coordinate chart is
\begin{equation}
   dV = |g|^{1/2}  \; dV_{R^4}
\end{equation}
where $|g| = |det([g_{ij}])|$.  Using $|g|_{,l} =
|g|\,g^{mn}g_{mn,l}$ and $g^{ij}_{\hspace{.1in},k} = -
g^{ia}g^{jb}g_{ab,k}$ we can integrate by parts over an open set $U$
(whose closure is compact) in the coordinate chart to get
\begin{eqnarray}\nonumber
   \int_U R \; dV  &=&
   \int_{\partial(\Phi(U))} (g^{ik}g^{jl} - g^{ij}g^{kl}) g_{ij,k}
    \nu_l  |g|^{1/2} dA_{R^4}
   +\int_{\Phi(U) } g_{ij,k}g_{ab,c}\cdot \\&&
   \left(-\frac14 g^{ia}g^{jb}g^{kc} + \frac12 g^{ia}g^{jc}g^{kb}
   + \frac14 g^{ij}g^{ab}g^{kc} -\frac12
   g^{ij}g^{ac}g^{kb} \right) |g|^{1/2}  \; dV_{R^4}
\end{eqnarray}
where $\nu_l = \langle \frac{\partial}{\partial x^l}, \nu
\rangle_{R^4}$ and $\nu$ is the outward unit normal to $\Phi(U)$ in
the $R^4$ coordinate chart.

Note that our formula for the Einstein-Hilbert action is of the form
\begin{equation}\label{eqn:scalarcurvquad}
   \int_{U} R \; dV = \int_{\Phi(U)}
   \mbox{Quad}_M(M')
   \; dV_{R^4} \;\;\;+\;\;\; \mbox{boundary
   term},
\end{equation}
where
\[
M = \{g_{ij}\} \;\;\;\mbox{ and }\;\;\;  M' = \{g_{ij,k}\}
\]
are the components of the metric in the coordinate chart and all of
the first derivatives of these components in the coordinate chart,
and where
\begin{equation}
   \mbox{Quad}_{Y}(\{x_\alpha\}) = \sum_{\alpha,\beta}
   F^{\alpha\beta}(Y) x_\alpha x_\beta
\end{equation}
for some functions $\{F^{\alpha\beta}\}$ is a quadratic expression
of the $\{x_\alpha\}$ with coefficients in $Y$. What is interesting
about this particular quadratic expression is that while its value
at each point depends on the choice of coordinate chart, by equation
\ref{eqn:scalarcurvquad} its integral over $\Phi(U)$ is invariant
with respect to smooth variations of $\Phi$ compactly supported in
the interior of $U$ (which hence do not affect the boundary term).
Even more intriguing, this quadratic expression is the unique one
with this property, up to multiplication by a constant, which
follows from the works of Cartan \cite{Car22}, Weyl \cite{Wey22},
and Vermeil \cite{Ver17} (and pursued further by Lovelock
\cite{Lov72}).

Hence, if we define
\begin{equation}
   F_{\Phi,U}(g) = \int_{\Phi(U)}
   \mbox{Quad}_M(M')
   \; dV_{R^4}
\end{equation}
for that same quadratic expression, then
\begin{equation}
{H}_U(g)= \,F_{\Phi,U}(g) \;+\; \mbox{boundary term}
\end{equation}
by equation \ref{eqn:scalarcurvquad}. Hence, the Euler-Lagrange
equation for $F_{\Phi,U}(g)$, for variations away from the boundary,
must be the same as the one for ${H}_U(g)$, which is $G=0$.

More generally, let
\begin{equation}
   F_{\Phi,U}(g) = \int_{\Phi(U)}
   \mbox{Quad}_M(M')
   \; dV_{R^4}
\end{equation}
for \emph{any} nontrivial quadratic expression, but require that $g$
is at a critical point of $F_{\Phi,U}(g)$ for \emph{all} coordinate
charts $\Phi$ and all open sets $U$ (whose closure is compact and in
the interior of $\Omega$).  Then the resulting Euler-Lagrange
equation can not depend on any one particular choice of coordinate
chart, and so by uniqueness as proved by Cartan \cite{Car22}, Weyl
\cite{Wey22}, and Vermeil \cite{Ver17} up to a multiplicative
constant, we must have
\begin{equation}
c \cdot {H}_U(g)= \,F_{\Phi,U}(g) \;+\; \mbox{boundary term}
\end{equation}
for some nonzero $c$. Hence, again the Euler-Lagrange equation for
$F_{\Phi,U}(g)$, for variations away from the boundary, must be the
same as the one for ${H}_U(g)$, which again is $G=0$.

Now suppose we let
\begin{equation}
   F_{\Phi,U}(g) = \int_{\Phi(U)}
   \mbox{Quad}_M(M' \cup M)
   \; dV_{R^4}
\end{equation}
for any quadratic expression (which is nontrivial on the $M'-M'$
part), and continue to require that $g$ is at a critical point of
$F_{\Phi,U}(g)$ for all coordinate charts $\Phi$ and all open sets
$U$ (whose closure is compact and in the interior of $\Omega$). Then
the resulting Euler-Lagrange equation still can not depend on any
one particular choice of coordinate chart. Furthermore, the
uniqueness theorems of Cartan \cite{Car22}, Weyl \cite{Wey22}, and
Vermeil \cite{Ver17} imply the even more general result that
\begin{equation}
c \int_{U} (R - 2 \Lambda) \; dV= \,F_{\Phi,U}(g) \;+\;
\mbox{boundary term}
\end{equation}
for two constants $c \ne 0$ and $\Lambda$.  Hence, the
Euler-Lagrange equation for $F_{\Phi,U}(g)$, for variations away
from the boundary, will be the same as the expression on the left.
The expression on the left is called the Einstein-Hilbert action
with cosmological constant and is well known to have Euler-Lagrange
equation
\begin{equation}\label{eqn:cosmoeqn}
   G + \Lambda g = 0,
\end{equation}
which is the equation for vacuum general relativity with a
cosmological constant $\Lambda$.

When the cosmological constant term is moved to the other side of
the equation and interpreted as matter as in $G = -\Lambda g = 8\pi
T$, this matter is given the name ``dark energy'' (not to be
confused with dark matter) and has the same stress energy tensor as
a fluid with positive energy density and negative pressure (for
$\Lambda > 0$). The idea that a fluid can have negative pressure is
confusing to many, but the point is that dark energy is not a fluid.
Dark energy is simply the phenomenon resulting from the constant in
equation \ref{eqn:cosmoeqn} above. Standard theory today assigns a
very small but positive value for $\Lambda$ to explain the
accelerating expansion of the universe. While its density is very
small, it is everywhere, and so ends up accounting for roughly
$73\%$ of the mass of the universe in the standard $\Lambda CDM$
model \cite{WMAPobservations}.

The above discussion provides the motivation for Axiom \ref{A1} from
section \ref{GeometricMotivation}.  It is also helpful for
understanding the implications of Axiom \ref{A1}, which we discuss
in the next appendix.

\section{Derivation of the Einstein-Klein-Gordon Equations with a
Cosmological Constant from Axiom \ref{A1}.}\label{geomcalcs}

We begin by noting that the connection $\nabla$ on the tangent
bundle of a spacetime $N$ is a fundamental geometric object,
arguably second only to the metric $g$ in importance.  While the
metric is used to measure length, area, volume, etc., the connection
is used to differentiate a vector field in the direction of another
vector field.  The curvature tensors are then usually defined in
terms of the connection, for example.

Given any metric, there exists a unique standard connection called
the Levi-Civita connection $\bar\nabla$ which, in addition to the
linearity and Leibniz properties that all connections satisfy, is
both metric compatible and torsion free.  It is a short exercise to
show that given a general connection $\nabla$, its difference with
the standard connection (or, for that matter, any other connection)
\begin{equation}
   D(X,Y,Z) = \langle \nabla_X Y , Z \rangle - \langle \bar\nabla_X Y , Z \rangle
\end{equation}
is a tensor, meaning that $D$ is multilinear in each of its three
slots.  Note that we have used the convenient notation $g(X,Y) =
\langle X, Y \rangle$. We refer the reader to \cite{ONeill} for more
about connections.

Given a fixed coordinate chart, let $\{\partial_i\}$, $0 \le i \le
3$, be the tangent vector fields to $N$ corresponding to the
standard basis vector fields of the coordinate chart. Let $g_{ij} =
g(\partial_i, \partial_j)$ and $\Gamma_{ijk} = \langle
\nabla_{\partial_i}
\partial_j,
\partial_k \rangle$.  Then the above tensor equation, in
coordinates, becomes
\begin{equation}\label{eqn:defDlowered}
  D_{ijk} = \Gamma_{ijk} - \bar\Gamma_{ijk}.
\end{equation}
Furthermore, the Koszul formula for a connection \cite{ONeill}
implies that
\begin{equation}\label{eqn:ChristoffelLowered}
   \bar{\Gamma}_{ijk} = \frac12 \left(
   g_{ik,j} + g_{jk,i} - g_{ij,k} \right),
\end{equation}
which we note has first derivatives of the metric in its expression
(which will be important later).

Axiom \ref{A1} directs us to look for functionals of the form
\begin{equation}
   F_{\Phi,U}(g,\nabla) = \int_{\Phi(U)}
   \mbox{Quad}_M(M' \cup M \cup C' \cup C)
   \; dV_{R^4}
\end{equation}
where
\begin{equation}
M = \{g_{ij}\} \;\;\;\mbox{ and }\;\;\; C = \{\Gamma_{ijk}\}
\;\;\;\mbox{ and }\;\;\; M' = \{g_{ij,k}\} \;\;\;\mbox{ and }\;\;\;
C' = \{\Gamma_{ijk,l}\}
\end{equation}
are the components of the metric and the connection in the
coordinate chart and all of the first derivatives of these
components in the coordinate chart.  From the previous appendix we
know that
\begin{equation}
   F_{\Phi,U}(g,\nabla) = \int_{U} (R - 2\Lambda) \; dV
\end{equation}
falls into this form, at least up to a boundary term which we do not
write down since it is irrelevant for the ultimate Euler-Lagrange
equations which will be produced.  Also,
\begin{equation}
   F_{\Phi,U}(g,\nabla) = \int_{U} (R - 2\Lambda - \mbox{Quad}_g(D) )\; dV
\end{equation}
will be of the correct form as well by equations
\ref{eqn:defDlowered} and \ref{eqn:ChristoffelLowered}, again, up to
the usual boundary term.  Here $\mbox{Quad}_g(D)$ means some
quadratic functional of $D$ with coefficients in the metric $g$.
Note that the above two expressions have no dependence on $\Phi$.

However, Axiom \ref{A1} does not allow us to add in a general
expression of the form $\mbox{Quad}_g(\nabla D)$ to the integrand.
This is because the $\bar\Gamma$ terms of equation
\ref{eqn:defDlowered} already contain first derivatives of the
metric, so taking another derivative and squaring would introduce
quadratic terms of second derivatives of the metric, which is not
allowed by the axioms. However, the $\bar\Gamma$ terms go away if we
look at the torsion part of the $D$ tensor defined to be the $D$
tensor antisymmetrized in the first two indices.  The torsion tensor
(with the third index lowered) has components
\begin{eqnarray}
   T_{ijk} &=& D_{ijk} - D_{jik} \\
   &=& (\Gamma_{ijk} - \bar\Gamma_{ijk})
   - (\Gamma_{jik} - \bar\Gamma_{jik}) \\
   &=& \Gamma_{ijk} - \Gamma_{jik}
\end{eqnarray}
by the symmetry of the $\bar\Gamma$ in the first two indices.

Again, we would like to consider some kind of derivative of this
torsion tensor which we could then square.  However, Levi-Civita
covariant derivatives generically would introduce $g'\Gamma$ terms
which, when squared, would not be allowed by our axioms. Similarly,
taking a $\nabla$ covariant derivative would introduce $\Gamma^2$
terms before we squared, which is also not allowed by the axioms. As
before, we need to find a way to take a derivative of some part of
the torsion tensor which does not require a covariant derivative.

Define
\begin{eqnarray}
   \gamma_{ijk} &=& \frac16 (T_{ijk} + T_{jki} + T_{kij}) \\
   &=& \frac16 (D_{ijk} - D_{jik} + D_{jki} - D_{kji} + D_{kij} -
   D_{ikj}) \\
   &=& \frac16 (\Gamma_{ijk} - \Gamma_{jik} + \Gamma_{jki} - \Gamma_{kji} + \Gamma_{kij} - \Gamma_{ikj})
\end{eqnarray}
to be the fully antisymmetric part of the difference tensor $D$ (and
half the fully antisymmetric part of the torsion tensor $T$. Thus,
$\gamma_{ijk}$ are the components of a three form. Hence, we can
take the exterior derivative of $\gamma$ to get
\begin{equation}
  d\gamma_{ijkl} = \gamma_{jkl,i} -\gamma_{kli,j} +
  \gamma_{lij,k} - \gamma_{ijk,l}
\end{equation}
which are the antisymmetric coefficients of the tensor $d\gamma$
which do not involve derivatives of the metric, just derivatives of
$\Gamma$. Hence, functionals of the form
\begin{equation}\label{generalform}
   F_{\Phi,U}(g,\nabla) = \int_U (cR - 2\Lambda -
   \frac{c_3}{24}|d\gamma|^2 -
   \mbox{Quad}_g(D))  \; dV,
\end{equation}
are allowed by the axioms, up to a boundary term which is irrelevant
for the Euler-Lagrange equations produced.

\begin{conjecture}\label{ClassificationConjecture}
All functionals for which there exists a smooth metric and smooth
connection which satisfy Axiom \ref{A1} with that functional are of
the form of equation \ref{generalform}.
\end{conjecture}
We leave this as a very interesting geometric conjecture to study.  We will take $c \ne 0$
so that the resulting Euler-Lagrange equations for the metric are nondegenerate.  Without loss of 
generality, we may as well take $c=1$.

We comment that for generic quadratic expressions
$\mbox{Quad}_g(D)$, all of the components of $D$ other than the
fully antisymmetric part $\gamma$ (and irreducible components of the
same type) will have to be zero to satisfy their zeroth order
Euler-Lagrange equations. The irreducible components of the
difference tensor $D$ of the same type as $\gamma$ are the three
traces of $D$.  For our purposes here, we will consider the case
where the quadratic expression does not have any cross terms so that
these three traces are zero as well.  We comment that two of these
traces are in the metric compatibility part of the difference tensor
and one is in the torsion part of the difference tensor.  We
consider this case because it is representative of the general case
in that a generic quadratic expression will also lead to the
Einstein-Klein-Gordon equation with a cosmological constant.

Hence, in this representative case,
\begin{equation}\label{eqn:D=gamma}
   D_{ijk} = \gamma_{ijk},
\end{equation}
which defines the connection $\nabla$ in terms of $\gamma$ and $g$
by equations \ref{eqn:defDlowered} and \ref{eqn:ChristoffelLowered}.
The action functional for $\gamma$ and $g$ then reduces to
\begin{eqnarray}
   F_{\Phi,U}(g,\nabla)
   &=&  \int_U (R - 2\Lambda - \frac{c_3}{24}|d\gamma|^2 -
   \frac{c_4}{6} |\gamma|^2)  \; dV \\
   &=& \int_U (R - 2\Lambda - c_3|d\gamma|_{4-form}^2 -
   c_4 |\gamma|_{3-form}^2)  \; dV
\end{eqnarray}
where we note that the norm of a fully antisymmetric $k$-tensor
$\omega$ is $k!$ times that of the definition of the norm of
$\omega$ as a $k$-form. Next let
\begin{equation} \label{eqn:defv}
   \gamma = *(v^*)
\end{equation}
where $v$ is a vector field, $v^*$ is the $1$ form dual to $v$, and
$*$ is the Hodge star operator which takes $k$ forms to $(4-k)$
forms in dimension $4$.  We note that $|\gamma|_{3-form}^2 = -
|v|^2$ and $|d\gamma|_{4-form}^2 = -(\nabla \cdot v)^2$ because of
the Lorentzian metric.  With the above substitution, the action
functional becomes
\begin{equation}\label{eqn:action}
 F_{\Phi,U}(g,\nabla) = \int_U (R - 2\Lambda + c_3(\nabla \cdot v)^2
 + c_4 |v|^2)  \; dV,
\end{equation}
where $\nabla \cdot v$ denotes the divergence of $v$.  This amounts
to a change of variables for the action from $(g,\gamma)$ to
$(g,v)$.

Computing the Euler-Lagrange equation for variations of the vector
field $v$ is straightforward using the divergence theorem, but
computing the Euler-Lagrange equation for variations in the metric
$g$ requires the following formulas.  Let $g(s)$ be a family of
metrics, and let $\frac{d}{ds} g = h$. Then direct calculation shows
that
\begin{eqnarray}
   \frac{d}{ds}\; R &=& -\langle Ric, h \rangle + \nabla \cdot \left(
   \nabla \cdot h - \nabla(\langle g, h \rangle)\right) \\
   \frac{d}{ds}\; dV &=& \frac12 \langle g, h \rangle \; dV \\
   \frac{d}{ds}\; \nabla \cdot v &=& \frac12 \langle v, \nabla \langle g, h \rangle
   \rangle \\
   \frac{d}{ds}\; |v|^2 &=& \langle \nu \otimes \nu, h \rangle
\end{eqnarray}
where $\nu = v^*$ is the $1$-form dual to the vector field $v$.  It
follows that the Euler-Lagrange equations for the action in equation
\ref{eqn:action} are
\begin{eqnarray}
   G + \Lambda g &=& c_4 \; \nu \otimes \nu - \frac12\left(c_3 (\nabla \cdot v)^2 + c_4
   |v|^2 \right)g \\
   \nabla (\nabla \cdot v) &=& \frac{c_4}{c_3} v.  \label{eqn:EL2}
\end{eqnarray}

For the dominant energy condition to be satisfied, we need $c_3, c_4
\ge 0$.  To arrive at a nontrivial equation for $v$ we need $c_3 \ne
0$ and to arrive at a deterministic equation for $v$ we need $c_4
\ne 0$. Hence, let's take $c_3, c_4 > 0$. In this case, these
equations are more easily understood if we define a new function $f$
such that
\begin{equation}
   f = \left(\frac{c_3}{c_4}\right)^{1/2} \nabla \cdot v.
\end{equation}
so that by equation \ref{eqn:EL2}
\begin{equation}\label{eqn:v=gradf}
   v = \left(\frac{c_3}{c_4} \right)^{1/2} \nabla f.
\end{equation}
Using the above two equations and substituting into the original
Euler-Lagrange system of equations, we get that
\begin{eqnarray}\label{eqn:1}
   G + \Lambda g &=& c_3 \; \left\{ df \otimes df - \frac12\left(|df|^2 + \frac{c_4}{c_3} f^2 \right)g \right\} \\
   \Box f &=& \frac{c_4}{c_3} f  \label{eqn:2}
\end{eqnarray}
has a solution if and only if the original system does.  Note that
$\Box$ is the Laplacian with respect to the Lorentzian metric $g$
so that, modulo reparameterizations of the metric, the above
equations are hyperbolic for the scalar function $f$ and the metric
$g$.  We note that the above system has its own action
\begin{equation}\label{eqn:alt
action}
 \tilde F_{\Phi,U}(g,\nabla) = \int_U (R - 2\Lambda - c_3 |df|^2
 - c_4 f^2)  \; dV .
\end{equation}
Note that $\tilde F$ is not quite the same action as $F$ since the
last two terms have opposite signs.

By equations \ref{eqn:defDlowered}, \ref{eqn:ChristoffelLowered},
\ref{eqn:D=gamma}, \ref{eqn:defv}, and \ref{eqn:v=gradf}, the
connection will have components
\begin{equation}\label{ConnectionFormula}
\Gamma_{ijk} = \left(\frac{c_3}{c_4}\right)^{1/2} (*df)_{ijk} +
   \frac12 \left(g_{ik,j} + g_{jk,i} - g_{ij,k} \right)
\end{equation}
in this representative case.  We comment that if we had chosen a
different quadratic expression of $D$ before ``with cross terms''
between the fully antisymmetric $\gamma$ part of $D$ and the trace
parts of $D$, this formula would be modified. However, in all of
these cases, we still get the Einstein-Klein-Gordon equations with a
cosmological constant being the relevant Euler-Lagrange equations.
The only difference is the interpretation of what the connection is
in equation \ref{ConnectionFormula}. However, so far it is not clear
how the connection manifests itself physically, other than
gravitationally, which is effectively the same in all of these
cases.

As a last step, we introduce new constants $\Upsilon$ and $\mu_0$
defined such that
\begin{equation}
   \frac{c_4}{c_3} = \Upsilon^2 \;\;\;\mbox{ and }\;\;\;
   c_4 = 16\pi \mu_0.
\end{equation}
Plugging these new constants into equations \ref{eqn:1} and
\ref{eqn:2} give us the Einstein-Klein-Gordon equations with a
cosmological constant in geometrized units with the gravitational
constant and the speed of light set to one, which are
\begin{eqnarray}\label{eqn:EE2}
   G + \Lambda g &=& 8 \pi \mu_0 \; \left\{ 2 \frac{ df \otimes df}{\Upsilon^2}
   - \left(\frac{|df|^2}{\Upsilon^2} + f^2 \right)g \right\} \\
   \Box f &=& \Upsilon^2 f    \label{eqn:KG2}
\end{eqnarray}
where $G$ is the Einstein curvature tensor, $f$ is the scalar field
representing dark matter, $\Lambda$ is the cosmological constant,
and $\Upsilon$ is a new fundamental constant of nature whose value
has yet to be determined. The other constant $\mu_0$ is not a
fundamental constant of nature as it can easily be absorbed into
$f$, but simply is present for convenience and represents the energy
density of an oscillating scalar field of magnitude one which is
solely a function of $t$.  It is perfectly fine to set $\mu_0$ equal
to one, just as we have done to the speed of light and the
gravitational constant.  Also, note that since
\begin{equation}
   \Box f = \nabla \cdot \nabla f = \frac{1}{\sqrt{|g|}} \partial_i
   \left(\sqrt{|g|} \; g^{ij} \partial_j f \right) = g^{ij} \left(
   \partial_i \partial_j f - \bar\Gamma_{ij}^{\hspace{.1in}k} \partial_k f \right)
\end{equation}
equation \ref{eqn:KG2} is hyperbolic in $f$ when the metric has
signature $(-+++)$.

\section{A Bigger Geometric Theory}\label{BiggerTheory}

There are compelling geometric and physical motivations to
generalize Axiom \ref{A1} as follows.

\begin{axiom}\label{A2}
For all coordinate charts $\Phi : \Omega \subset N \rightarrow R^4$
and open sets $U$ whose closure is compact and in the interior of
$\Omega$, $(g,\nabla)$ is a critical point of the functional
\begin{equation}
   F_{\Phi,U}(g,\nabla) = \int_{\Phi(U)}
   \mbox{Quad}_{M \cup C}(M' \cup M \cup C' \cup C)
   \; dV_{R^4}
\end{equation}
with respect to smooth variations of the metric and connection
compactly supported in $U$, for some fixed quadratic functional
$Quad_{M \cup C}$ with coefficients in $M \cup C$.
\end{axiom}
We remind the reader that
\begin{equation}
M = \{g_{ij}\} \;\;\;\mbox{ and }\;\;\; C = \{\Gamma_{ijk}\}
\;\;\;\mbox{ and }\;\;\; M' = \{g_{ij,k}\} \;\;\;\mbox{ and }\;\;\;
C' = \{\Gamma_{ijk,l}\}
\end{equation}
as before.

The geometric motivation for this generalization is that Axiom
\ref{A1} actually changes completely if we redefine $C =
\{\Gamma_{ij}^{\;\;\;k}\}$  and $C'$ similarly.  That is, we get
different theories resulting from the axiom depending on which of
the tensor indices (the first and the third) of $\Gamma$ we raise or
lower.  The theory with all lowered indices seems most interesting,
so that has been the topic of this paper.  The other theories are
interesting too, but result in a slight modification of the
Einstein-Maxwell equations with a cosmological constant, which is
less exciting than a geometric theory of dark matter. Of course
these other theories with raised indices are interesting to study as
well.

On the other hand, if the quadratic expression $\mbox{Quad}_{M \cup
C}$ is allowed to have coefficients in $C$ as well as $M$, then this
issue of raising and lowering indices becomes irrelevant as the same
theory is produced in all cases.  Furthermore, this generalized
axiom also generalizes these other theories into one bigger theory.

Hence, it is very natural geometrically to prefer this second axiom.
We would have introduced this axiom earlier, but we did not want to
complicate things unnecessarily.  Of course everything we have
already studied about scalar field dark matter will still be a part
of this generalized theory, but there will be (at least) two
additional matter fields as well, which is a physical motivation for
being intrigued by this latest axiom.

We observe that actions of the form
\begin{equation}\label{conjecture2}
   F_{\Phi,U}(g,\nabla) = \int_U (c R - 2\Lambda - \mbox{Quad}_{g}(\bar\nabla T)
   - \tilde {\mbox{Quad}}_{g}(D)) \; dV,
\end{equation}
up to an irrelevant boundary term, are allowed by these axioms, and
we conjecture that these are all of the actions compatible with
Axiom \ref{A2}. We leave this as another interesting conjecture to
study.

If the above conjecture is correct, then the resulting matter fields
generically will be the irreducible components of the torsion tensor
$T$. The fully antisymmetric part, studied in this paper, leads to a
scalar field model of dark matter. The trace part, which will be a
$1$ form, is reminiscent of the vector potential $1$ form $A$ of
electromagnetism.  The final matter field, which is everything left
over, can be thought of as a $3$ tensor which is antisymmetric in
the first two indices, antisymmetrizes to zero, and has zero trace.
We have no idea how to interpret this physically, but this is
certainly an interesting direction to pursue.

\section{The Coldness of Scalar Field Dark Matter in the
Homogeneous and Isotropic Case}\label{SFCDM}

In this appendix we prove the following theorem, stated originally
in section \ref{cosmologicalpredictions}.

\begin{theorem}
Suppose that the spacetime metric is both homogeneous and isotropic,
and hence is the Friedmann-Lema\^{i}tre-Robertson-Walker metric
$-dt^2 + a(t)^2 ds_\kappa^2$, where $ds_\kappa^2$ is the constant
curvature metric of curvature $\kappa$. If $f(t,\vec{x})$ is a
real-valued solution to the Klein-Gordon equation (equation
\ref{eqn:KG} with mass term $\Upsilon$) with a stress-energy tensor
which is isotropic, then $f$ is solely a function of $t$.
Furthermore, if we let $H(t) = a'(t)/a(t)$ be the Hubble constant
(which of course is actually a function of $t$), and $\rho(t)$ and
$P(t)$ be the energy density and pressure of the scalar field at
each point, then
\begin{equation}
   \frac{\bar{P}}{\bar{\rho}} = \frac{\epsilon}{1 + \epsilon}
\end{equation}
where
\begin{equation}
   \epsilon = - \frac{3\overline{H'}}{4\Upsilon^2}
\end{equation}
and
\begin{equation}
   \bar{\rho} = \frac{1}{b-a} \int_a^b \rho(t) \;dt , \hspace{.2in}
   \bar{P} = \frac{1}{b-a} \int_a^b P(t) \;dt, \hspace{.2in}
   \overline{H'} = \frac{\int_a^b H'(t) f(t)^2 \;dt}
   {\int_a^b f(t)^2 \;dt} ,
\end{equation}
where $a,b$ are two zeros of $f$ (for example, two consecutive
zeros).
\end{theorem}

\emph{Proof:} We first prove that $f$ is only a function of $t$.  By
equation \ref{eqn:EE}, the stress-energy tensor $T$ of the real
scalar field is given by
\begin{equation}
   T / \mu_0 = 2 \frac{ df \otimes df}{\Upsilon^2}
   - \left(\frac{|df|^2}{\Upsilon^2} + f^2 \right)g .
\end{equation}
For this tensor to be isotropic, the first term in the above
expression must be isotropic.  Hence, $df$ must be a multiple of
$dt$.  Integrating $df$ along the level sets of $t$ proves that $f$
is constant on the level sets of $t$.  Hence, $f$ is a function of
$t$ (which proves that the stress-energy tensor of the scalar field
is homogeneous as well).

Then since $df = f'(t)dt$ and hence $|df|^2 = -f'(t)^2$, we get
\begin{equation}
   \rho(t) / \mu_0 = f(t)^2 + \frac{f'(t)^2}{\Upsilon^2}
   \;\;\mbox{ and }\;\;
   P(t) / \mu_0 = -f(t)^2 + \frac{f'(t)^2}{\Upsilon^2}
\end{equation}
where $\rho = T(\partial_t,\partial_t)$ is the energy density and $P
= T(e,e)$ is the pressure of the scalar field, where $e$ is any unit
vector perpendicular to the time direction.

By equation \ref{eqn:KG} and the formula for the metric Laplacian,
\begin{eqnarray}
\Upsilon^2 f(t) &=& \Box_g f \\
   &=& |g|^{-1/2} \partial_i
   \left(|g|^{1/2} \; g^{ij} \partial_j f \right) \\
   &=& a(t)^{-3} \partial_t ( -a(t)^3 f'(t)) \\
   &=& - f''(t) - 3 H(t) f'(t)
\end{eqnarray}
since $|g| = |\det(g)| = a(t)^6$ and $H(t) = a'(t)/a(t)$.  We can
then use this result in the calculation below,
\begin{eqnarray}
\int_a^b \frac{f'(t)^2}{\Upsilon^2} \;dt &=& \int_a^b
-\frac{f(t)f''(t)}{\Upsilon^2} \;dt \\
&=& \int_a^b f(t) \left( f(t) + \frac{3}{\Upsilon^2} H(t) f'(t) \right) \;dt \\
&=& \int_a^b \left(1 - \frac{3}{2\Upsilon^2} H'(t) \right) f(t)^2 \;dt \\
&=& \left( \int_a^b f(t)^2 \;dt \right) \left( 1 -
\frac{3\overline{H'}}{2\Upsilon^2}  \right)
\end{eqnarray}
where we have integrated by parts twice and have not picked up any
boundary terms since $f(a) = 0 = f(b)$.  It follows that
\begin{equation}
   \frac{b-a}{\mu_0} \cdot \bar{\rho} = \left( \int_a^b f(t)^2 \;dt \right) \left( 2 -
\frac{3\overline{H'}}{2\Upsilon^2}  \right)
\end{equation}
and
\begin{equation}
   \frac{b-a}{\mu_0} \cdot \bar{P} = \left( \int_a^b f(t)^2 \;dt \right) \left(  -
\frac{3\overline{H'}}{2\Upsilon^2}  \right) ,
\end{equation}
proving the theorem.

We comment that the complex scalar field case is very similar to the
above statement since a complex scalar field with quadratic
potential is precisely equivalent to two independent real scalar
fields (given by the real and complex parts of the complex scalar
field).   Complex scalar fields can be defined which have much less
oscillations of the pressure (which occurs when the real and complex
parts of the field are half a period out of phase with one another
with the same magnitudes), but since real fields are special cases
of complex fields, for example, this phenomenon is not generically
the case.

We also comment about out assumption that the stress-energy tensor
of the real scalar field is isotropic.  If this scalar field were
the only matter in the universe, then it would follow from the
Einstein equation that the stress energy tensor of the scalar field
was both homogeneous and isotropic, assuming the spacetime was both
of these.  However, we want our theorem to be compatible with other
matter as well.  In this case, as long as the other matter fields
have isotropic stress-energy tensors, it still follows from the
Einstein equation that the stress-energy tensor of the scalar field
dark matter is isotropic. Specifically, since dark energy due to the
cosmological constant is always isotropic (and homogeneous), it
follows that a model universe with just dark matter and dark energy
which was homogeneous and isotropic would imply that the scalar
field dark matter's stress-energy tensor was isotropic. Hence, this
is a very general assumption when modeling homogeneous, isotropic
spacetimes.

\end{appendix}


\end{document}